\documentclass[a4paper,11pt]{article}

\usepackage[utf8]{inputenc}
\usepackage[T1]{fontenc}

\usepackage{jcapmod}

\usepackage{subcaption}
\usepackage{amsmath,amsfonts,amssymb,macros}
\usepackage{physics}
\usepackage{mathtools}
\usepackage{bm}
\usepackage{cancel}
\usepackage[table]{xcolor}
\usepackage[most]{tcolorbox}
\usepackage[unicode=true,pdfusetitle,
 bookmarks=true,bookmarksnumbered=false,bookmarksopen=false,
 breaklinks=false,pdfborder={0 0 1},backref=false,colorlinks=true]
 {hyperref}
\hypersetup{
 linkcolor=blue, citecolor=cyan, urlcolor=red, filecolor=blue}

\usepackage{float}
\usepackage{acronym}
\usepackage{multicol}
\usepackage{array}
\usepackage{multirow}

\usepackage{graphicx}

\usepackage[a4paper,margin=2.5cm]{geometry}

\newcommand{\re}{\mathrm{e}}
\numberwithin{equation}{section}

\usepackage{tikz}
\usetikzlibrary{shapes.geometric,arrows}
\tikzstyle{arrow} = [thick,->,>=stealth]

\usepackage[normalem]{ulem}

\begin{document}

\pagenumbering{roman}

\begin{titlepage}

\baselineskip=15.5pt \thispagestyle{empty}

\begin{center}
    {\fontsize{19}{24}\selectfont \bfseries Modified gravity approaches to the cosmological constant problem}
\end{center}

\vspace{0.1cm}

\begin{center}
    {\textbf{Foundational Aspects of Dark Energy (FADE) Collaboration}}
\end{center}

\renewcommand*{\thefootnote}{\fnsymbol{footnote}}

\begin{center}
    {\fontsize{12}{18}\selectfont Heliudson Bernardo,$^{1}$ \footnote{heliudson.deoliveirabernardo@mcgill.ca}  Benjamin Bose,$^{2,3,4}$ \footnote{ben.bose@ed.ac.uk}  Guilherme Franzmann,$^{4,5}$ \footnote{guilherme.franzmann@su.se}  Steffen Hagstotz,$^{6,7}$ \footnote{steffen.hagstotz@lmu.de}  Yutong He,$^{5,8}$  \footnote{yutong.he@su.se} Aliki Litsa,$^{8}$ \footnote{aliki.litsa@fysik.su.se}  and Florian Niedermann$^{5}$ \footnote{florian.niedermann@su.se} }
\end{center}

\renewcommand*{\thefootnote}{\arabic{footnote}}
\setcounter{footnote}{1}


\vspace{-0.3cm}
\begin{center}
\rule{14cm}{0.08cm}
\end{center}

    \vskip8pt
   \noindent \textsl{$^1$ Department of Physics, Ernest Rutherford Physics Building, McGill University,\\
        3600 Rue Universit\'e, Montr\'eal, Qu\'ebec H3A 2T8, Canada} 
        
    \vskip8pt
   \noindent 
    \textsl{$^2$ Département de Physique Théorique, Université de Genève, \\ 24 quai Ernest Ansermet,
    1211 Genève 4, Switzerland}
    
    \vskip8pt
   \noindent 
    \textsl{$^3$ Institute for Astronomy, University of Edinburgh, Royal Observatory, Blackford Hill, Edinburgh, EH9 3HJ, U.K}

    \vskip8pt
   \noindent 
    \textsl{$^4$ Basic Research Community for Physics e.V., Mariannenstraße 89, Leipzig, Germany}

\vskip8pt
   \noindent 
    \textsl{$^5$ Nordita, \\
KTH Royal Institute of Technology and Stockholm University,\\
Hannes Alfvéns v\"ag 12, SE-106 91 Stockholm, Sweden}

    \vskip8pt
   \noindent 
    \textsl{$^6$  Universitäts-Sternwarte, Fakultät für Physik, Ludwig-Maximilians Universität München, \\
Scheinerstraße 1, D-81679 München, Germany}

\vskip8pt
   \noindent 
    \textsl{$^7$  Excellence Cluster ORIGINS, Boltzmannstraße 2, D-85748 Garching, Germany}
    
    \vskip8pt
   \noindent 
    \textsl{$^{8}$  The Oskar Klein Centre for Cosmoparticle Physics, Stockholm University \\ Roslagstullsbacken 21A, SE-106 91 Stockholm, Sweden}
    

\vspace{2cm}

\hrule
\begin{center}
    Invited review for Special Issue \href{https://www.mdpi.com/journal/universe/special_issues/cosmological_constant_problem}{``Cosmological Constant''} in  \href{https://www.mdpi.com/journal/universe}{\textit{Universe.}}
\end{center}
\hrule

\newpage
\setcounter{page}{1}

\hrule
\vspace{0.3cm}

\noindent {\bf Abstract} 

The cosmological constant and its phenomenology remain among the greatest puzzles in theoretical physics. We review how modifications of Einstein's general relativity could alleviate the different problems associated with it that result from the interplay of classical gravity and quantum field theory. We introduce a modern and concise language to describe the problems associated with its phenomenology, and inspect no-go theorems and their loopholes to motivate the approaches discussed here. Constrained gravity approaches exploit minimal departures from general relativity; massive gravity introduces mass to the graviton; Horndeski theories lead to the breaking of translational invariance of the vacuum; and models with extra dimensions change the symmetries of the vacuum. We also review screening mechanisms that have to be present in some of these theories if they aim to recover the success of general relativity on small scales as well. Finally, we summarise the statuses of these models in their attempt to solve the different cosmological constant problems while being able to account for current astrophysical and cosmological observations.

\vskip10pt
\hrule
\vskip10pt

\noindent \textbf{Acronyms}

\begin{multicols}{2}

\begin{acronym}[SMPP]
\acro{CCP}{Cosmological Constant Problem}
\acro{UV}{Ultraviolet}
\acro{EFT}{Effective Field Theory}
\acro{GR}{General Relativity} 
\acro{EMT}{Energy-Momentum Tensor}
\acro{CDM}{Cold Dark Matter}
\acro{CC}{Cosmological Constant}
\acro{DE}{Dark Energy}
\acro{SM}{Standard Model of Particle Physics}
\acro{QFT}{Quantum Field Theory}
\acro{QCD}{Quantum Chromodynamics} 
\acro{DEP}{Dark Energy Problem}
\acro{IR}{Infrared}
\acro{DGP}{Dvali-Gabadadze-Porrati gravity}
\acro{CMB}{Cosmic Microwave Background}
\acro{CHC}{Cosmic History Constraint}
\acro{AC}{Astrophysical Constraints}
\acro{SLED}{Supersymmetric Large Extra Dimension}

\end{acronym}
\end{multicols}

\vskip10pt
\hrule
\vskip10pt

\noindent \textbf{Disclaimer}: This review does not intend, in any way, to present an exhaustive collection of modified gravity approaches. 
Instead, the models were chosen to exemplify the different ways one can bypass no-go theorems surrounding the cosmological constant problem, and they are subjected to the authors' personal biases and preferences on the subject.

\end{titlepage}

\tableofcontents
\setcounter{page}{2}

\clearpage



\pagenumbering{arabic}
\setcounter{page}{1}

\section{Introduction}
\label{sec:introduction}

The cosmological constant problem (CCP) is one of the most persistent puzzles in theoretical physics. It appears at the interface between quantum mechanics and gravity, 
and it seemingly contradicts one of the major building blocks of modern physics, which is that scales in nature decouple. We hope that it is possible to understand physics at low energies without a detailed knowledge of the ultraviolet (UV)-complete theory. Formally, this decoupling of scales is expressed through the enormously successful framework of effective field theories (EFT), which is challenged by the CCP.

Viewed solely as a metric theory for spacetime, Einstein's general relativity (GR) allows for a cosmological constant $\Lambda$ that is left unfixed by the principles and symmetries of the theory. Such a parameter may compete with the energy-momentum tensor (EMT) of local sources in the equations of motion to produce certain solutions, as was the case of Einstein's cosmological static solution \cite{Einstein1917, ORaifeartaigh:2017uct}. In fact, the weak strength of the gravitational interaction is such that all local tests of GR are compatible with $\Lambda=0$, singling out cosmological observations to constrain $\Lambda$.

The current cosmological concordance model, $\Lambda$CDM, is fully based on classical GR and employs a cosmological constant (CC) to describe the present period of accelerated expansion necessary to fit both early- \cite{Planck:2018vyg} and late-time (e.g. \cite{Bautista:2020ahg,BOSS:2016wmc,Hildebrandt:2016iqg,DES:2022qpf,Brout:2022vxf,DES:2020cbm,SPT:2016izt}, and see \cite{Huterer:2017buf} for a recent review) observational data. Although there are other cosmological models in which the driver of the current epoch of accelerated expansion, herein referred to as dark energy (DE), is not described by a cosmological constant, standard GR with a CC is still the preferred model once many data sets are analysed together~\cite{Huterer:2017buf}. 

However, problems appear once we consider the standard model of particle physics (SM) together with GR. The SM is based on quantum field theory (QFT) in Minkowski spacetime. Within the QFT paradigm the local
energy-momentum tensor of field systems
might be non-zero even in the vacuum state, in which case they have the same structure as the EMT of a CC term~\cite{Birrell:1982ix}. Hence, the existence of the SM fields, inferred from local experiments, implies that there are quantum-matter induced contributions to the cosmological constant. This means that \emph{if} DE is a CC, then the $\Lambda$-fitted value (a.k.a. effective cosmological constant) also includes contributions from the SM quantum fields, generically referred to as \emph{the vacuum energy density}, $\rho_{\text{vac}}$.

Typically, any vacuum energy contribution is enormous and would completely dominate the gravitational dynamics of the Universe. This was first noted by Nernst in 1916~\cite{Nernst:1916}, and famously Pauli quipped that ``the Universe would not even reach to the moon'' \cite{Enz_Thellung:1960}\footnote{Lenz~\cite{Lenz:1926} was the first to make such a comparison (see e.g. \cite{Peruzzi_Realdi:2011,Kragh2012, kragh2014weight} and references therein for historical notes).} given the large vacuum energy from quantum fields. The problem was formally analysed in a cosmological context by Zel'dovich in the 70s \cite{Zeldovich:1967gd, Zeldovich:1968ehl}. By the late 80s, due to the perturbative UV incompleteness of GR \cite{tHooft:1974toh, Goroff:1985th}, it was expected that a quantum version of the theory would fix $\Lambda$. Prior to the discovery of the accelerated cosmological expansion, popular attempts of doing so were based on supergravity theories, which typically do not allow a positive cosmological constant. The seminal review \cite{Weinberg:1988cp} summarises the status of the problem before the early 90s.

Perhaps the most straightforward procedure for solving the mismatch is to postulate a bare, purely classical contribution inherent to GR to the cosmological constant, $\Lamb$, that would cancel part of $\rho_{\text{vac}}/M^2_{\text{Pl}}$. However, the modern view of QFT within the effective field theory approach is that such a proposal is radiatively unstable or extremely fine-tuned, as discussed later.
There are many reviews on other approaches (see, for instance, \cite{Martin:2012bt,Copeland:2006wr}), and the present work aims at reviewing modified gravity theories to alleviate or solve the cosmological constant problem. As we will discuss in this review, the problem actually has different facets, and it is \emph{a priori} unclear whether all of them can be addressed at once or not.

Since the problem occurs when trying to calculate the gravitational effect of QFT vacuum energies, one approach to tackle the issue is to modify gravity itself. As shown in Sec. \ref{sec:problems} and \ref{sec:how_to_modify_gravity} there is little space to solve the CCP within classical GR. Setting aside anthropic arguments and discussions about fine-tuning \cite{Barrow:1986nmg,gribbin1989cosmic,Weinberg:1987dv,10.1093/mnras/274.1.L73,Martel:1997vi,Garriga:1999hu,Peacock:2007cw,SobralBlanco:2020too}, modifying GR not only offers a new view on the CCP but also provides new phenomenological signatures. Generally speaking, modified gravity theories introduce extra fields and/or constraints on GR such that the coupling of matter to the metric is modified and/or there are extra universal couplings with the extra fields in a way that is consistent with observations. It is natural then to ask if and how the CCP manifests itself in such theories, and this is one of the motivations for developing them.

Another motivation for studying modified gravity approaches is the fact that some of these theories are typical examples of how to evade Weinberg's no-go theorem, as discussed in Sec.~\ref{sec:no-go}, and its complement, reviewed in Sec.~\ref{sec:spectral}. We highlight possible loopholes in the assumptions of the no-go theorems in Sec.~\ref{sec:loop} and use them as a systematic guiding principle for the modified gravity theories considered throughout the review. The idea of this review is not to cover all modified gravity literature. Instead, we make a narrow choice of models that we deem promising and exemplify the discussed loopholes with an emphasis on how the CCP is addressed in each one of them. For other reviews on modified gravity and the cosmological constant issues, see \cite{Clifton:2011jh, Hinterbichler:2011tt, Koyama:2015vza, deRham:2014zqa, Weinberg:1988cp,Nobbenhuis:2004wn,Martin:2012bt,Polchinski:2006gy,Bousso:2007gp,Padilla:2015aaa}.  

We start by carefully introducing the different contributions and aspects of the cosmological constant problem in Sec.~\ref{sec:problems}. Any possible solution to the cosmological constant problem is severely constrained by powerful no-go theorems, which we recap in Sec.~\ref{sec:how_to_modify_gravity} together with potential ways to circumvent the theorems while taking into consideration various observational constraints. In Sec.~\ref{sec:wishlist}, we summarise the task at hand from both a theoretical and phenomenological point of view before diving into various approaches in Sec.~\ref{sec:mod_gravity_approaches}. We discuss the extent to which the various modified gravity theories can solve parts of the problem in Sec.~\ref{sec:disc}.

\paragraph{Conventions:} 

Unless otherwise stated, we set $c = \hbar = 1$ and use the metric signature $(-+++)$. The gravitational coupling is denoted by $\MPl^2$ in all sections, apart from Sec.~\ref{ssec:global_seq} and~\ref{sec:local_seq} on sequestering, where $\kap^2$ and $\kap^2(x)$ are used to indicate the promotion of the Planck mass to a variable. We use $\Lm$ or $S_{\rm m}$ for a generic matter component, and only specify its content as $\lll(...)$ or $S(...)$ when necessary.

\section{What is the problem after all?} \label{sec:problems}

Much discussion has been focused on the CCP in the past. There is some confusion in the literature about what the problem is, as there are different distinct issues that need to be tackled. To make it clear what the challenges are, and how modifications of gravity can help sort them out, we briefly review them below:

\begin{itemize}
    \item \textbf{The weight of vacuum.} The gravitating vacuum energy at the level of the classical Einstein equations receive contributions from the vacuum energy of the fields in the SM, $\rhovac$. QFT calculations of the latter for a given energy scale $\mu$, defined by the renormalization scale at which we trust our theories, seem to result in a very high value for the vacuum energy that scales as $\sim m^4 \ln (m^2/\mu^2)$, where $m$ is the mass of the particle for which the vacuum energy is being computed \cite{Martin:2012bt}. This typically differs by many orders of magnitude from the actual value associated with the accelerated expansion of the Universe at cosmological scales, which is roughly $(10^{-3} \, \text{eV})^4$ (the mass of the top quark, for instance, is roughly $10^{11} \, \text{eV}$, leading to a 56-orders-of-magnitude gap). One of the problems is to reconcile this discrepancy. Following standard practice, we call this the \textit{old cosmological constant problem} (old-CCP). \label{par:pr1}

    \item \textbf{Phase transitions.} The potential energy for a constant field configuration contributes as vacuum energy. But, for instance, the Higgs field's effective potential depends on the  background temperature. As the Universe cools down, the potential changes its shape, shifting its global minimum. Therefore, its contribution to the vacuum energy also changes. This problem is a classical instability due to the change in the contribution coming from the potential energy of the Higgs field before and after its phase transition. In fact, this is just one example of what happens generally for any phase transition, including the one due to quantum chromodynamics (QCD) and possibly others from unification theories. This has been called \textit{the classical cosmological constant problem} \cite{Martin:2012bt} (class-CCP). \label{par:p3}
    
    \item \textbf{Dark energy.} The Universe is undergoing a period of accelerated expansion \cite{SupernovaSearchTeam:1998fmf,SupernovaCosmologyProject:1998vns} that can be explained by a non-zero vacuum energy in the form of a positive CC. Thus, we need to explain where this positive-cosmological-constant-like term is coming from at the cosmological level when using Einstein equations. We will call this the \textit{dark energy problem} (DEP). \label{par:p4} 
    
    \item \textbf{UV Sensitivity.} Finally, another problem comes from the fact that the vacuum energy computed in QFT is UV sensitive, despite being possibly the most infrared (IR) quantity one could conceive (as it is a constant throughout spacetime). In particular, there are two ways in which the UV sensitivity of $\rhovac$ manifests: \label{par:pr2}
    
    \begin{itemize}
        \item It is directly connected with the Higgs' mass UV sensitivity. As we saw above, the vacuum energy scales with the mass of particles, and the Higgs' mass squared is itself highly UV sensitive (quadratic in cutoff). This is the hierarchy problem (see \cite{Koren:2020biu} for a recent review), and it manifests as an even worse sensitivity in the computation of the vacuum energy (since it is quartic in the cutoff); 
        
        \item As we change the QFT cutoff by increasing it to higher-energy scales, we might be able to excite new fields with higher masses, disturbing again the fixing of the CC done at lower scales. Thus, even if all the masses were not fine-tuned in the SM, $\rhovac$ would still be sensitive to new fields showing up at new higher-energy EFTs.
    \end{itemize}
    
    Thus, in short, once we change the energy scale in which we are computing the vacuum energy, the radiative corrections from higher-order loop corrections in QFT will shift the value of $\rhovac$. We will refer to this issue as the \textit{new cosmological constant problem} (new-CCP).
    
\end{itemize}

Note that these problems do not have the same nature. The weight of vacuum can be easily fixed with the presence of a bare CC in Einstein equations, which is then fixed by measurements (as any other fundamental constant of nature is fixed by applying the renormalisation program). This cancellation between the bare CC and the vacuum energy could be seen as a coincidence, but poses no problem to the theory. Thus, we take the stance that this has never been an issue, despite it still being widely referred to as such, both in the physics literature, but also in popularised accounts about the CCP.

However, fixing the value of the bare CC remains a potential issue as phase transitions unfold and shift its value, and the final value of the effective vacuum energy can be adjusted only \textit{once} by a bare CC. On the other hand, if an approach can guarantee that phase transitions do not spoil this adjustment, then naturally the final value of the effective CC could correspond to the one providing the correct phenomenology associated with DE, sorting out the DE problem. Thus, the old, the classical, and the dark energy problems can all be understood at the classical level of Einstein equations and are directly related to phenomenology. 

Note, however, that the dark energy problem may or may not be solved in these approaches. That depends on whether these models allow to have some residual-like CC term that would induce an accelerated expansion. Nonetheless, in principle this problem can be tackled independently by combining these approaches with quintessential models that would drive the dark energy dynamics (see \cite{Tsujikawa:2013fta} for a review). 

Finally, the UV sensitivity is a deeper issue. Within GR, even if one manages to classically solve the other problems, this sensitivity would force the solution to be fine-tuned (see also \cite{Burgess:2013ara}). As we consider models that modify gravity for which there is a self-tuning mechanism that effectively prevents the vacuum energy from gravitating, these models only \textit{directly} solve the old- and the classical-CCPs. By doing so, they tacitly accept that the UV sensitivity is not necessarily an issue, in the sense that they do not attempt at fixing it, leaving QFT calculations untouched. Instead, gravity is modified such that the breakdown of the UV-IR decoupling does not lead to dramatic observable effects, thereby disposing of the need for fine-tuning. 

\section{How to modify gravity}
\label{sec:how_to_modify_gravity}

There are two features shared by almost all modified gravity approaches to the CCPs.  
First, all models discussed in this review rely on the idea of making gravity less sensitive to the presence of vacuum energy. We will refer to this general idea as self-tuning and introduce it in Sec.~\ref{sec:self-tuning_intro}.
Second, self-tuning is typically implemented in terms of additional degrees of freedom, which in general mediate a gravitational force. This requires the presence of a mechanism that suppresses this additional force's phenomenology at observationally accessible scales. We provide a review of common screening mechanisms in Sec.~\ref{sec:screening}.

\subsection{Self-tuning} \label{sec:self-tuning_intro}
The key observation is that, due to the equivalence principle, gravity is sourced by vacuum energy, raising the need to cancel it against a highly fine-tuned cosmological constant (see Sec.~\ref{sec:problems}). Therefore, if we modify our gravitational theory in such a way that vacuum energy either drops out of Einstein equations or is strongly suppressed, this avoids the need to fine-tune the CC to make its value compatible with observations. Instead, we assume that the CC takes on a generic value set by the cutoff of the theory.  In other words, we accept that the cosmological constant is highly sensitive to unknown UV physics but have our gravitational theory turn a blind eye\footnote{An often echoed concern is that self-tuning would also spell the end for inflation. However, this assumes that sources behaving similar to a cosmological constant (such as an inflaton field in slow-roll) are equally decoupled from gravity, which does not need to be true. Sequestering in Sec.~\ref{sec:constrained_gravity} provides the cleanest example of a mechanism that only affects a true cosmological constant.}.

This rather general idea comes in many flavors and under different names. Common notions are ``self-tuning'', ``shielding of the CC'' or ``self-adjustement mechanism''. A model-independent implementation is provided by the ``degravitation proposal'' which promotes the Newtonian constant to a high-pass filter (see Sec.~\ref{sec:degravitation}). For simplicity, we will refer to the general idea as self-tuning but resort to the notion commonly used in the literature when reviewing individual models.

To explain the concept of self-tuning and get an idea of common obstacles, we will first consider a somewhat naive and insufficient implementation of self-tuning in Sec.~\ref{sec:toymodel}. In a more general context, the reason for its failure is formalised by Weinberg's no-go theorem reviewed in Sec.~\ref{sec:no-go}. A complementary argument based on a spectral decomposition will be provided in Sec.~\ref{sec:spectral}. This will guide us towards a small handful of loopholes that can tell us how to modify gravity in order to implement self-tuning. They are later used to motivate the different models reviewed in Sec.~\ref{sec:mod_gravity_approaches}. 

\subsubsection{A failed start}\label{sec:toymodel}

As a warm-up, we consider a simple scalar field toy model that illustrates the idea of self-tuning and highlights the essential difficulties we face when implementing it. We will then generalise the discussion, using model-independent arguments, when reviewing two complementary no-go's in the literature in Sec.~\ref{sec:no-go} and \ref{sec:spectral}.

Consider an action containing GR, a (dimensionless) scalar field $\phi$ and a matter sector $S_{\text{m}}[e^{\beta \phi} g_\munu,\Psi]$ with field content collectively denoted by $\Psi$ that exhibits a conformal coupling to $\phi$ (controlled by $\beta$),\footnote{We could also look at a general conformal coupling, but as it turns out the exponential case will be sufficiently interesting.}

\begin{equation}
S = \frac{\MPl^2}{2}\int d^4x\sqrt{-g}\, R
+ \int d^4 x\sqrt{-g}\Big(-\frac{M^2}{2}\p_\mu\phi\p^\mu\phi - V(\phi)\Big)
+ S_{\text{m}}[e^{\beta \phi} g_\munu,\Psi]\,,
\label{eqn:action_toy}
\end{equation}
where $M_\mathrm{Pl}$ is the Planck mass and $M$ sets the coupling strength of the scalar $\phi$. In other words, we can think of $\phi$ as mediating an additional gravitational force. Note that we define the theory in the `Einstein frame' where the action for the metric is the Einstein-Hilbert action. The gravitational equations of motion are
\begin{subequations} \label{eom_toy_mdoel}
\begin{align}
    M^2 \Box\phi &= V'(\phi) - \frac{\beta}{2} T \,,\label{eqn:toy_del_phi}\\
    \MPl^2 G_\munu  &= T_\munu^{(\phi)} + T_\munu \,, \label{eqn:toy_del_g}
\end{align}
\end{subequations}
where we defined the EMT of the scalar field as
\begin{equation} \label{eq:EMT}
T_\munu^{(\phi)} = M^2 \p_\mu\phi\p_\nu\phi + g_\munu\Big(-\frac{1}{2} M^2 g^{\alp\bet}\p_\alp\phi\p_\bet\phi - V(\phi)\Big)\,,
\end{equation}
and that of the matter sector as
\begin{equation} \label{eq:EMT_mat}
T_\munu = -\frac{2}{\sqrt{-g}}\frac{\del S_\mathrm{m}}{\del g^\munu}\,.
\end{equation}
Taking the trace of Eq.~\eqref{eqn:toy_del_g} returns
\begin{equation}\label{eq:trace}
\MPl^2 R  + T^{(\phi)} + T = 0\,.
\end{equation}
As usual, we expect the vacuum energy $\rhovac$ to contribute to the matter energy density. To make this more explicit, we can decompose
\begin{align}\label{eq:matterT}
T_\munu  = \tau_\munu -   \rho_\mathrm{vac}\, e^{2 \beta \phi} g_\munu \,,   
\end{align}
where $\tau_\munu $ corresponds to a localised source with asymptotic fall-off. In particular, $\tau_\munu =0 $ in the vacuum.

The old and new CCP in this language boil down to the statement that $\rho_\mathrm{vac}$, due to Eq.~\eqref{eq:trace}, needs to be finely tuned to a small value (irrespective of the cutoff of the theory or the loop-order) to avoid making $R$ unacceptably large. The classical CCP, on the other hand, is the observation that this tuning would be spoiled in a phase transition (unless the phase transition itself is tuned not to make a contribution to $\rho_\mathrm{vac}$). Of course, this reasoning only holds if there is no cancellation between $T^{(\phi)}$ and $T$ at play.  
Now, self-tuning is based on the the idea that such a cancellation does take place. As we will explain, it can be enforced through the ansatz (at this stage other definitions differing by gradient terms are possible too)
\begin{align}\label{self_tuning_toy_model}
T  + T^{(\phi)} = - V'(\phi)+\frac{\beta}{2} T\,.
\end{align}
For the mechanism to work, a sufficient assumption is that the scalar field vacuum is translationally invariant,
\begin{align} \label{toy_trans_inv}
\partial_\mu \phi_\mathrm{vac} = 0 \,.
\end{align}
The reason is that when the system settles to its vacuum configuration, the LHS of  Eq.~\eqref{eqn:toy_del_phi} vanishes, enforcing  $(V'(\phi) - \beta/2 T)  \to 0$. The self-tuning Eq.~\eqref{self_tuning_toy_model} along with Eq.~\eqref{eq:matterT} then imposes the condition $4 \rho_\mathrm{vac}\, e^{2\beta \phi} - T^{(\phi)}|_{\phi=\phi_\mathrm{vac}}=0$, which, in turn, cancels vacuum energy from Eq.~\eqref{eq:trace} [and similarly from Eq.~\eqref{eqn:toy_del_g}]. In particular, if the vacuum energy changes during a phase transition or by integrating in or out new particles, leading to $\rho_\mathrm{vac}\, \to \hat \rho_\mathrm{vac}\, =\rho_\mathrm{vac} + \Delta \rho_\mathrm{vac} $, the field $\phi$ is pushed away from its stationary point and settles to a new configuration $\hat \phi_\mathrm{vac}$, which now imposes $ 4 \hat \rho_\mathrm{vac}\,e^{\beta \hat \phi_\mathrm{vac}} - T^{(\phi)}|_{\phi=\hat \phi_\mathrm{vac}}=0$. As a consequence, in the absence of localised matter sources, i.e.,~for $\tau_\munu =0$, the Minkowski metric remains a solution despite the presence of huge vacuum energy. This is indeed the definition of \textit{self-tuning}.

In short, we have -- somewhat naively -- imposed self-tuning by supplementing the system~\eqref{eom_toy_mdoel} with the ad-hoc Eq.~\eqref{self_tuning_toy_model}. Can this be done consistently without over-constraining the system? To find out, we try to derive a potential $V(\phi)$ such that all equations can be satisfied simultaneously.
Translational invariance in Eq.~\eqref{toy_trans_inv} implies that close to the vacuum configuration we can neglect gradients and time derivatives; specifically $\partial_\mu \phi \partial_\nu \phi \ll V(\phi)$ and thus $T^{(\phi)} \simeq -4V(\phi)$. Substituting this back into the self-tuning Eq.~\eqref{self_tuning_toy_model} and using Eq.~\eqref{eq:matterT} yields
\begin{equation}\label{eq:self-tuning2}
4V(\phi)-V'(\phi) \approx - 4 \left( 1- \frac{\beta}{2} \right) \rho_\mathrm{vac}\,e^{2 \beta \phi}  \,,
\end{equation}
which is a differential equation for $V(\phi)$. Its general solution takes the form
\begin{equation}\label{pot_self_tiuning}
V(\phi)\simeq
\begin{cases}
 - \rho_\mathrm{vac}\, e^{2 \beta \phi} + V_0 e^{4\phi} &\quad \text{for} \quad \beta \neq 2\\
  V_0 e^{4\phi} &\quad \text{for} \quad \beta = 2
\end{cases}\,,
\end{equation}
where $V_0$ is an integration constant. There are two interesting cases: 

\begin{itemize}
\item For $\beta \neq 2$, the potential directly depends on $\rho_\mathrm{vac}\,$ in such a way that it exactly cancels the vacuum energy in both equations in Eq.~\eqref{eom_toy_mdoel} [or Eq.~\eqref{eq:trace} equivalently]. This is fine-tuning again and it would not be robust under a change of the loop-order or a shift in the EFT cut-off. And it also means the cancellation is no longer dynamical. Therefore this model does not address any of the CCPs discussed in Sec.~\ref{sec:problems}.

\item For $\beta = 2$, the potential does not depend on $\rho_\mathrm{vac}$. Unfortunately, we still have arrived at an impasse.  Our self-tuning mechanism assumed that $\phi$ settles down to a constant vacuum configuration $\phi_\mathrm{vac}$. However, the potential in Eq.~\eqref{pot_self_tiuning} has no minimum at finite field values. Instead, it is a run-away potential, which approaches its minimum only asymptotically as $\phi\map-\infty$. Setting $V_0=0$ to avoid this conclusion would constitute a fine-tuning again. The reason is that $V_0$ receives huge (UV-sensitive) radiative corrections because only the combination $V_0+ \rho_\mathrm{vac}$ contributes to Eq.~\eqref{eom_toy_mdoel}.
\end{itemize}

We could hope that the running itself is not a problem. After all, it seems to exponentially suppress the vacuum energy contribution to $T  \supset  \rho_\mathrm{vac}\, e^{4 \phi} g_\munu$ as $\phi\map-\infty$. The problem with that is that not only the vacuum energy is getting suppressed. For example, the conformal coupling with $\phi$ implies that the Higgs mass term is replaced through $\mH^2 H^2 \to \mH^2 H^2 e^{4\phi}$. Now, as $\phi$ runs down its potential, the Higgs mass goes to zero, which -- unless the running can be made very slow -- is not compatible with observations.  So neither setting $V_0=0$ nor accepting the running seems to be a valid option. In the next section, we will find that this failure can be explained by a very general symmetry argument.

\subsubsection{Weinberg's argument} \label{sec:no-go} 

Weinberg in his seminal paper on the cosmological constant problem investigates self-tuning  and finds that under a few key assumptions, self-tuning is not possible without fine-tuning~\cite{Weinberg:1988cp}. Here, we review his argument (also see~\cite{Burgess:2013ara,Padilla:2015aaa}).

We consider the Lagrangian $\lll(g_\munu,\phi_i)$
of a metric field $g_\munu$ and a set of self-tuning scalars $\phi_i$. The presence of additional matter fields $\Psi$ that give rise to vacuum energy is understood. This approach covers the toy model from the previous section but is far more general. In particular, it does not assume the gravitational sector to be described by GR or canonical kinetic terms for $\phi_i$.
What we do assume is that the vacuum of the theory is translationally invariant,
\begin{align}\label{eq:key_assumption}
(\phi_\mathrm{vac})_i = \mathrm{const} \quad  \text{and} \quad (g_\mathrm{vac})_\munu = \mathrm{const}\,.
\end{align}
While this is in agreement with our expectation from typical field theories,\footnote{For example, the Higgs vacuum is $H_\mathrm{vac} = (0, v_H/\sqrt{2})^T = \mathrm{const}$, with vacuum expectation value $v_H = 246\,\mathrm{GeV}$.} we will discuss an example of self-tuning in Sec.~\ref{sec:massive_gravity} and Sec.~\ref{sec:self-tuning} where this assumption is violated. In any event, general coordinate invariance tells us that $\mathcal{L}$ has to transform as a density (to make $\mathrm{d}^4 x \mathcal{L}$ invariant). This is guaranteed if $\mathcal{L}$ takes on the form~\cite{Padilla:2015aaa}\footnote{We note that this is  not the only object that transforms as a density. Instead, we can use $\epsilon^{\mu\nu\alpha\beta} A_{\mu\nu\alpha\beta}$ as the integration measure, where $\epsilon^{\mu\nu\alpha\beta}$ is the Levi-Civita symbol and $A^{\mu\nu\alpha\beta}$ a 4-form field. This is the loophole exploited by some of the approaches discussed in Sec.~\ref{sec:constrained_gravity}.}
\begin{align} \label{eq:L_vac}
    \mathcal{L}_\mathrm{vac} \equiv \mathcal{L}((g_\mathrm{vac})_\munu,(\phi_\mathrm{vac})_i) = \sqrt{-g_\mathrm{vac}}\, \rho_\mathrm{vac}((\phi_\mathrm{vac})_i, \,\ldots)\,,
\end{align}
where the ellipsis indicates the dependence on further SM vacuum expectation values. Note that $\rho_\mathrm{vac}$ as defined here also includes the contribution from a bare cosmological constant and the self-tuning field's potential.

In this general context, self-tuning can be realised through the equation
\begin{align}\label{eq:self-tuning_general}
   2g_\munu\pfrac{\lll}{g_\munu} = \sum_i f_i(\phi)\pfrac{\lll}{\phi_i}\,,
\end{align}
where $f_i$ are a set of general functions that depend on the self-tuning fields $\phi_i$. As a quick sanity check, with these definitions, we recover Eq.~\eqref{eq:self-tuning2} from Eq.~\eqref{eqn:action_toy} for $f_1 = -4$. The idea is the same as before. The $\phi_i$ are subject to their equations of motion,
\begin{align}
    \pfrac{\lll}{\phi_i}- \partial_\mu  \pfrac{\lll}{(\partial_\mu \phi_i)} =0\,.
\end{align}
Therefore, as the $\phi_i$ approach their constant vacuum value, $\partial \mathcal{L} / \partial \phi_i \to 0$, which due to Eq.~\eqref{eq:self-tuning_general} implies $g_\munu \partial{\mathcal{L}_\mathrm{vac}}/ \partial g_\munu = 0$. From Eq.~\eqref{eq:L_vac}, we see that this requires the vanishing of $\rho_\mathrm{vac}$. It indeed looks as if the $\phi_i$ have ``self-tuned'' to make the vacuum energy vanish. 

To learn more about the viability of the mechanism, we further  constrain the functional form of $\mathcal{L}_\mathrm{vac}$. The key observation is that Eq.~\eqref{eq:self-tuning_general} implies the scaling symmetry (with $\epsilon \ll 1$)
\begin{equation}
\del_\eps (g_\mathrm{vac})_\munu = 2\eps (g_\mathrm{vac})_\munu \quad \text{and} \quad
\del_\eps (\phi_\mathrm{vac})_i = -\eps f_i(\phi_\mathrm{vac})\,,
\label{eqn:no_go_new_scale_sym}
\end{equation}
under which the vacuum theory is invariant. Indeed, it is easy to check that $\delta_\epsilon \mathcal{L}_\mathrm{vac} = 0$ if we use Eq.~\eqref{eq:self-tuning_general} along with Eq.~\eqref{eq:key_assumption}. Applying an appropriate field transformation $\phi_i\map\tilde\phi_i$,
we can write Eq.~\eqref{eqn:no_go_new_scale_sym} as (suppressing the subscript ``vac'')
\begin{equation}
\del_\eps g_\munu = 2\eps g_\munu\,, \quad \quad
\del_\eps \tilde \phi_0 = -\eps \,, \quad \text{and} \quad \del_\eps \tilde \phi_{i \neq 0} = 0 \,.
\end{equation}
As $\delta_\epsilon \left[e^{2 \tilde{\phi}_0} g_\munu\right] = 0$, we conclude that $\mathcal{L}_\mathrm{vac}= \mathcal{L}(e^{2 \tilde \phi_0} g_\munu )$. This can be made compatible with Eq.~\eqref{eq:L_vac}, if we factorise $\rho_\mathrm{vac}( \phi_{i}) \equiv e^{4 \tilde \phi_0} \tilde \rho_\mathrm{vac}(\tilde \phi_{i \neq 0})$. We therefore find,
\begin{align} \label{eq:Weinberg_L}
\mathcal{L}_\mathrm{vac} = \sqrt{- \mathrm{det}(g_\munu)} \,e^{4 \tilde \phi_0} \tilde \rho_\mathrm{vac}(\tilde \phi_{i \neq 0})\,,
\end{align}
which indeed takes the right form if we use $\sqrt{- \mathrm{det}(g_\munu)} \,e^{4 \tilde \phi_0} = \sqrt{- \mathrm{det}(e^{2 \tilde \phi_0}g_\munu)}$.
Moreover, this recovers our previous result in Sec.~\ref{sec:toymodel} (for $\beta=2$) if we identify $\tilde \rho_\mathrm{vac} = V_0 + \rho_\mathrm{vac}$. As before, setting $\tilde \rho_\mathrm{vac} = 0$ (or setting it to its observed value) is fine-tuning.
For $\tilde \rho_\mathrm{vac} \neq 0$, on the other hand, we again find that Eq.~\eqref{eq:Weinberg_L} describes a run-away potential, which contradicts the initial assumption in Eq.~\eqref{eq:key_assumption} and therefore concludes the no-go argument.

We summarize the main assumptions used for the argument:
\begin{itemize}
    \item[(i)] The scalar field vacuum is constant and thus invariant under translations [see Eq.~\eqref{eq:key_assumption}];
    \item[(ii)] The vacuum geometry is constant (this assumption is stronger than maximal symmetry);
    \item[(iii)] The theory can be formulated in terms of a Lagrange density $\mathcal{L}$. The covariant integration measure is provided by $d^4 x \sqrt{-g}$ as in \eqref{eq:L_vac}.
\end{itemize}
\noindent
We will discuss loopholes in Sec.~\ref{sec:loop}. 

\subsubsection{Beyond Weinberg's argument} \label{sec:spectral}

Weinberg's no-go theorem can be complemented by an independent argument that does not assume the translational invariance of the scalar field vacuum, can be generalized to maximally symmetric vacua with positive curvature, and allows contact with phenomenology~\cite{Niedermann:2017cel}. It  builds on a spectral decomposition of the gravitational exchange amplitude,
\begin{align} \label{exchange_amplitude}
\mathcal{A} = \sum_{s= 0,2}  \int_0^{\infty} \!\!\!\mathrm{d}\mu\, \rho_s(\mu) \int_{x,y} \bar{T}(x)^{\alpha\beta} G^{(s)}_{\alpha\beta\gamma\delta}(x,y;\mu) T(y)^{\gamma\delta}\,.
\end{align}
Here, $\bar{T}_{\mu\nu}$ is a localised gravitational probe such as the Moon, light rays in the gravitational field of the Sun or one of the two masses in a torsion balance. Its counterpart ${T}_{\mu\nu} $ represents a generic source. The gravitational exchange between both sources is described in terms of a propagator $G^{(s)}_{\alpha\beta\gamma\delta}(\mu)$. The spectral density $\rho_s(\mu)$ then gives a particular weight to contributions with different mass-squared $\mu=m^2$ and bosonic spin $s \in \{0,2 \}$.\footnote{Spin-1 fields linearly coupled to a conserved source cannot contribute to $\mathcal{A}$.} It can be decomposed into massive and massless contributions as $\rho_s(\mu)= \rho_{s,m}(\mu) + \bar{\rho}_s \delta (\mu) $. For example, GR only contains a massless spin-2 field, which is realised through $\bar{\rho}_2 = 1/ M^2_\mathrm{Pl} $, $\bar{\rho}_0=0$, and $\rho_{s,m}(\mu) = 0$, and gives rise to the familiar exchange amplitude (valid for localised sources)
\begin{align}\label{A_GR}
   \mathcal{A}_\mathrm{GR} = \frac{1}{M^2_\mathrm{Pl}} \int \left[ \bar{T}^{\alpha\beta} \frac{1}{-\Box}T_{\alpha\beta} - \frac{1}{2} \bar{T}\frac{1}{-\Box}T \right]\,, 
\end{align}
where we use $\frac{1}{\Box} f$ as a shorthand for the convolution of $f$ with the Green function of the d'Alembertian. On the other hand, a single massive spin-2 particle with mass $m$ and Planckian coupling is introduced through $\rho_{2,m} = \delta(m^2-\mu)/M^2_\mathrm{Pl}$. It contributes to Eq.~\eqref{exchange_amplitude} as
\begin{align}
    \mathcal{A}_\mathrm{FP} =  \frac{1}{M^2_\mathrm{Pl}} \int \left[ \bar{T}^{\alpha\beta} \frac{1}{-\Box + m^2}T_{\alpha\beta} - \frac{1}{3} \bar{T}\frac{1}{-\Box+m^2}T \right]\,, 
\end{align}
where the factor $1/3$ corresponds to the notorious Fierz-Pauli tensor structure~\cite{Fierz:1939ix} (see also Sec.~\ref{sec:linear_massive_gravity}). Finally, a scalar mediator will lead to a contribution with trivial tensor structure  $\propto \bar{T} \frac{1}{-\bar{\Box} + m^2} T$.

This general formulation covers a wide range of gravitational models that rely on introducing additional gravitational degrees of freedom. It builds on a minimal set of assumptions:
\begin{enumerate}
    \item[(i)] \label{test} We assume classical and quantum stability, which requires  $\mu \geq 0$ (absence of tachyons) and $\rho_s(\mu) > 0$ (unitarity);
    \item[(ii)] The vacuum geometry is either Minkowski or (quasi) de Sitter;
    \item[(iii)] A weakly coupling assumption underlies the explicit derivation of $G^{(s)}_{\alpha\beta\gamma\delta}(\mu)$ and limits the approach to linear source couplings;
    \item[(iv)] The spin-0 and spin-2 propagators take on their canonical form.
\end{enumerate}

Now, self-tuning describes the absence of a gravitational response in the presence of vacuum energy, explicitly 
\begin{subequations}\label{constraints}
\begin{align}\label{IR_constraint}
    \mathcal{A} \Big|_{T_{\mu\nu} = -\rho_\mathrm{vac} g_{\mu\nu} }\overset{!}{=} 0\,.
\end{align}
This is a constraint that applies in the deep infrared as vacuum energy is a spacetime constant and therefore only couples to the zero (four-)momentum mode. On a Minkowski background, this follows from the fact that the momentum space representation of vacuum energy is $T_{\mu\nu} \propto \eta_{\mu\nu}  \delta^{(4)}(p)$. This infrared condition has to be supplemented with a phenomenological constraint that enforces the approximate recovery of Eq.~\eqref{A_GR} for localised sources and hence applies in the ultraviolet where $-\Box \to \infty$\footnote{These formal relations can be defined rigorously in momentum space.}. To be specific, we demand\footnote{As an aside, screening mechanisms discussed in Sec.~\ref{sec:screening} fulfill this condition based on a violation of the weak coupling assumption (iii).}
\begin{align}\label{UV_ constraint}
    \mathcal{A} \xrightarrow{-\Box \to \infty} \mathcal{A}^{\epsilon}_\mathrm{GR} = \frac{1}{M^2_\mathrm{Pl}} \int \left[ \bar{T}^{\alpha\beta} \frac{1}{-\Box}T_{\alpha\beta} - \frac{1}{2} (1- \epsilon) \bar{T}\frac{1}{-\Box}T \right]\,,
\end{align}
\end{subequations}
where the parameter $\epsilon < 10^{-5}$ parametrises a small  deviation from the GR tensor structure in Eq.~\eqref{A_GR} allowed by solar system observations \cite{Will:2014kxa}. This approach is complementary to Weinberg's argument because it does not require the scalar field vacuum to be invariant under translations. Also, it does not rely on a Lagrangian formulation, nor is it restricted to the classical limit. 

Both conditions in Eq.~\eqref{constraints} can be readily evaluated on a Minkowski background. The self-tuning constraint yields\footnote{To derive this condition, Eq.~\eqref{A_GR} needs to be slightly generalised to also account for non-localised sources such as vacuum energy.} $\bar{\rho}_2 = 6 \bar{\rho}_0$. We note that it does not depend on the massive particle spectrum, which, in our language, shows that Fierz-Pauli massive gravity has a built-in self-tuning mechanism (see Sec.~\ref{sec:massive_gravity}). In any event, the phenomenological constraint yields
\begin{subequations}
\begin{align}
    &\frac{\bar{\rho}_{2}}{p^2} + \int_{0^+}^{\infty} \mathrm{d}\mu \frac{\rho_{2,m}(\mu)}{p^2 + \mu} \xrightarrow{p \to \infty} \frac{1}{p^2 M_\mathrm{Pl}}\,,\label{UV1}\\
    & \frac{1}{3} \int_0^{\infty} \mathrm{d}\mu \frac{\rho_{2,m}(\mu)}{p^2+\mu} + 2 \frac{\bar{\rho}_0}{p^2} + 2 \int_{0^+}^{\infty} \mathrm{d}\mu \frac{\rho_0(\mu)}{p^2+\mu} \xrightarrow{p \to \infty} \frac{\epsilon}{p^2 M_\mathrm{Pl}} \label{UV2}\,.
\end{align}
\end{subequations}
Using the positivity of each term in~\eqref{UV2}, we find $\bar{\rho}_0 \lesssim \epsilon/M_\mathrm{Pl}^2$ and $p^2 \int_\mu \frac{\rho_{2,m}}{p^2+\mu} \lesssim \epsilon/M_\mathrm{Pl}^2$. With this we conclude from Eq.~\eqref{UV1} that $\bar{\rho}_2 \simeq 1/M_\mathrm{Pl}^2$. The conditions on $\bar{\rho}_0$ and $\bar{\rho}_2$ are however in contradiction with the self-tuning relation  $\bar{\rho}_2 = 6 \bar{\rho}_0$. In other words, self-tuning under the above assumptions is incompatible with phenomenology. This concludes the argument. As shown in~\cite{Niedermann:2017cel}, the same conclusion can be reached for de Sitter vacua.

\subsubsection{Loopholes} \label{sec:loop}

Is this generalised approach, alongside Weinberg's argument, heralding the end of self-tuning models? It certainly rules out many direct attempts at implementing the mechanism. But as is true with every no-go theorem, these too are limited by their assumptions. They allow us to formulate loopholes that themselves can serve as the seeds of future research. Here we single out three possibilities. 

\begin{enumerate}
    \item Models that rely on non-linear couplings to restore GR [violating assumption (iii) of the amplitude argument]. In particular, this includes models that build on the Vainshtein mechanism presented in Sec.~\ref{sec:derivative_screening}. In this review we discuss two examples: massive gravity in Sec.~\ref{sec:massive_gravity} and scalar-tensor theories such as Fab-4 in Sec.~\ref{sec:self-tuning}. Whether the no-go arguments can be generalised to this case remains to be seen. It should also be noted that Vainshtein screening comes with its own set of problems pertaining to its UV sensitivity in the non-linear regime~\cite{Burrage:2012ja,deRham:2014wfa,Kaloper:2014vqa}. Both models also get around Weinberg's complementary no-go by relying on scalar field vacuum that breaks translational invariance [violating assumption (i) in Sec.~\ref{sec:no-go}]. In the case of massive gravity, this happens in the Stuckelberg sector of the theory; 
    
    \item Another key assumption in both theorems is that vacuum energy, and thus also the corresponding geometry, is maximally symmetric. In particular, the vacuum does not break local Lorentz invariance. As we will argue in Sec.~\ref{sec:xdim}, in braneworld models the vacuum energy that arises from SM matter loops (spontaneously) breaks Lorentz invariance in the directions orthogonal to the brane [violating assumption (ii) of both arguments];
    
    \item For the generalised no-go we assumed that propagators take on their canonical form as could be derived in canonical local and ghost-free field theories of a given spin. Accordingly, this loophole relies on introducing non-standard propagators [violating assumption (iv) of the amplitude argument]. This is for example how sequestering in Sec.~\ref{sec:constrained_gravity} avoids the no-go. It also avoids Weinberg's argument by  introducing either global variables or employing a 4-form contribution to the volume measure [violating assumption (iii) in Sec.~\ref{sec:no-go}].  
\end{enumerate}

To provide an explicit example of how the last loophole can be exploited, we follow ~\cite{Niedermann:2017cel} and discuss the decapitation idea first introduced in a string theory context~\cite{Adams:2002ft}. To that end, we consider an exchange amplitude of the form
\begin{align}\label{A_decap}
    \mathcal{A} = \mathcal{A}_\mathrm{GR} + \frac{1}{6 M_\mathrm{Pl}^2} \int \bar{T} \frac{1}{- \Box} T - \mathcal{A}_\mathrm{ghost} \,.
\end{align}

The second term is the contribution of a canonical scalar field. Its numerical coefficient has been chosen such that it cancels with $\mathcal{A}_\mathrm{GR}$ for a vacuum energy source. As we have seen before this is not possible without violating the observational constraint in Eq.~\eqref{UV_ constraint} when considering localised sources. This is where the third term $\mathcal{A}_\mathrm{ghost}$ comes to the rescue. For localised sources it exactly cancels the second term while it vanishes for vacuum energy:
\begin{align}\label{A_ghost}
    \mathcal{A}_\mathrm{ghost} = 
    \begin{cases}
    \frac{1}{6 M_\mathrm{Pl}^2} \int \bar{T} \frac{1}{- \Box} T &\quad \mathrm{for} \quad T_{\mu\nu}(x) \quad \mathrm{local}\\
    0 &\quad \mathrm{for} \quad T_{\mu\nu} = - \rho_\mathrm{vac} g_{\mu\nu} 
    \end{cases} \, .
\end{align}
As a result, we recover exact GR for every source that is localised, but realise self-tuning for  vacuum energy sources. We call it a ghost term because it leads to a negative contribution in Eq.~\eqref{A_decap}. However, it does not lead to an instability as could be expected for generic ghost fields because it is either vanishing or exactly compensated by the healthy scalar mode (this is somewhat analogous to Faddeev–Popov ghosts in gauge quantum field theory, which also do not lead to instabilities due to an cancellation with healthy but unphysical modes). Writing $\mathcal{A}_\mathrm{ghost} = \frac{1}{6 M_\mathrm{Pl}^2}\int_{x,y} \bar{T}(x) G_\mathrm{decap}(x,y) T(y) $, the nontrivial question is what type of propagator $G_\mathrm{decap}$ will lead to this exotic behavior. As it turns out, a non-local construction can do it, explicitly~\cite{Adams:2002ft}
\begin{align}\label{G_decap}
   G_\mathrm{decap}(x,y) = G_0(x,y) - \frac{1}{V} \int_z G_0(z,y)- \frac{1}{V} \int_z G_0(x,z) + \frac{1}{V^2} \int_{z_1,z_2} G_0(z_1,z_2) \,,
\end{align}
where $G_0$ denotes the standard Green function of the d'Alembertian and $V$ is the (appropriately regularised) spacetime volume. It is referred to as a  decapitated propagator because it does not couple to vacuum loops. To show that it fulfills Eq.~\eqref{A_ghost}, the property $\int_x G_\mathrm{decap}(x,y)=\int_y G_\mathrm{decap}(x,y)=0$ can be used. It has been demonstrated in \cite{Niedermann:2017cel} that the sequestering model (discussed in Sec.~\ref{ssec:global_seq}) provides a possible field theory realization of the decapitation idea.

Of course, the list of loopholes is not complete and other models are conceivable that for example incorporate a mild tachyonic instability, consider an anti-de Sitter vacuum or find different ways of deviating from the canonical field theory framework.

\subsection{Screening Mechanisms} \label{sec:screening}

As we have seen, by challenging the assumption of GR we can hope to address aspects of the CCPs. A crucial constraint coming from observations on such avenues is that the proposed modification to gravity be undetectable in our local environment. Specifically, the numerous precision tests of gravity in the Solar System only allow extremely mild deviations from GR at these scales (see for example \cite{Bertotti:2003rm,PhysRevD.53.6730,Anderson:1995df,Dickey:1994zz,Talmadge:1988qz}). This poses a problem for scalar tensor theories which generically predict a supplementary force on top of the standard GR prediction, usually referred to as the `fifth force'. This is of particular relevance for scalar tensor theories such as the Fab-4 (see Sec.~\ref{sec:self-tuning}) or the Dvali-Gabadadze-Porrati braneworld model (DGP) \cite{Dvali:2000hr,Dvali:2000xg} (see Sec.~\ref{sec:xdim}), which can be written as a scalar tensor theory in the limit that the 5-dimensional degrees of freedom decouple (see \cite{Luty:2003vm} for example). 

The scalar tensor theories we will consider here and in this review will fall under the broad class of Horndeski theories \cite{Horndeski:1974wa} which can be written as %
\begin{align}
 S_{\mathrm H}  & = \int d^{4}x \sqrt{-g} \big[ G_2(\phi,X) - G_3(\phi,X)\Box\phi +G_4(\phi,X)R  \nonumber \\
& + G_{4,X}(\phi,X)[(\Box \phi)^2-(\nabla_\mu \nabla_\nu \phi)^2]  + G_5(\phi,X)G_{\mu \nu}\nabla^\mu\nabla^\nu\phi  \nonumber \\
&- \frac{1}{6}G_{5,X}(\phi,X) [(\Box\phi)^3-3\Box\phi(\nabla_\mu\nabla_\nu\phi)^2+2(\nabla_\mu\nabla_\nu\phi)^3] \Big] \, ,
\label{eq:horndeski}
\end{align}
where each $G_i(\phi,X)$, $i=2,3,4,5$ is a free function of the scalar field $\phi$ and its canonical kinetic term $X = -(\partial \phi)^2/2$, and $G_{i,X}(\phi,X) = \partial G_{i}/\partial X$, with $\Box$ being the D'Alembertian operator. $R$ and $G_{\mu\nu}$ are the Ricci scalar and Einstein tensor components, respectively. This class ensures the resulting equations of motion are at most second order and non-degenerate which makes them Ostrogradski ghost-free  (see \cite{Woodard:2006nt} for a discussion) and theoretically viable. We note that one can have healthy theories with higher-than-2nd order derivatives by considering degenerate theories or certain field symmetries \cite{Motohashi:2014opa,Langlois:2015cwa,Langlois:2015skt,Crisostomi:2016czh,Nicolis:2008in}.

As an illustrative example of the problem, we can consider the archetypal Jordan-Brans-Dicke theory \cite{jordan1959present,brans1961mach} which is contained in the Horndeski class. We can write this theory in the Einstein frame, with a minimal coupling of the scalar field to the gravitational sector, i.e., it does not multiply the Ricci scalar, but a modified matter sector as 
\begin{equation}
    \mathcal{L}_{\rm JBD} = \left[\frac{\MPl^2 R}{2} - \frac{1}{2}g^{\mu \nu} \partial_\mu \phi \partial_\nu \phi - V(\phi)\right] + \Lm(A^2(\phi) g_{\mu \nu}, \Psi_{\rm M}) \, , \label{eq:JBD}
\end{equation}
where $A(\phi)$ is the conformal transformation between minimally and non-minimally coupled frames,  $\phi$ is the scalar degree of freedom and $\Psi_{\rm M}$ are the collective matter fields. We note that both frames violate the equivalence principle and refer the reader to \cite{Hui:2009kc} for a discussion \footnote{Note that this fact can and has been utilised to test such theories (see for example \cite{Sakstein:2017bws}).}. 

If we consider now a perfect fluid 
\begin{equation}
T^{\mu \nu} = \Big( \rhom + \frac{p}{c^2} \Big) u^\mu u^\nu + p g^{\mu \nu} \, , 
\end{equation} 
with rest frame mass density $\rhom$ and we assume a small pressure $p\ll \rho c^2$, in a perturbed Minkowski spacetime, 
\begin{equation}
    ds^2 = -(1+2\Phi)dt^2 + (1-2\Psi)(dx^2 +dy^2 +dz^2) \, , 
\end{equation}
we can write the 00-th Einstein field equation in the Newtonian limit as 
\begin{equation}
    \nabla^2 \Phi = \frac{1}{2 \MPl^2} \rhom - \frac{1}{2} \nabla^2 \phi  \, , \label{eq:modpoisson} 
\end{equation}
where $\nabla^2$ is the Euclidean Laplacian operator. Note we have ignored pressure contributions and have only considered first order contributions in the metric perturbation $\Phi$. We can also write down the geodesic equation in the non-relativistic limit ($d x_i/dt \ll 1$), which gives the acceleration a test particle experiences
\begin{equation}
    \mathbf{a} = -\nabla \Phi - \beta \nabla \phi \equiv \mathbf{a}_{\mathrm N} + \mathbf{a}_{5}\, , \label{eq:modgeo}
\end{equation}
where we have defined $\beta(\phi) \equiv [\ln A(\phi)]'$, a prime denoting a scalar field derivative, and defined the Newtonian acceleration and an additional acceleration associated with a ``5th force'' sourced by $\phi$. One basic criterion for any viable gravitational theory is that the scalar field's contribution to the acceleration is negligible, or {\it screened}, in environments like the Solar System\footnote{Cosmic structure formation measurements are also constraining but to a much lesser degree \cite{Koyama:2015vza}.}. We will discuss two generic ways of ensuring this holds true: screening through a well chosen potential for the field and screening that employs derivatives of the field. We refer the interested reader to  \cite{Koyama:2015vza,Joyce:2016vqv,Sakstein:2014jrq} for more detailed reviews on screening mechanisms.

\subsubsection{Potential screening}

This method of screening relies on a well chosen potential $V(\phi)$ for the scalar field. Assuming negligible time variation of $\phi$ and a pressureless perfect fluid, the equation of motion for the scalar field in Eq.~\eqref{eq:JBD} is given by 
\begin{equation}
    \nabla^2 \phi = \frac{1}{\MPl^2} \beta(\phi) \rhom - V'(\phi) \equiv V_\eff'(\phi) \, , \label{eq:kgess}
\end{equation}
where we have defined an effective potential $V_\eff$. Screening is realised when the effective potential finds a minimum, i.e., $\nabla^2 \phi = 0$. With appropriate choices of $V(\phi)$ and $\beta(\phi)$ [or equivalently $A(\phi)$], this can be made to happen in regions of high density such as the Solar System. With this choice, we note that moving beyond the Solar System, to say larger scales and lower densities, objects may begin to feel the additional force contribution, which will have an impact on structure formation. Typically such theories aim to reproduce $\Lambda$CDM at high redshift where measurements of the cosmic microwave background (CMB) leave little room for deviation \cite{Planck:2018vyg}. This leaves the possibility of low redshift, large scale impacts of the scalar field (see for example \cite{Cataneo:2018mil, Carroll:2006jn, Pogosian:2016pwr, Peirone:2017ywi}).

Returning to Eq.~\eqref{eq:kgess}, it would appear that we have two free functions with an infinite number of new degrees of freedom with which to match the data. While in principle this is true, these would be ruled out from a Bayesian standpoint. A simple and popular example of such a theory that promised to both exhibit screening and was able to provide acceleration without the CC \footnote{This theory now still requires the CC for cosmic acceleration, having been strongly constrained by data (see for example \cite{Burrage:2017qrf}).} is the Hu-Sawicki model of $f(R)$ \cite{PhysRevD.76.064004},
\begin{equation}
    f(R) = -m^2 \frac{c_1 (R/m^2)^n}{1+c_2(R/m^2)^n} \, . \label{eq:fofr}
\end{equation}
Here $c_1$ or $c_2$ is fixed to necessarily match a $\Lambda$CDM expansion history. $m^2$ is also typically fixed leaving two free parameters, $n$ and $c_1$ or $c_2$, the latter usually being recast in terms of the value of the scalar field today $f_{\mathrm R0} \equiv \frac{d f}{dR}|_{z=0}$. This functional choice also fixes both $V(\phi)$ and $\beta(\phi)$ in the equivalent Einstein frame cast scalar tensor theory. 

To see how screening works, let us consider a spherically symmetric matter distribution with radius $\mathcal{R}$ and total density $\rho_m$, embedded in a background of density $\bar{\rho}$. By the symmetry of the distribution we should be able to find some  $r_S < \mathcal{R}$ such that the effective potential $V_\eff'(\phi) \approx 0$ which defines the screened regime. Further, for $r>r_S$ we can find a region such that $\phi = \bar{\phi} + \delta \phi(r)$, where $\delta \phi$ is a small perturbation. The equation of motion for this perturbation is then
\begin{align}
  \nabla^2 \delta \phi  & =  \nabla^2 \phi - \nabla^2 \bar{\phi}   \nonumber \\ 
  & \approx \frac{1}{\MPl^2} (\rhom - \bar{\rho}) \bar{\beta} - (V'(\phi)- V'(\bar{\phi}) )  \nonumber \\ 
  & \approx  \frac{1}{\MPl^2} \bar{\beta} \delta \bar{\rho} - \delta \phi V''(\bar{\phi}) \, , \label{eq:sfpeof} 
\end{align}
where we have defined the density contrast $\delta \bar{\rho} \equiv \rhom - \bar{\rho}$, where a bar denotes a background quantity. In the first line we have only considered the background term in $\beta$, to stay at consistent order in the perturbations $\beta(\phi) \approx \beta(\bar{\phi}) \equiv \bar{\beta}$, which can be restated as 
\begin{equation}
    \frac{d \beta}{d \phi} \delta \phi \ll \bar{\beta} \qquad \mathrm{for} \qquad r>r_S  \,.
\end{equation}
 The second approximation assumes $V'(\phi)$ is slowly varying. For cosmologically relevant fields, we typically choose their background mass $m_0^2 \equiv V''(\bar{\phi}) \sim H_0$, where $H_0$ is the Hubble constant, so that we only see modified gravity effects on very large scales. Near the screening regime we can then ignore the $V''$ term in Eq.~\eqref{eq:sfpeof}. Integrating  Eq.~\eqref{eq:sfpeof} for $r>r_S$ gives the acceleration associated with the fifth force
\begin{equation}
    \mathbf{a}_5 \equiv -\bar{\beta} \nabla \phi =  -\bar{\beta} \nabla \delta \phi =  - 2 \bar{\beta}^2 \frac{G \mathcal{M}(r)}{r^2} \left[ 1 - \frac{\mathcal{M}(r_S)}{\mathcal{M}(r)} \right] \qquad \mathrm{for}  \qquad r>r_S \, ,\label{eq:fifthforce}
\end{equation}
where the density contrast sourced mass within radius $r$ is given by 
\begin{equation}
    \mathcal{M}(r) = 4 \pi \int_0^r  x^2 \delta \bar{\rho}(x) dx \, .
\end{equation}
From Eq.~\eqref{eq:fifthforce} we see that if $r_S = \mathcal{R}$ then the object is fully screened ($F_5 = 0$), but conversely if $r_S =0$ the object is fully unscreened. Typically we would have $0 < r_S < \mathcal{R}$, which would offer a partially screened object with the fifth force being sourced by a shell of mass between $r_S<r<\mathcal{R}$. The screening radius $r_S$ can be derived from the field profile once the potential and conformal transformation have been defined. These can be chosen such that $r_S$ is only slightly smaller than $\mathcal{R}$ so that the fifth force is only sourced by a thin-shell of mass and is consequently limited in size and scales of effect. Of course, one could also find fully screened configurations, but these do not offer interesting signatures which can be looked for experimentally. 

Chameleon models are a class of such models exhibiting this type of screening and have been extensively studied in the literature \cite{Khoury:2003rn,Navarro:2006mw,Faulkner:2006ub}, (see \cite{Burrage:2017qrf,Brax:2021wcv} for reviews of fairly recent constraints). The choice for potential and conformal transformation in these models usually take the form 
\begin{equation}
    V(\phi) = \frac{m_0^{n+4}}{\phi^n} \, , \qquad A(\phi) = e^{B \phi} \, , \label{eq:chamex}
\end{equation}
$B$ being a constant. Note that we can also get chameleon screening from minimally coupled matter in $f(R)$ gravity [see for example Eq.~\eqref{eq:fofr}] after the field redefinition necessary to get to the Einstein frame. Given the freedom these models allow, data constraints are fairly restricted to specific models or parametrisations [see \cite{Brax:2021wcv} for constraints on the model in Eq.~\eqref{eq:chamex}].

\subsubsection{Derivative screening} \label{sec:derivative_screening}
To see how this type of screening works, we can proceed analagously to the self-interaction screening case. First let us return to the modified geodesic equation, Eq.~\eqref{eq:modgeo}. One way of shutting off the scalar field term on the RHS is to include additional derivative interactions in the action. As an example we can take the 5-dimensional DGP braneworld model \cite{Dvali:2000hr} in the decoupling limit, which has been argued reduces to the cubic Galileon theory \cite{Luty:2003vm,Nicolis:2004qq} 
\begin{equation}
        \mathcal{L}_{\mathrm G3} = \left[\frac{\MPl^2 R}{2} - \frac{1}{2}g^{\mu \nu} \partial_\mu \phi \partial_\nu \phi - \frac{1}{\lambda_0^2} g^{\mu \nu} \partial_\mu \phi \partial_\nu \phi \Box \phi + \alpha_{\rm V} \phi \frac{1}{\MPl^2} T \right]  \, , \label{eq:cubicgal} 
\end{equation}
where $\lambda_0^2$ gives the energy scale of the theory and we have made the coupling to the trace of the matter sector EMT explicit through the dimensionless constant $\alpha_{\rm V}$. We note this can be achieved similarly by setting  $A^2(\phi) = e^{2 \phi \alpha_{\rm V}}$ in Eq.~\eqref{eq:JBD}. This action produces the following equation of motion 
\begin{equation} 
\Box \phi +  \frac{2}{\lambda_0^2} \left[ (\Box \phi)^2 - \nabla_\nu \nabla_\mu \phi \nabla^\nu \nabla^\mu \phi \right]  = \frac{1}{\MPl^2} \alpha_{\rm V} \rhom  \, , \label{eq:kgeg3}
\end{equation} 
where $\nabla_\mu$ is the covariant derivative. In DGP gravity we have \cite{Koyama:2007ih} $\lambda_0^2 = \frac{3}{2 r_c^2} \beta_\DGP$ and $\alpha_{\rm V} = \frac{1}{3\beta_\DGP}$, where $\beta_\DGP$ is associated with the cosmological background in the model and $r_c$ is the cross-over scale below which gravity becomes 4-dimensional. 

As in the chameleon example, if we consider small enough scales where time derivatives are small compared to spatial ones and consider a spherically symmetric mass distribution of total mass $M$, Eq.~\eqref{eq:kgeg3} can be written as (see \cite{Barreira:2013eea} for example)
\begin{equation}
    \frac{1}{r^2} \frac{d}{dr} \left[r^2 \frac{d\phi}{dr}+\frac{4r}{\lambda_0^2} \left(\frac{d\phi}{dr} \right)^2 \right]  =  \frac{1}{\MPl^2} \alpha_{\rm V} \rhom \, . 
\end{equation}
Integrating this equation yields
\begin{equation}
   r^2 \frac{d\phi}{dr}+\frac{4r}{\lambda_0^2} \left(\frac{d\phi}{dr} \right)^2  = \frac{\alpha_{\rm V} M}{4 \pi \MPl^2} \equiv 2 \alpha_{\rm V} r^2  F_{\rm N}  \, ,  \label{eq:forceratv} 
\end{equation}
    where $F_{\rm N}$ is the Newtonian gravitational force per unit mass. Taking the small $r$ limit of Eq.~\eqref{eq:forceratv}, when the 2nd term on the LHS dominates, yields
\begin{equation}
     \frac{F_5}{F_{\rm N}} = 2 \alpha_{\rm V} \left(\frac{r}{r_{\rm V}}\right)^{3/2} \, ,
\end{equation}
where $F_5=d\phi/dr$ is the fifth force per unit mass, and we have defined the so-called Vainshtein radius $r_{\rm V} \equiv \left( \frac{8 \GN M \alpha_{\rm V}}{\lambda_0^2} \right)^{1/3}$, named after Arkady Vainshtein who first proposed this mechanism  to protect Fierz-Pauli theory (see Sec.~\ref{sec:linear_massive_gravity}) from Solar System constraints \cite{Vainshtein:1972sx}. We see $F_5 \propto r^{3/2}$, and so for large Vainshtein radii we get a large suppression of the fifth force when $r \ll r_{\rm V}$. This mechanism is highly efficient, for example in the case that $M = M_\odot$ and $\alpha_{\rm V}$ of $\mathcal{O}(1)$ we find that $r_{\rm V}$ is 7 orders of magnitude larger than the Solar System. On the other hand, in the large scale limit where the first term of  Eq.~\eqref{eq:forceratv} dominates, we have a modification of $F_5/F_{\rm N} \approx 2 \alpha_{\rm V}$ which can offer interesting phenomenology to go out and test experimentally (again, see \cite{Brax:2021wcv} for recent constraints).

\bigbreak

\begin{figure*}
    \centering
    \includegraphics[scale=0.75]{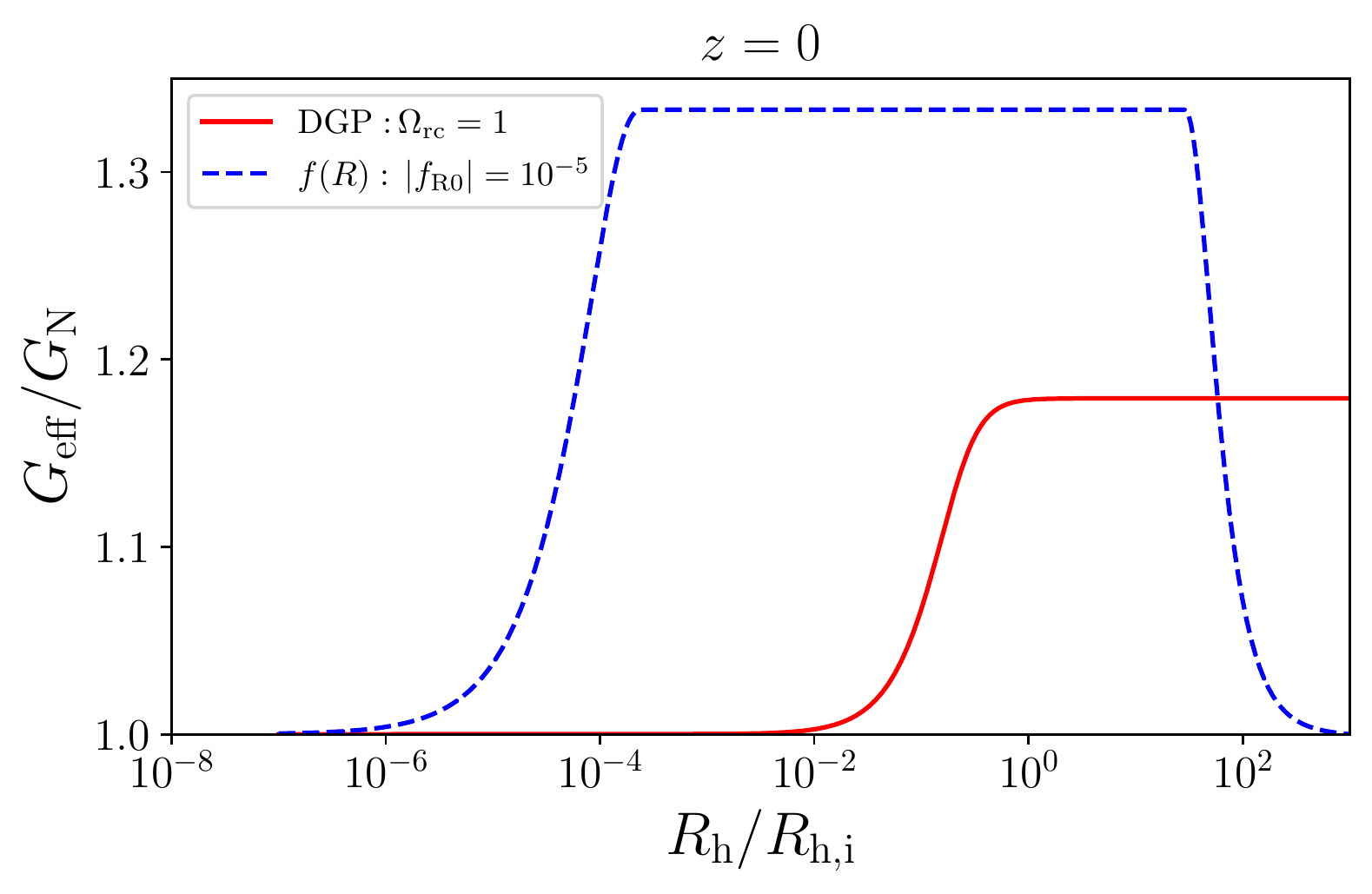}
    \caption{The normalised effective Newton's constant as a function of dark matter halo radius $R_{\mathrm h}$ in the chameleon screened Hu-Sawicki $f(R)$ gravity ({\bf blue, dashed}) and the Vainshtein screened DGP gravity ({\bf red, solid}). $R_{\mathrm h,i}$ is the initial size of the halo and $\Omega_{\rm rc} \equiv 1/(4 H_0^2 r_c^2)$. The  amplitudes in the unscreened regime reflect the specific values of $f_{\rm R0}$ and $r_c$ chosen.} 
    \label{fig:scr}
\end{figure*}

As an illustration of both types of screening mechanism, we can consider the effective modification to Newton's gravitational constant $G_\eff$, defined using Eq.~\eqref{eq:modpoisson} as 
\begin{equation}
   \nabla^2 \Phi \equiv \frac{\rhom }{2 \MPl^2} \frac{G_\eff}{\GN} \, .
\end{equation}
This quantity has been derived under various assumptions, such as a spherically symmetric density distribution, in both the chameleon screened model of Eq.~\eqref{eq:fofr} \cite{Lombriser:2013wta}  and in the Vainshtein screened DGP model \cite{Schmidt:2009yj}. In Fig.~\ref{fig:scr} we plot $G_\eff/\GN$ against the normalised halo radius. We can clearly see that at small scales the screening mechanism kicks in and $G_\eff \rightarrow \GN$ as required. Interestingly, the DGP model (red, solid) also modifies the large scale growth of structure while $f(R)$ (blue, dashed) experiences as Yukawa suppression at large scales, giving distinct and interesting phenomenology. Further, we see the fifth force acting at some intermediate range of scales which affects the growth of structure. Detection of fifth force impact on structure and signatures of modified gravity is a prime science goal of upcoming and ongoing large scale structure experiments \cite{Amendola:2016saw,DES:2021zdr,Alam:2020jdv}.

\bigbreak 

We conclude this section with some general remarks on the theoretical viability of screened models. Note that screening effects kick in at scales where the low-energy theory should begin to fail. In the case of self-interaction screening, the field potential should begin to receive non-negligible high-energy corrections in the screening regime which poses a problem for these models. Similar considerations were made for non-linear interaction models, such as Vainshtein screened ones, \cite{Burrage:2012ja,Kaloper:2014vqa} although for these models it has been argued that they may be protected against high-energy corrections as they typically satisfy non-renormalisation theorems \cite{Brax:2016jjt,deRham:2014wfa}. We also direct the reader to Sec.~6 of \cite{Joyce:2014kja} and the references therein for more on this discussion.

\section{Modified Gravity Approaches}\label{sec:mod_gravity_approaches}

Now that we have clarified what are the issues surrounding the cosmological constant and reviewed the no-go theorems constraining possible modifications of gravity, we are apt to discuss some of the approaches that exploit the loopholes of these theorems, as discussed in Sec.~\ref{sec:loop}. Including our tacit discussion about semi-classical gravity which combines GR and QFT, we will discuss nine different proposals that attempt at instantiating the self-tuning mechanism. As we already mentioned in the introduction, this is in no way exhaustive and only constitutes a representative sample of the rich literature available on modifications of gravity\footnote{In particular, the literature goes beyond proposals simply trying to tackle the CCPs, as modifications of gravity have also been extensively considered as an explanation for the late-time accelerated expansion of the Universe and an alternative to dark matter \cite{Clifton:2011jh}.}.

We start in Sec.~\ref{sec:wishlist} by briefly recalling the challenges introduced in Sec.~\ref{sec:problems} and defining two additional requirements related to astrophysical and cosmological data that a successful proposal has to fulfill. Then, in Sec.~\ref{sec:constrained_gravity} we discuss a general class of models that makes use of top forms to constrain the gravitational dynamics; in Sec.~\ref{sec:massive_gravity} we discuss the degravitation mechanism and its realization in the context of massive gravity, and we finish by reviewing how self-tuning can be implemented in the context of Horndeski theories in Sec.~\ref{sec:self-tuning}.

\subsection{What do we want?} \label{sec:wishlist}

As we have seen in Sec.~\ref{sec:problems}, there are at least three problems that a model can attempt at solving: the classical and new CCP, and the DEP. However, even if a model successfully solves one or more of these problems by modifying gravity, we need to guarantee that the model does not spoil phenomenology that has been successfully accounted for with only GR. 

Nowadays, we have strict constrains on the ensuing dynamics of the early Universe that led to the CMB \cite{Planck:2018vyg} and we have tight constrains on Solar System physics \cite{Will:2014kxa,Bertotti:2003rm,PhysRevD.53.6730,Anderson:1995df,Dickey:1994zz,Talmadge:1988qz}. Thus, to assess the statuses of the models discussed below, we need to also take into account whether their proposed solutions to any of the CCPs is able to accommodate the CMB data together with the success of the $\Lambda$CDM model, and whether these models contain a screening mechanism, as described in Sec.~\ref{sec:screening}, that leaves local physics on small astrophysical scales untouched. We will call these requirements \textit{Cosmic History Constraint (CHC)} and \textit{Astrophysical Constraints (AC)}\label{par:data_cons}. 

Combining all the requirements above, we thus have a wishlist with five conditions that a successful model should satisfy. As now we shift our discussion towards how different modified gravity proposals can address the CCPs, these are the conditions that we should keep in mind while assessing how successful these different proposals are. Later, in Sec.~\ref{sec:disc} we summarise this wishlist and we briefly review which requirements each approach is able to tackle (see  Table~\ref{table:1}).

\subsection{Constraining Gravity} \label{sec:constrained_gravity} 

There is a class of approaches that modify GR by constraining its dynamics such that the vacuum energy does not gravitate. We will discuss four of them: global and local sequestering mechanisms \cite{Kaloper:2013zca,Kaloper:2015jra,DAmico:2017ngr}, unimodular gravity \cite{inbook,Ng:1990xz}, and the non-local approach introduced in \cite{Carroll:2017gqo}. 

As we will see, local sequestering is the the most successful proposal addressing the CCPs discussed in Sec.~\ref{sec:problems}. Thus, this section starts by discussing the global sequestering mechanism where the physical intuition is clearer as a way to further motivate the introduction of the local mechanism. Finally, since both unimodular gravity and the non-local approach can be easily related to the sequestering mechanism, they are discussed comparatively. 

Naturally, this does not exhaust the list of proposals in the same vein. More recent proposals include \cite{Lombriser:2019jia}, that promotes the Plank mass to a Lagrange multiplier, thus constraining the averaged Ricci scalar and preventing the vacuum energy from gravitating. Meanwhile \cite{Kaloper:2022jpv,Kaloper:2022utc} promote the CC to an integration constant, but instead of canceling the vacuum energy, they present a mechanism that relaxes it through multiple membrane nucleations (each localised in time).

\subsubsection{The Global Vacuum Energy Sequestering}\label{ssec:global_seq}

One way to avoid the no-go theorems discussed in Sec.~\ref{sec:no-go} was introduced in \cite{Kaloper:2013zca}, where the vacuum energy generated by matter loops is gravitationally decoupled. This mechanism is called \textit{global vacuum energy sequestering} and is based on promoting the bare cosmological constant, $\Lambda_{\text{b}}$, and the Planck mass, $M_{\rm Pl}$, to Lagrange multipliers. It further introduces a global interaction term $\sigma$ as a function of $\Lambda_{\text{b}}$ and a mass scale $\mu$ around the QFT cut-off scale. Thus, this is a minimal modification of GR as it does not introduce any new propagating degrees of freedom. For clarity, we denote the Planck mass by $\kappa$ in this and next sections to indicate that it is now a variable.

In the Jordan conformal gauge, the action reads \cite{DAmico:2017ngr}

\begin{equation} \label{eq:global_seq}
    S= \int d^4x \sqrt{-g} \left[\frac{\kappa^2}{2}R - \Lambda_{\text{b}} + \Lm \right] + \sigma \left(\frac{\Lambda_{\text{b}}}{\mu^4}\right)\, ,
\end{equation}
where $\mathcal{L}_{\text{m}}$ stands for the matter Lagrangian and note that $\sigma$ is outside the integral. Varying it w.r.t $\Lambda_{\text{b}}$ and $\kappa^2$, respectively, yields two global constraints

\begin{equation} \label{eq:const_seq}
    \frac{1}{\mu^4} \frac{d\sigma}{d\Lambda_{\text{b}}} = \int d^4x \sqrt{-g} \,, \quad \quad \quad \int d^4x \sqrt{-g} \, R = 0\, .
\end{equation}
Due to the presence of the smooth function $\sigma$, the first constraint does not force the world volume to vanish. Nonetheless, in order to have a large and old universe, it is required that $\sigma$ cannot be a linear function; otherwise $\mu$ would have to be very small, lying at scales many orders of magnitude below particle physics scales and the cut-off of this effective field theory.

Meanwhile, the second equation can be recast as $\langle R \rangle =0$ where the 4-volume average is defined as $\langle (\ldots) \rangle \equiv  \int  d^4x \sqrt{-g} (\ldots) /  \int  d^4x \sqrt{-g}$. Thus, the scalar curvature is always averaged to zero. Naturally, this is only well-defined if the 4-volume is non-vanishing, imposing that $d\sigma/d\Lambda_{\text{b}}$ is non-zero.

Now, varying the action w.r.t. to the metric produces the gravitational field equations, which upon taking the trace, averaging, and using the second global constraint in Eq.~\eqref{eq:const_seq} leads to,
\begin{equation} \label{eq:const_1_seq}
    \Lambda_{\text{b}} = \frac{1}{4} \langle T^\mu_{\phantom{\mu}\mu}  \rangle \,, 
\end{equation}
completely fixing  the bare cosmological constant in terms of the matter sources\footnote{In contrast to unimodular gravity, for example - see Sec.~\ref{subsec:unimodular}.}. This is only possible because $\Lambda_{\text{b}}$ is considered as a dynamical variable whose value is determined by Eq.~\eqref{eq:const_1_seq}. Then, we can finally rewrite the Einstein field equations as 

\begin{equation} \label{eq:EE_seq}
    \kappa^2 G^\mu_{\phantom{\mu}\nu} = T^\mu_{\phantom{\mu}\nu}  - \frac{1}{4} \delta^\mu_{\phantom{\mu}\nu} \langle T^\alpha_{\phantom{\alpha}\alpha}  \rangle  \,, 
\end{equation}
where the bare cosmological constant is a global dynamical field fixed by the field equations. Notably, as the matter source is shifted by the last term on the RHS, both classical and quantum contributions that take the form of a cosmological constant will not gravitate\footnote{There is a caveat here concerning gravitational loop corrections,  which would introduce $\kappa$ dependent terms to Eq.~\eqref{eq:global_seq} that are not sequestered. This problem can be avoided if topological curvature invariants are also considered, as it is discussed in \cite{Kaloper:2016jsd} and \cite{El-Menoufi:2019qva}.}. In particular, the latter is guaranteed due to general covariance, which imposes that the loop corrections contribute in the same way at any order of the QFT loop expansion \cite{Martin:2012bt}. 

More explicitly, as it is shown in \cite{DAmico:2017ngr}, we can consider the matter Lagrangian at any given order in loops to be split between the renormalised quantum vacuum energy contributions, $\rho_{\rm vac}$, and local excitations, which leads to $T^\mu_{\phantom{\mu}\nu} = -\rho_{\rm vac} \delta^\mu_{\phantom{\mu}\nu} + \tau^\mu_{\phantom{\mu}\nu}$. Then, the field equations Eq.~\eqref{eq:EE_seq} become

\begin{equation}\label{Einstein_sequester}
    \kappa^2 G^\mu_{\phantom{\mu}\nu} = \tau^\mu_{\phantom{\mu}\nu}  - \frac{1}{4} \delta^\mu_{\phantom{\mu}\nu} \langle \tau^\alpha_{\phantom{\alpha}\alpha}  \rangle  \,, 
\end{equation}
and the regularised vacuum energy drops as a source. Nonetheless, we still have a residual cosmological constant above given by the second term on the RHS which arises as the historic average of the trace of matter excitations. Thus, it is a non-local source both in space and time, and it is typically small in universes which grow large and old due to the spacetime averaging. Therefore, global sequestering could potentially also address the DE problem.

Something similar happens with contributions coming from phase transitions. They are not completely suppressed in the field equations Eq.~\eqref{Einstein_sequester}, but for large and old universes they become far smaller than the current critical density after the transition  \cite{Kaloper:2013zca}. The intuition here is that the contribution is suppressed by the spacetime volume, and typically the Universe spends a relatively short time in the false vacuum \cite{Kaloper:2016yfa}. Note that before the transition, despite these contributions not being sequestered, within reasonable assumptions they remain consistent with early universe cosmological data \cite{Kaloper:2014dqa}.

We make some additional remarks about the above construction: 

\begin{itemize}

    \item The first constraint in Eq.~\eqref{eq:const_seq} imposes that the world volume is finite as we assume a smooth global interaction $\sigma$. This necessarily selects a universe with spatially compact sections  that is temporally finite (see \cite{Kaloper:2014fca} for a model realizing this condition); 
    
    \item The action Eq.~\eqref{eq:global_seq} is not additive over spacetime due to the global interaction term $\sigma$, leading to subtleties for its quantization and its embedding in a complete UV theory.
    \end{itemize}
Both these challenges can be evaded by the local version of the sequestering mechanics discussed below. For more discussions about the global proposal and its cosmological implications, see~\cite{Kaloper:2014dqa}.

\subsubsection{Local Sequestering}\label{sec:local_seq}

In the original formulation of vacuum energy sequestering, the smooth function $\sigma(x)$ is added directly to the gravitational action rather than to its Lagrangian, while $\kappa^2$ and $\Lambda$ are rigid Lagrangian multipliers. A local formulation of energy sequestering was put forward in \cite{Kaloper:2015jra}, where $\sigma(x)$ has a local form and the Lagrangian multipliers are promoted to fields. This is accomplished after modifying the action Eq.~\eqref{eq:global_seq} to
\begin{equation}
    S= \int d^4x \sqrt{-g} \left[\frac{\kappa^2(x)}{2}R - \Lambda(x) + \Lm \right] + \int \sigma \left(\frac{\Lambda(x)}{\mu^4}\right)F_{(4)} + \int \hat{\sigma}\left(\frac{\kappa^2(x)}{M^2_{\text{Pl}}}\right)\hat{F}_{(4)}\, ,
    \label{eqn:action_local_seq}
\end{equation}
where $F_{(4)}$ and $\hat{F}_{(4)}$ are two four-forms that satisfy usual Bianchi identities\footnote{We denote the rank of a $p$-form by the subscript $(p)$. }, i.e., locally $F_{(4)} =d A_{(3)}$ and $\hat{F}_{(4)} = d\hat{A}_{(3)}$, and the functions $\sigma$ and $\hat{\sigma}$ are assumed to be smooth. Note that $\Lambda(x)$ and $\kappa^2(x)$ are spacetime functions whose equations of motion are, respectively, 
\begin{equation}
    \frac{\sigma'}{\mu^4}F_{\mu\nu\rho\sigma} = \sqrt{-g} \varepsilon_{\mu\nu\rho\sigma}, \quad \frac{\hat{\sigma}'}{M_{\text{Pl}}^2}\hat{F}_{\mu\nu\rho\sigma} = -\frac{1}{2}R \sqrt{-g}\varepsilon_{\mu\nu\rho\sigma}\,,
\end{equation}
where $\varepsilon_{\mu\nu\rho\sigma}$ is the four-dimensional Levi-Civita symbol. The equation of motion for the metric is 
\begin{equation}
    \kappa^2 G^\mu_{\;\;\nu} = T^\mu_{\;\;\nu}+(\nabla^\mu \nabla_\nu - \delta^\mu_\nu \nabla^2)\kappa^2(x) - \Lambda(x) \delta^{\mu}_\nu\,,
\end{equation}
where $T_{\mu\nu}$ is the matter energy-momentum tensor. Due to the topological nature of the four-form actions, there are no flux contributions on the RHS in the above. 

On the other hand, it follows from the equations of motion for $A_{(3)}$ and $\hat{A}_{(3)}$,
\begin{equation}
    \frac{\sigma'}{\mu^4} \partial_\mu \Lambda = 0 = \frac{\hat{\sigma}'}{M_{\text{Pl}}^2} \partial_\mu \kappa^2\,,
\end{equation}
that $\Lambda$ and $\kappa^2$ are constants on-shell. This makes $\kappa$ set the bare Planck-scale value and $\Lambda$ play a similar role as the $\Lambda_b$ in the global case. Meanwhile, the traceful part of the metric equation gives
\begin{equation}
    \Lambda = \frac{1}{4}\langle T^\alpha_{\;\;\alpha}\rangle + \frac{1}{4}\kappa^2 \langle R\rangle \, ,
\end{equation}
where the last term can also be written in terms of the four-form fluxes as 
\begin{equation}
    \frac{1}{4}\kappa^2 \langle R \rangle = -\frac{\mu^4\kappa^2}{2 M^2_{\text{Pl}}}\frac{\hat{\sigma}'}{\sigma'}\frac{\int \hat{F}_{(4)}}{\int F_{(4)}} \equiv \Delta \Lambda \, .
\end{equation}
Hence, compared to the global case, Eq.~\eqref{eq:EE_seq} gets modified to 
\begin{equation}
    \kappa^2 G^\mu_{\;\;\nu} = T^\mu_{\;\;\nu} -\frac{1}{4}\delta^\mu_{\;\;\nu}\langle T^\alpha_{\;\;\alpha}\rangle -\Delta \Lambda \delta^\mu_{\;\;\nu} \, .
\end{equation}
The cancellation of the matter loop corrections to the cosmological constant operates as in the global case, but there is now an extra term $\Delta \Lambda$ in the residual cosmological constant. This is the main difference between the global and local approaches to vacuum energy sequestering. The other difference is that the first constraint in Eq.~\eqref{eq:const_seq} is absent such that infinite spacetime volumes can be considered.

Although the $\Lambda$ and $\kappa^2$ dependence of $\Delta \Lambda$ makes it UV-sensitive, the authors of \cite{Kaloper:2015jra} argue that the smoothness of $\sigma$ and $\hat{\sigma}$ guarantees that the variation of the prefactors in $\Delta \Lambda$, that depend actually on the dimensionless variables $\kappa^2/M^2_{\text{Pl}}$ and $\Lambda/\mu^4$, are bounded by $\mathcal{O}(1)$ numbers. Moreover, note that the fluxes are UV-insensitive to the choice of UV cutoff, being dominated by the IR scale in the integrals. For instance, for a constant $F_{(4)}$ and in a spacetime where $R$ is bounded, both $F_{(4)}$ and $\hat{F}_{(4)}$ fluxes diverge with the spacetime volume, such that their ratio can be finite. In summary, the value of $\Delta \Lambda$ can be finite, small, it is UV stable, and should be ultimately determined by observations. Finding support in the latter is a challenge for sequestering, due to its similarity with GR and hence difficulties in finding unique phenomenological signatures. Another challenge is to come up with a UV embedding (see \cite{Kaloper:2018kma,Padilla:2018hvp} for attempts in the context of axion monodromy).

\subsubsection{Non-local Approach}\label{ssec:carroll}

Another constrained approach, related to sequestering, was proposed by Carroll and Remmen~\cite{Carroll:2017gqo} (CR),
where a non-local constraint is applied to the action, and the averaged Lagrangian density is forced to vanish on-shell. The action being considered is
\begin{equation}
S_{\rm CR} = \eta\int d^4x\sqrt{-g}\Bigg[\frac{\MPl^2}{2}(R - 2\Lamb) + \Lm - \frac{1}{48}F^2_{(4)} + \frac{1}{6}\nabla\cdot(F_{(4)}A_{(3)})\Bigg]\,,
\label{eqn:Carrolls_nloc}
\end{equation}
where the parameter $\eta$ is a constant that acts as a global Lagrange multiplier.
It generalises the conventional measure through $\sqrt{-g} d^4x \to \sqrt{-g} \eta d^4x$.
In very general terms, the idea of introducing $\eta$ here is to enforce a cancellation between the four-form $F_{(4)}$ and $\Lamb$ in the equation of motion\footnote{On a more fundamental level, the occurrence of $F_{(4)}$ can be linked to the presence of membranes. But it has been argued that such an explanation might compromise the mechanism~\cite{DAmico:2017ngr}.}.

Although Eq.~\eqref{eqn:Carrolls_nloc} is different from Eqs.~\eqref{eq:global_seq} and~\eqref{eqn:action_local_seq}, 
its connection to both global and local sequestering can be seen by recasting it in an equivalent form (ignoring surface terms) as~\cite{DAmico:2017ngr}
\begin{equation}\label{eq:CR_Sequstering_relation}
S_{\rm CR} = \eta\int d^4x\sqrt{-g}\Bigg[\frac{\MPl^2}{2}R - \tilde\Lam + \Lm\Bigg] + \eta\sigma\Big(\frac{\tilde\Lam}{\mu^4}\Big)\int F_{(4)}\,,
\end{equation}
where
$\tilde\Lam = -\theta^2/2$ is defined as a new field variable  
with $\theta$ the magnetic dual of $F_{(4)}$,
$\Lamb$ is absorbed by $\Lm$,
we introduced a function $\sigma^2(z) = -2z$, and the four-form is re-scaled as $F_{(4)} \to F_{(4)}/\mu^2$.
Comparing with the sequestering proposals, we see that Eq.~\eqref{eq:CR_Sequstering_relation} 
bears resemblance with both the global and local sequestering mechanism.
One of the flux terms used in local sequestering is recovered here, but $\eta$ remains a global constraint and $\tilde\Lam$ is also a global variable.
Although $\eta$ can be made into a local parameter~\cite{Oda:2017qce}, it would not resolve the main problem of the CR model, i.e., radiative instability, that will be discussed below.

Going back to CR's approach, 
we proceed from Eq.~\eqref{eqn:Carrolls_nloc} 
to see how the CC is forced to vanish.
The constant scalar field $\eta$ admits, upon regularization $\int d^4x\sqrt{-g}\equiv V$,
an equation of motion that forces the averaged Lagrangian to vanish, i.e.,
\begin{equation}
\frac{1}{V}\int d^4x\sqrt{-g}\lll = \lan\lll\ran = 0 \, .
\label{eqn:EOM_eta_avg}
\end{equation}
Now introducing a constant scalar $\theta$ such that $F_{\mu\nu\rho\sigma} = \theta \sqrt{-g}\veps_{\mu\nu\rho\sigma}$ (obviously fulfilling the flux equations), we could rewrite the last two terms in Eq.~\eqref{eqn:Carrolls_nloc} as
\begin{equation}
-\frac{1}{48}F^2_{\mu\nu\rho\sigma} = \frac{1}{2}\theta^2,\quad \quad
\frac{1}{6}\nabla_\mu(F^{\mu\nu\rho\sigma}A_{\nu\rho\sigma}) = -\theta^2\,.
\label{eqn:LF_LDJ_theta}
\end{equation}
These can be substituted into the constraint Eq.~\eqref{eqn:EOM_eta_avg} to give
\begin{equation}
\frac{\MPl^2}{2}\big(\lan R\ran - 2\Lamb\big) + \lan\Lm\ran - \frac{1}{2}\theta^2 = 0\,,
\label{eqn:avg_EFE_constraint}
\end{equation}
which can be read as fixing $\theta$ in terms of $\Lamb$. The Einstein equation in this model reads
\begin{equation}
0 = \frac{\MPl^2}{2}\Big(R_\munu - \frac{1}{2}Rg_\munu + \Lamb g_\munu\Big) - \frac{1}{2}T_\munu + \frac{1}{4}g_\munu\theta^2\,.
\label{eqn:EOM_gmunu_aft_dual}
\end{equation}
The crucial observation is that $\Lamb$ is cancelled when we use Eq.~\eqref{eqn:avg_EFE_constraint} to substitute for $\theta$
\begin{equation}
R_\munu - \frac{1}{2}Rg_\munu + \frac{1}{2}\lan R\ran g_\munu = \MPl^{-2}\Big(T_\munu - \lan\Lm\ran g_\munu\Big)\,.
\label{eqn:avg_EFE_nloc}
\end{equation}
Comparing Eq.~\eqref{eqn:EOM_gmunu_aft_dual} and Eq.~\eqref{eqn:avg_EFE_nloc}, an effective CC can be identified in terms of the averaged quantities as
\begin{equation}
\Lam_\eff = \Big\lan\frac{1}{2}R + \MPl^{-2}\Lm\Big\ran\,.
\label{eqn:Carrolls_nloc_lam_eff}
\end{equation}

We see that the CR model indeed achieves a cancellation of the tree level CC and hence features a form of self-tuning.  At the same time, it remains compatible with all known observations, since standard GR solutions including the Friedmann equations are unchanged in this model.
It also circumvents Weinberg's no-go, as the global parameter $\eta$ alters the spacetime measure, similar to what is done in sequestering.
In terms of its non-locality and the implication to the DE problem and phase transitions, it is similar to the case of global sequestering (see discussions in Sec.~\ref{ssec:global_seq}).

On the other hand, CR's mechanism is not stable under radiative corrections. A simple and elegant argument was provided in \cite{DAmico:2017ngr}: as $\eta$ multiplies the whole action, it can be identified with $1/\hbar$, corresponding to the 0-th order in an $\hbar$ expansion. Now, since the loop-corrected action contains higher powers in $\hbar$ it will also come with higher powers of $1/\eta$. This however, changes the structure of the whole theory. In particular, a variation with respect to $\eta$ will no longer yield Eq.~\eqref{eqn:avg_EFE_nloc}.

A remedy to this problem is to have a purely geometrical global constraint, which can be achieved, for example, by an action such as
\begin{equation}
S_{\rm hybrid} = \int d^4x\sqrt{-g}\Bigg[\eta\frac{\MPl^2}{2}R - \MPl^2\Lamb + \Lm - \frac{1}{48}F^2_{(4)} + \frac{1}{6}\nabla\cdot( F_{(4)}A_{(3)})\Bigg]\,,
\label{eqn:S_hybrid}
\end{equation}
where the Lagrange multiplier is only attached to $R$. This however, takes us back to sequestering. To be precise, when we identify $\eta\MPl^2 = \kap^2$ and perform the same steps that led to \eqref{eq:CR_Sequstering_relation}, we obtain a hybrid between the local and global sequestering mechanisms.

\subsubsection{Unimodular Gravity} \label{subsec:unimodular}

\textit{Unimodular gravity} is a modification to GR already proposed by Einstein in 1919 \cite{inbook}. Similar to sequestering models where non-local constraints are imposed, in unimodular gravity the determinant of spacetime is constrained to be constant, $\sqrt{-g} = \const$. This condition is known as the \textit{unimodular condition} and, as we are going to see below, leads to the same field equations as those of GR with a CC. In the following we are going to present the basic premises of unimodular gravity at a classical level, as well as its connection to the CCP. An extensive overview of the details of quantum unimodular gravity lies beyond the scope of the present review. For further details on the quantisation of unimodular gravity see \cite{Smolin:2009ti, Padilla:2014yea, Percacci:2017fsy, Eichhorn:2013xr} and for criticisms see \cite{Kuchar:1991xd, Padilla:2014yea}. 

Let us start from the Einstein-Hilbert action (without a CC) in the presence of matter and including a Lagrange multiplier $\lambda(x)$ \cite{Padilla:2014yea},
\begin{equation}
    S_{\rm EH} = \int d^4 x \left[\sqrt{-g}\frac{M_{\rm Pl}^2}{2}R -\lambda(x)\left(\sqrt{-g}-\epsilon_0\right) \right] + S_{\rm m}\,.
\label{eq:SEH}
\end{equation}
Here, $S_{\rm m}$ refers to the action containing matter fields and $\epsilon_0$ is a constant volume-element. Varying Eq.~\eqref{eq:SEH} w.r.t. the metric leads to the field equations
\begin{equation}
    R_{\mu\nu} -\frac{1}{2}g_{\mu\nu}R = \frac{1}{M_{\rm Pl}^2}T_{\mu\nu} - \frac{\lambda(x)}{2}g_{\mu\nu} \,,
\label{eq:Field_eqs}
\end{equation}                 
while varying w.r.t. the Lagrange multiplier leads to $\sqrt{-g} = \epsilon_0$. 
Note that this unimodular condition breaks the full diffeomorphism invariance of the theory. In Eq.~\eqref{eq:Field_eqs}, the EMT is given by the variation of the matter action, $S_{\rm m}$, as shown in Eq.~\eqref{eq:EMT_mat}.
Taking the divergence of Eq.~\eqref{eq:Field_eqs} and accounting for energy-momentum conservation, $\nabla_{\mu} T^{\mu\nu} = 0$, leads to
\begin{equation}
    \partial_{\mu}\lambda(x) = 0\,,
\end{equation}
with solution
\begin{equation}
    \lambda = \mathrm{constant} \, . 
\end{equation}
In other words, the Lagrange multiplier is fixed to be a constant and the CC can be identified with $\Lambda = \lambda/2$.

An alternative derivation of the field equations is performed in the Henneaux-Teitelboim formulation of unimodular gravity \cite{Smolin:2009ti, Henneaux:1989zc, Fiol:2008vk}, where full diffeomorphism invariance remains unbroken. The action of the theory is written as \cite{Fiol:2008vk}
\begin{equation}
    S_{\rm HT} = \int d^4 x \left\{\sqrt{-g}\left[\frac{M_{\rm Pl}^2}{2}\left(R - 2\phi\right) + \mathcal{L}_{\mathrm m}\right] + M_{\rm Pl}^2\phi\partial _{\mu}\tilde{A}^{\mu}\right\}\,,
\end{equation}
where $\tilde{A}^{\alpha} = \frac{1}{6}\epsilon^{\alpha\beta\gamma\delta}A_{\beta\gamma\delta}$ is the vector density associated with a three-form $A_{\alpha\beta\gamma}$. In this formalism, the unimodular condition arises from the equation of motion of the scalar field $\phi(x)$, as $\sqrt{-g} = \partial_{\mu}\tilde{A}^{\mu}$. Additionally, the equation of motion of the vector density field $\tilde{A}^{\mu}$ is $\partial _\mu \phi(x) = 0$. The latter, again, shows that $\phi = \mathrm{const.}$ can be identified as the cosmological constant $\Lambda$. Finally, varying the action w.r.t. the metric $g_{\mu\nu}$ leads to the Einstein equations with a CC.

Thus, it becomes clear that, within unimodular gravity, the CC arises as an integration constant in the Einstein field equations. It has been argued that this provides a conceptual exit from the CCP \cite{Ng:1990xz, Smolin:2009ti}. A way to understand this argument is by stating that any change in the EMT of the form $T_{\mu\nu}\rightarrow T^{'}_{\mu\nu} = T_{\mu\nu} + g_{\mu\nu} C$ can be absorbed into the integration constant $\Lambda$ by a shift $\Lambda\rightarrow \Lambda + C/M_{\rm Pl}^2$. This is interpreted as the curvature of spacetime not coupling to quantum corrections of the form $g_{\mu\nu}C$. The claim, then, is that such a shift can address the CCP by cancelling off a given amount of vacuum energy. Here, we should note that the same logic can be applied in GR, where $\Lambda$ again is a free constant whose value can be chosen arbitrarily.

The argument above could, indeed, constitute a solution to the CCP, if the latter was limited to its old formulation, as explained in detail in Sec.~\ref{sec:problems}. In other words, if the issue merely consisted of the fact that accounting for the vacuum energy contribution 
to the EMT leads to a much larger CC than the one measured by observations today. Such a problem could be resolved by using the freedom of fine-tuning an integration constant to match the observed value. However, a solution cannot be provided when one considers the new-CCP. In the latter, the renormalisation prescription of the effective theory is unstable under higher-order loop corrections, requiring the bare value of the cosmological constant to be re-tuned every time an additional loop contribution is taken into account. Since in unimodular gravity (like in GR) the integration constant playing the role of the CC can only be fixed once, the conceptual problem remains \cite{Padilla:2014yea, Nojiri:2015sfd}. In fact, the same difficulty is encountered when attempting to solve the classical CCP in the unimodular gravity framework. This should not come as a surprise since unimodular gravity constitutes a minimal modification to GR that is also not dynamical in any way. Hence, unimodular gravity is only able to resolve the CCP problem to the same extent as standard GR itself. 

\subsection{Massive gravity}\label{sec:massive_gravity}

Massive gravity theories are a natural extension of GR, and the formulation is of inherent theoretical interest. However, massive gravity also provides an interesting angle to the cosmological constant problems via the degravitation mechanism, where the graviton mass acts as a filtering scale in the coupling between gravity and a cosmological constant term. In that sense, the right question to ask about the CCPs might not be \textit{why the cosmological constant term is so small} but rather \textit{why it gravitates so little}. Within GR, diffeomorphism invariance guarantees that gravitation couples universally to all sources, but in more generalised theories of gravity there is a meaningful difference. In the following, we will briefly recap how this mechanism works, show how it emerges naturally in the context of massive gravity, and review the development and current status of massive gravity theories.

For extensive reviews about massive gravity, see \cite{Hinterbichler:2011tt,deRham:2014zqa,Schmidt-May:2015vnx,deRham:2016nuf,Heisenberg:2018vsk}.

\subsubsection{Degravitation} \label{sec:degravitation}

The \textit{degravitation} mechanism \cite{Arkani-Hamed:2002ukf,Dvali:2007kt,deRham:2007rw} can be seen as an IR modification of GR that prevents sources characterized by wavelengths $\lambda_S$ larger than a given IR scale $L$ from gravitating. At the level of Einstein equations, it is phenomenologically implemented by promoting Newton's constant to a scale-dependent differential operator
\begin{equation}
    \, G_{\mu\nu} = 8\pi  \GN(L^2 \Box ) \, T_{\mu\nu}\,.
    \label{Eq.:pheno_degravitation}
\end{equation}
Expanding the sources in terms of its mode functions, the gravitational coupling behaves like
\begin{align}
     \GN (L^2 \lambda_S^{-2}) = 
    \begin{cases}
     \GN  &\quad \mathrm{for} \quad \lambda_S \ll L\\
    0 &\quad \mathrm{for} \quad \lambda_S \gg L
    \end{cases} \,,
\end{align}
acting as a \textit{high-pass filter}, thus preventing deep-IR modes, $\lambda > L$, from gravitating. The cosmological constant is a particular case with an infinite characteristic wavelength, thus it degravitates. Therefore, regardless of the value of $\rho_{\rm vac}$, as long as it behaves as a cosmological constant it does not source Einstein equations.

In the context of gravitational theories, this mechanism can be implement by having massive gravitons or a resonance\footnote{Superposition of small-mass massive spin-2 states.}~\cite{Dvali:2007kt,deRham:2007rw}. DGP, briefly discussed both in the context of derivative screening in Sec.~\ref{sec:derivative_screening} and of braneworld models in Sec.~\ref{sec:xdim}, can be considered a special case of the resonant graviton. In the rest of this section, we focus on the linear and non-linear constructions of massive gravity.

\subsubsection{Linear Massive Gravity}\label{sec:linear_massive_gravity}

The search for a consistent generalisation of GR that describes the behaviour of massive spin-2 fields is an old problem. The Lagrangian for a linear theory of a massless spin-2 field $h_{\mu \nu}$ can be written as
\begin{equation}
    \mathcal{L}_\mathrm{lin} = h^{\mu \nu} \mathcal{E}_{\mu \nu}^{\alpha \beta} h_{\alpha \beta} - M^{-1}_{\rm Pl} h^{\mu \nu} T_{\mu \nu} \, ,
\end{equation}
where the kinetic terms are given by
\begin{equation}
    \mathcal{E}_{\mu \nu}^{\;\;\;\;\alpha \beta} = \frac{1}{2} 
    \left(
    \eta_\mu^{\;\;\alpha} \eta_\nu^{\;\;\beta} \Box - \eta_\nu^{\;\;\beta} \partial_\mu \partial^\alpha  - \eta_\mu^{\;\;\beta} \partial_\nu \partial^\alpha 
     + \eta_{\mu \nu} \partial^\alpha \partial^\beta + \eta^{\alpha \beta} \partial_\mu \partial_\nu - \eta_{\mu \nu} \eta^{\alpha \beta} \Box 
    \right) \, .
\end{equation}
Fierz and Pauli showed in 1939 \cite{Fierz:1939ix} that the linear theory of a massive spin-2 field is uniquely given by 
\begin{equation}
\label{eq:linear_massive_gravity}
     \mathcal{L}_\mathrm{FP} = h^{\mu \nu} \mathcal{E}_{\mu \nu}^{\alpha \beta} h_{\alpha \beta} - \frac{m_\mathrm{FP}}{2} \left(h^{\mu \nu} h_{\mu \nu} - h^2 \right) - M^{-1}_{\rm Pl} h^{\mu \nu} T_{\mu \nu} \, ,
\end{equation}
with the Fierz-Pauli mass $m_\mathrm{FP}$, while other possible mass terms lead to the existence of ghosts. From the trace and the divergence of the equations of motion, we can derive
\begin{equation}
\label{eq:FP_eom}
    \left(\Box - m_\mathrm{FP}^2 \right) h_{\mu \nu} = M^{-1}_{\rm Pl} \left[ T_{\mu \nu} - \frac{1}{3} \left( \eta_{\mu \nu} - \frac{1}{m^2_\mathrm{FP}} \partial_\mu \partial_\nu T\right) \right] \, ,
\end{equation}
which realises the degravitation condition outlined above, where the source $T_{\mu \nu}$ is seen through a high-pass filter $\left(1-\frac{m_\mathrm{FP}^2}{\Box} \right)^{-1} $. For sources with small associated scale $\lambda_S \ll m_\mathrm{FP}$, the filter plays a negligible role, while for sources $\lambda_S \gg m_\mathrm{FP}$ such as a cosmological constant the coupling is suppressed.
Considering non-relativistic sources in Eq.~\eqref{eq:FP_eom} leads to a Yukawa-type potential for the gravitational interaction \cite{Dvali:2007kt}
\begin{equation}
    V(r) \sim \frac{\exp\big(-m_\mathrm{FP}r\big)}{r} \, ,
\end{equation}
which again suppresses the strength of the gravitational interaction on large scales. For the theory to be compatible with a standard cosmic evolution, the graviton mass should not exceed $m_{\rm FP} \sim H_0 \sim 10^{-33} \, \mathrm{eV}$.

Unfortunately, not all is well with the linearised theory. It is already apparent in Eq.~\eqref{eq:FP_eom} that the limit $m_\mathrm{FP} \rightarrow 0$ is not smooth, as pointed out by van Dam, Veltman and Zakharov \citep[the vDVZ discontinuity,][]{vanDam:1970vg, Zakharov:1970cc}. A careful accounting of the propagating degrees of freedom in the theory (e.g. by using the Stückelberg formalism \cite{Stueckelberg:1957zz}) demonstrates that the massless limit of Fierz-Pauli gravity is indeed not GR, but GR plus an additional attractive scalar field. This leads to a mismatch with observation: if the coupling to non-relativistic sources is supposed to match Newtonian theory (where both the tensor and the new scalar field couple), then gravitational lensing in the new theory is weaker by a factor $3/4$ \cite{Hinterbichler:2011tt} since the scalar does not couple to the traceless energy-momentum tensor of light.

However, the additional interaction responsible for this discrepancy differs once higher orders of $h_{\mu \nu}$ are included in the action. Non-linear interactions lead to a suppression of the additional scalar degree of freedom inside the Vainshtein radius around massive sources \cite{Vainshtein:1972sx}, independent of the specific non-linear completion \cite{Arkani-Hamed:2002bjr} as explained in Sec.~\ref{sec:derivative_screening}. As a result, GR is expected to be restored close to massive sources such as the Earth (at the price of leaving the regime of applicability of the linear theory).

\subsubsection{Nonlinear Theories}\label{sec:nonlinmassgrav}

The question is then whether a non-linear completion of the Lagrangian given by Eq.~\eqref{eq:linear_massive_gravity} can be brought in agreement with observations and theoretical consistency requirements. After all, most non-linear extensions are plagued by the Boulware-Deser ghost, since the unique structure of the mass term that guarantess the consistency of the Fierz-Pauli theory is generally spoiled by higher-order corrections \cite{Boulware:1972yco}.

A solution was found by de Rham, Gabadaze and Tolley (dRGT) \cite{deRham:2010kj, deRham:2010ik} by constructing a non-linear generalisation of Eq.~\eqref{eq:linear_massive_gravity} around an arbitrary fixed background metric, and Hassan and Rosen proved the absence of ghosts in the theory to all orders \cite{Hassan:2011vm}. Nonetheless, even for ghost-free theories, demanding unitarity and analyticity of the S-matrix severely constrains the possible graviton mass scale \cite{Schmidt-May:2015vnx, Bellazzini:2017fep, deRham:2018qqo}. When combined with observational constraints from propagation of gravitational waves \cite{LIGOScientific:2017ync, Baker:2017hug} and solar system measurements \cite{Bernus:2019rgl} it seems that the theory becomes non-viable.

Making the reference metric of dRGT gravity itself dynamical leads to bimetric theories of gravity with interacting massless and massive spin-2 fields \cite{Hassan:2011vm, Schmidt-May:2015vnx} (and corresponding multi-gravity extensions with additional tensor fields \cite[e.g.][]{Khosravi:2011zi}). The physical metric responsible for gravitational effects is a mixture of the two tensor fields, and the behaviour in the linearised regime is governed by a mixing angle between massive and massless modes. As the mixing becomes small, one recovers the massless spin-2 predictions of GR, while a large mixing leads to the phenomenology of linear massive gravity.

The free mixing angle allows bimetric gravity to circumvent the existing limits on the graviton mass that are so problematic for dRGT. On the other hand, a large mixing is required for degravitation in the linear regime \cite{Platscher:2016adw} where the theory approaches the general phenomenology of linear massive gravity, with all advantages and problems connected to that: effects of a bare cosmological constant can be made to degravitate, but the additional fifth force from Fierz-Pauli theory described in Sec.~\ref{sec:linear_massive_gravity} reappears. Within the Vainshtein radius, the gravitational effect of $\Lambda$ does \emph{not} degravitate -- depending on how large the Vainshtein radius is chosen, this can slightly alleviate the problem, since upper bounds on vacuum energies from e.g. planetary orbits are 16 orders of magnitude larger than the cosmological bound \cite{Martin:2012bt}.

Setting constraints from cosmological structure formation remains difficult since perturbations in bigravity become non-linear very early on. While these instabilities do not necessarily rule out the theory \cite{Hogas:2019ywm}, there is no framework available yet to calculate non-linear predictions for cosmological scales in generality. It is however possible to use probes of the cosmic background expansion such as BBN \cite{Hogas:2021saw}, BAO, or supernova data \cite{Caravano:2021aum,Hogas:2021lns} to constrain the theory. There is a large parameter space available that produces a $\Lambda$CDM-like expansion history, but this parameter space cannot degravitate CC effects while also fitting the available data. 

In absence of degravitation, the problem of UV sensitivity of vacuum energy contributions remains. An interesting new angle is that if the problem of UV sensitivity is solved, interaction terms between the two metrics provide an additional source that behaves like a cosmological constant, but is itself protected from quantum corrections \cite{Hassan:2011vm}.

\subsection{Self-tuning with Horndeski theories} \label{sec:self-tuning}

The essence of self-tuning is to introduce a dynamical degree of freedom to GR, typically a scalar field $\phi(t)$, which is  used to cancel the contribution of the CC's energy density $\rhovac$  in the evolution equations of the Universe, effectively decoupling it from gravity in the cosmological context. We note here that we have broken Poincar\'e invariance by enforcing a non-trivial time dependence of $\phi$, which allows us to bypass Weinberg's theorem (see Sec.~\ref{sec:no-go}).
The cancellation relies on trivially satisfying the equation of motion for $\phi$, or making the equation redundant. Once this is achieved, the first Friedmann equation defines the relation $\dot{\phi} = f(\rhovac)$, where $f$ is a general function, which provides a dynamical constraint equation for $\phi$ in terms of $\rhovac$. This allows $\phi$ to dynamically match $\rhovac$ across cosmic evolution and through any phase transition the Universe may undergo. Such dynamical matching is what we will refer to in this section as {\it self-tuning}. Note this approach is not restricted to the cosmological setting as we shall shortly see, although its applicability to multiple settings simultaneously is not guaranteed. 

The models we look at here will all fall under the Horndeski scalar-tensor class of theories [see Eq.~\eqref{eq:horndeski}] which ensures the model is ghost-instability free. We note there have been significant extensions to this class \cite{Gleyzes:2014dya,Lin:2014jga,Gleyzes:2014qga,Gao:2014fra,Kase:2014cwa,Frusciante:2015maa,Horava:2009uw,Langlois:2015cwa,Langlois:2015skt,Crisostomi:2016czh,Motohashi:2016ftl} which we do not consider here. The Horndeski class as well as these beyond-Horndeski models have been well constrained by data, in particular by measurements of the propagation speed of gravitational waves (see \cite{Kobayashi:2019hrl} for a review on recent developments). Despite this, a large theory space remains open with interesting phenomenology which can be tested with cosmology.

Horndeski theory permits four free functions, $G_i(\phi, X)$, of $\phi$ and its canonical kinetic energy $X = -(\partial \phi)^2/2$ in the Lagrangian. In this context, for a flat  Friedman-Lemaître-Robertson-Walker (FLRW) background, we have the following three coupled differential equations coming from the 00-th component of the metric field equations, their trace and the variation of the Lagrangian with respect to the scalar field respectively
\begin{subequations}
\begin{align}
    F_1(H, \phi, \dot{\phi}, \rhom, \rhovac) = & 0  \, , \label{eq:st1} \\ 
    F_2(H, \dot{H}, \phi, \dot{\phi}, \ddot{\phi}, \rhom, \rhovac) = & 0 \, , \label{eq:st2}  \\ 
    F_3(H, \dot{H}, \phi, \dot{\phi}, \ddot{\phi}) = & 0 \, , \label{eq:st3}
\end{align}
\end{subequations}
where we have assumed a pressureless matter component. We then assume that the scale factor has an attractor solution. Given current observations one could choose this to be a de Sitter solution i.e., $a = e^{H_0 t}$, which describes the asymptotic future of a universe containing a cosmological constant. This being said, equally pertinent to the CCPs is the need for such self-tuning to occur in the Solar System (so within black hole solutions \cite{Babichev:2013cya} or the Minkowski solution \cite{Charmousis:2011bf} for example).

Considering the de Sitter attractor, a viable self-tuning model is restricted to the subset of Horndeski functions that fulfill the following (see Sec.~7 of \cite{Brax:2017idh} for example)
\begin{enumerate} 
    \item 
    The equation of motion for $\phi$ Eq.~\eqref{eq:st3} is trivially satisfied on the attractor or redundant with the other two equations; 
    \item  
    The Friedmann equation Eq.~\eqref{eq:st1} should depend on $\dot{\phi}$ for the tuning to be dynamic; 
    \item 
    The Friedmann equation must admit non-trivial expansion histories before hitting the attractor solution; 
    \item
    The theory incorporates a screening mechanism (see Sec.~\ref{sec:screening}). 
\end{enumerate}%
The first  condition leaves us with only two dynamical equations on the attractor. In particular, $F_{1}^\attractor$ is then allowed to set the value of $\dot{\phi}$ in terms of $\rhovac$, i.e., we have $\dot{\phi} = f(\rhovac)$. Moreover, any change in $\rhovac$, even a discontinuous one, appropriately changes the value of $\phi$, allowing for self-tuning over phase transitions.

Given these basic criteria, we now look at two different models in the literature that achieve self-tuning. These essentially differ in how they achieve the first criterion.

\subsubsection{The Fab-4}\label{sec:fab4}
One of the first instances of self-tuning in this context was the Fab-4, originally proposed in \cite{Charmousis:2011bf,Charmousis:2011ea} and named in homage to the Beatles. The Lagrangian functions were named after the band members 
\begin{subequations}
\begin{align}
    \mathcal{L}_\john &= \sqrt{-g} V_\john(\phi) G^{\mu \nu} \nabla_\mu \phi \nabla_\nu \phi \, ,  \\ 
    \mathcal{L}_\paul &= \sqrt{-g} V_\paul(\phi) P^{\mu \nu \alpha \beta } \nabla_\mu \phi \nabla_\nu \phi \nabla_\alpha \phi \nabla_\beta \phi \, , \\ 
    \mathcal{L}_\george &= \sqrt{-g} V_\george(\phi) R \, , \\
    \mathcal{L}_\ringo &= \sqrt{-g} V_\ringo(\phi) \hat{G} \, , 
\end{align}
\end{subequations}
where $R$ is the Ricci scalar, $G_{\mu \nu}$ the Einstein tensor, $P_{\mu \nu \alpha \beta}$ the double dual of the Riemann tensor and $\hat{G}$ the Gauss-Bonnet scalar. The various potentials are chosen to allow for self-tuning on the attractor, while the coupling to curvature terms is how the trivial solution to the Klein-Gordon equation is ensured. 

The original setup assumed a Minkowski vacuum attractor solution, which we consider for illustration. We find $F_3=0$ [see Eq.~\eqref{eq:st3}] identically on vanishing curvature backgrounds as each term is coupled to curvature (see \cite{Charmousis:2014mia} to see this explicitly). The Friedmann equation on the attractor is then found to be \cite{Charmousis:2014mia} 
\begin{equation}\label{Fab4-selftuning}
     F_1^\attractor(H, \phi, \dot{\phi}, \rhom, \rhovac) = V_\john(\phi) (\dot{\phi} H)^2 +  V_\paul(\phi) (\dot{\phi} H)^3 -  V'_\george(\phi) (\dot{\phi} H) + \rhovac = 0 \, ,  
\end{equation}
which puts a constraint on the forms of the possible potentials $V(\phi)$. These can also be chosen so as to reproduce the correct eras in our Universe's expansion history as according to observations \cite{Copeland:2012qf}. Further, the non-trivial derivative interactions of the {\it john} and {\it paul} terms leave the possibility for Vainshtein screening, although this has not been shown to be viable in tandem with self-tuning explicitly \cite{Niedermann:2017cel}, and is generally problematic for light scalar fields \cite{Chiba:2006jp}. 

Finally, we remark that the recent measurement of the speed of gravitational waves \cite{LIGOScientific:2017vwq,LIGOScientific:2017ync} has severely constrained the Fab-4 model \cite{Ezquiaga:2017ekz}, although the applicability of these constraints for such theories has been challenged in \citep{deRham:2018red,LISACosmologyWorkingGroup:2022wjo}.

\subsubsection{Well-tempered self-tuning}\label{sec:well-temp}
The second model we present arose to address some of the inital problems of the Fab-4, in particular its struggle in consistently reproducing all of the correct eras of the expansion history together while also cancelling a large value of $\rhovac$ \cite{Linder:2013zoa}. This being said, specific classes of scalar field potentials in Fab-4 also ameliorates this issue \cite{Copeland:2012qf}. In \cite{Appleby:2018yci,Appleby:2020njl,Bernardo:2021izq} the authors attempt to moderate the self-tuning in order to allow for the correct expansion history. The model of \cite{Appleby:2018yci} achieves the first self-tuning criterion by making the identification on the attractor
\begin{equation}
     F_2^\attractor(H,  \phi, \dot{\phi}, \ddot{\phi}, \rhom) = F_3^\attractor(H,  \phi, \dot{\phi}, \ddot{\phi}) \, , \label{eq:wtc}
\end{equation}
where we have removed the $\rhovac$ dependency using Eq.~\eqref{eq:st1} and assumed a de Sitter attractor with $\dot{H} = 0$. This idenfitication makes Eq.~\eqref{eq:st2} and Eq.~\eqref{eq:st3} degenerate and imposes a constraint on the Horndeski functions allowed in this scenario. The advantage of making these two equations degenerate only on the attractor, rather than have a trivial solution to Eq.~\eqref{eq:st3}, as in the Fab-4, is that it implicitly requires $\rhom = 0$ for the degeneracy to hold, i.e., the scalar field only screens $\rhovac$, allowing for a non-trivial expansion history. 

Further, the constraint on the Horndeski class imposed by Eq.~\eqref{eq:wtc} does not preclude a screening mechanism. In fact, the original proposal of \cite{Appleby:2018yci} considered the inclusion of $G_2(\phi,X)$ and $G_3(\phi,X)$ Lagrangian terms which give the capacity for Vainshtein and chameleon screening, although such a mechanism has not been shown to be possible explicitly. Lastly, we note that the functional forms of $G_2$ and $G_3$ should also be such that a $\dot{\phi}$ dependency remains in Eq.~\eqref{eq:st1}.

\subsubsection{Outlook}
Horndeski self-tuning can grant a removal of the CC contribution from various spacetimes including Minkowski and FLRW, and is robust to phase transitions. Further, it is able to recover a non-trivial expansion history in the cosmological context, which allows for late-time acceleration. The models also have the capacity to screen any deviations from GR at small physical scales in accordance with solar system tests. Lastly, the presence of a shift symmetry in some such models, for example that of the model described in \cite{Khan:2022bxs}
\begin{equation}
    \phi \rightarrow \phi + a , \qquad \qquad  \Lambda \rightarrow \Lambda + a c_1 M^3 \lambda \, ,  
\end{equation}
where $M$ is a mass of $\phi$ and $c_1$ the scalar field potential's Lagrangian coefficient, may be useful in addressing the impact of quantum corrections \cite{Padilla:2015aaa}. 

But despite the Horndeski self-tuning program being an interesting approach, it leaves many things to be desired. Self-tuning requires tuning of the mass scales and Lagrangian coefficients in order to recover the correct expansion history and effectively cancel the CC. These parameters are not guaranteed to be stable against radiative corrections (see \cite{Khan:2022bxs} for an example of such parameter tunings). 

On this point, in these models one may assume that $\rhovac$ is the value coming from the UV-complete theory, with all corrections to the CC accounted for. The solution for $\phi$ can then adapt dynamically to this value. This then says that our theory is `stable' against quantum corrections to the vacuum energy. The question is then whether or not the parameters of the theory, for example the scalar field potential or couplings to curvature, are stable against such corrections. One can force the coupling to the matter sector of the field to be weak (see for example \cite{Copeland:2021czt}) which helps protect the theory against radiative corrections and, at the same time, avoids fifth force constraints, although it may have implications for phase transitions as the response to changes in vacuum energy is slowed down.

An issue already pointed out is that the mechanism cancels $\rhovac$ on the attractor but  to address the large value of $\rho_{\mathrm \Lambda}$ at earlier times, which would change the expansion history significantly. Similarly, self-tuning as discussed here assumes some specific attractor solution at which the CC is cancelled by $\phi$ rather than at the level of the field equations as in the sequestering approach (see Sec.~\ref{sec:constrained_gravity}). This leaves the issue of the CC in all solutions but the attractor, which is clearly not sufficient as we do not observe a large CC contribution both on cosmological and solar system scales. Related to this is the issue of screening. As seen in Sec.~\ref{sec:screening} we require a choice of Horndeski functions that effectively cancel $\phi$'s contribution to the Poisson equation [see Eq.~\eqref{eq:modpoisson}] either through a well chosen potential or derivative term. Horndeski functions that screen  this contribution to the Poisson equation and self-tune the CC is yet to be shown explicitly. 

Finally we note that there have been many interesting proposals that take different approaches to self-tuning such as  \cite{Lacombe:2022cbq,SobralBlanco:2020too,Amariti:2019vfv,Charmousis:2017rof,Copeland:2021czt}, which we have not discussed here but are worth considering. They all involve some cancellation of the CC using an additional degree of freedom, with some such as  \cite{SobralBlanco:2020too,Lombriser:2019jia} making a connection with sequestering and others invoking higher dimensions such as \cite{Charmousis:2017rof}.

\subsection{Braneworld models} \label{sec:xdim}

The no-go theorems in Sec.~\ref{sec:no-go} assumed vacuum energy to be constant throughout space and time. In a nutshell, extra dimensional models, or more specifically braneworld models, relax that assumption by breaking the translational invariance of vacuum energy in the direction of the extra space.  This is possible because the Standard Model and hence \textit{our} Universe is assumed to be confined on a spatial hypersurface, referred to as a \textit{brane}, that is embedded in a higher dimensional bulk spacetime. In fact, in the simplest models, only gravity is allowed to propagate in the bulk spacetime.  As a consequence, the matter loops that give rise to vacuum energy (and its quantum corrections) are equally confined on the brane. Since vacuum energy is still constant along the intrinsic brane directions it acts as a surface  or brane tension $\lambda$. Crucially, this implies that it gravitates differently from a space-filling vacuum energy in a four-dimensional (4D) theory.

We will be mostly interested in the low-energy phenomenology of braneworld models; after all, the cosmological constant is an extreme IR source. It should still be noted that branes are important building blocks of string theory, which makes this class of solutions particularly interesting from a high-energy perspective. In fact, D-branes are non-perturbative states in string theory and appear both in tentative string constructions of SM \cite{Blumenhagen:2005mu, Maharana:2012tu} and four-dimensional maximally symmetric spaces such as $\text{dS}_4$ \cite{Kachru:2003aw, Balasubramanian:2005zx}. Since they source the $p$-form massless fields in the theory, D-branes have become an important ingredient in flux compactifications and string phenomenology \cite{Grana:2005jc, Blumenhagen:2006ci}. Discussing the fate of the CCP in string theory is beyond the scope of this review, and we will be interested in extra dimensional models regardless of their embedding into string theory. 
Moreover, the literature on braneworld models as solutions to the CC problem is rather abundant~\cite{Koyama:2007rx}. In this review, we will therefore focus on six-dimensional models that rely on a simple geometric mechanism, featuring conical deficit angles, to achieve self-tuning. For other models in five dimensions (and a discussion of their shortcomings) see~\cite{Arkani-Hamed:2000hpr,Kachru:2000hf,Forste:2000ps,Forste:2000ft,Csaki:2000wz,Binetruy:2000wn}, and more recently also~\cite{Charmousis:2017rof,Lacombe:2022cbq}\footnote{These models typically generalise the Randall-Sundrum model~\cite{Randall:1999vf}, which itself cannot address the CCP due to an immediate tuning between the brane tension and the bulk cosmological constant.}.

In Sec.~\ref{sec:comsic_string} we will start with a simple toy model to illustrate the general idea. In Sec.~\ref{subsec:sled} we will discuss a more complete model before providing a general outlook in Sec.~\ref{sec:xdim_outlook}.

\subsubsection{Our Universe as a cosmic string in six dimensions} \label{sec:comsic_string}

\begin{figure}[th]

\begin{subfigure}{0.45\textwidth}
\centering
\includegraphics[height=1.1\linewidth]{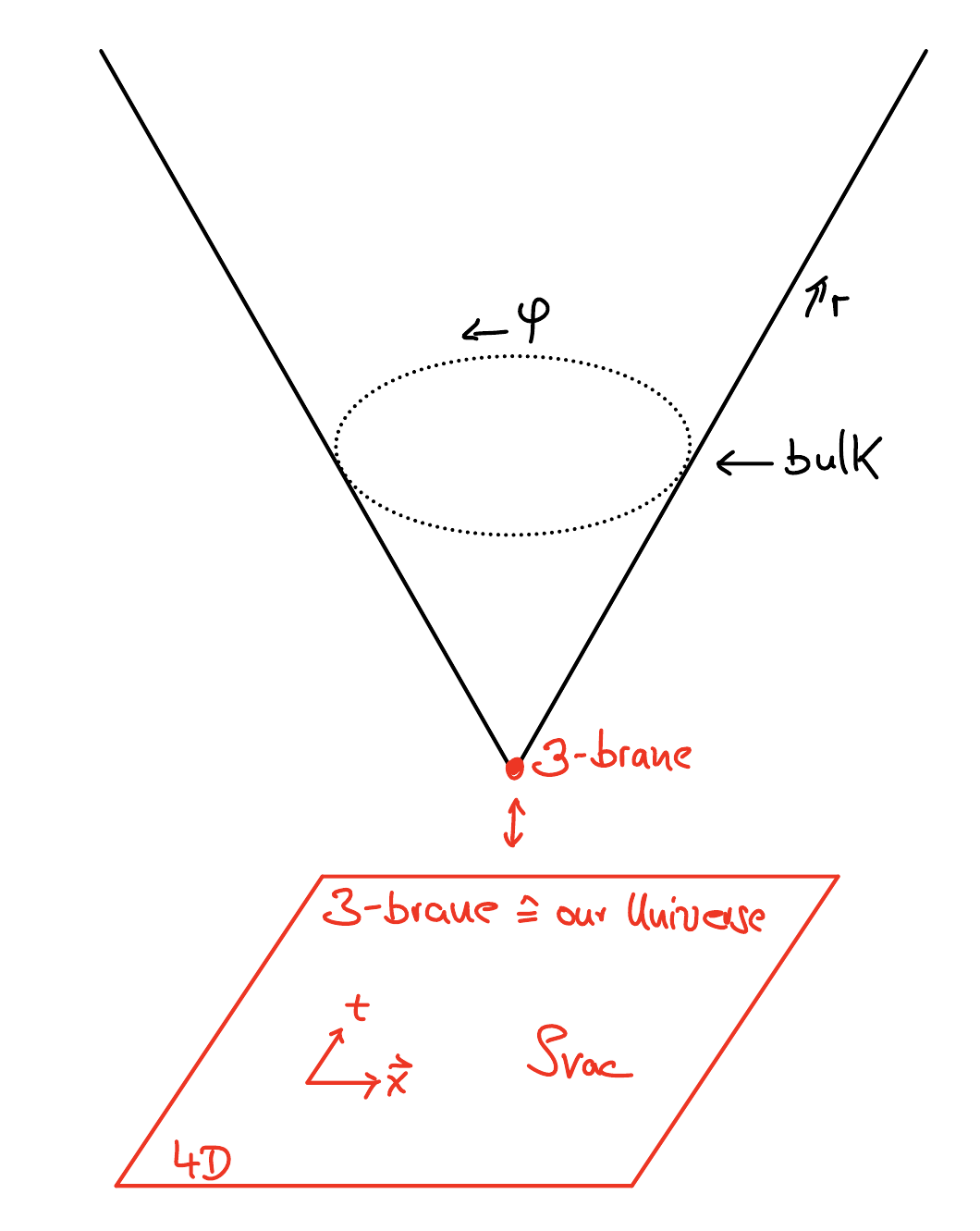} 
\caption{Conical geometry in Eq.~\eqref{ansatz_xdim}}
\label{fig:subim1}
\end{subfigure}
\quad
\begin{subfigure}{0.45\textwidth}
\centering
\includegraphics[height=1.1\linewidth]{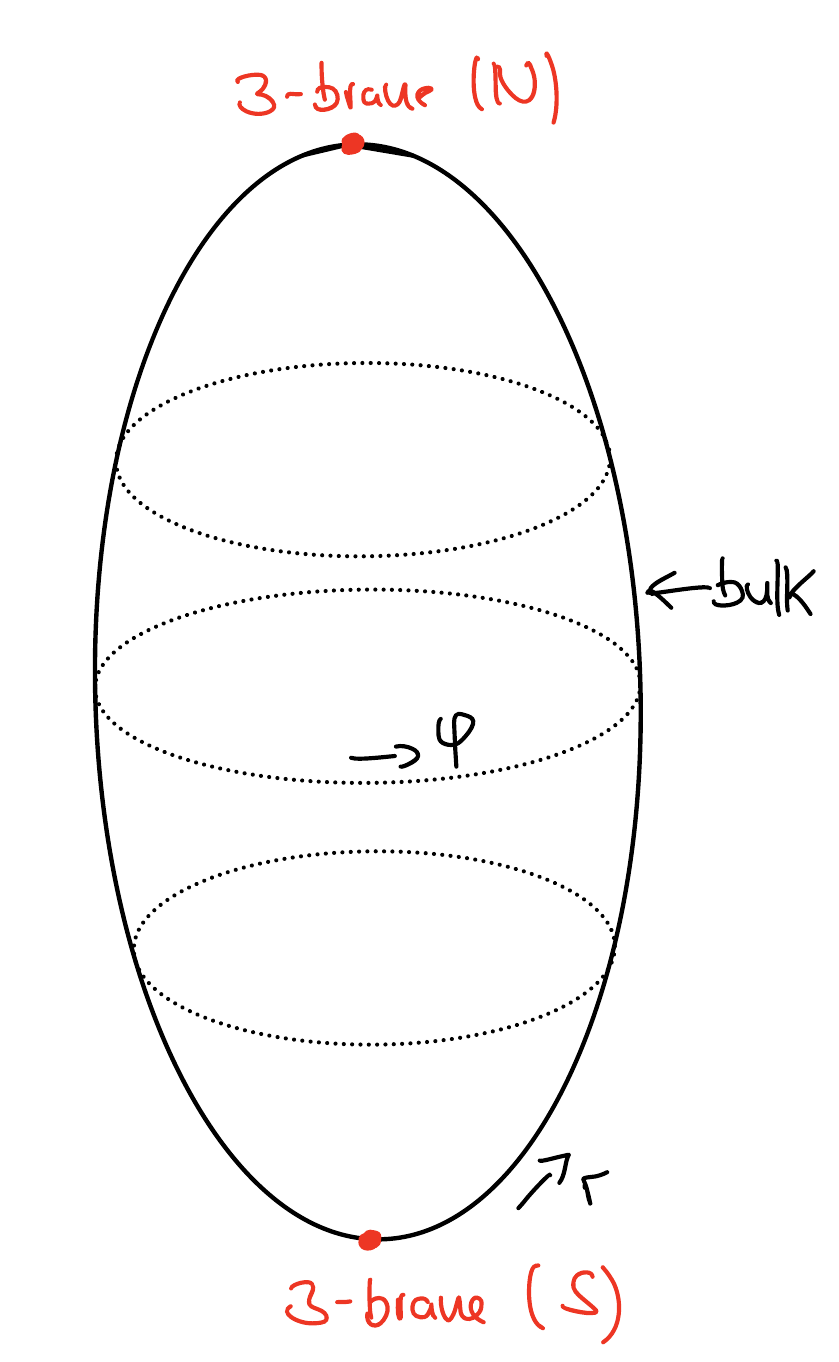}
\caption{Rugby ball geometry in Eq.~\eqref{eq:ansatz}}
\label{fig:subim2}
\end{subfigure}

\caption{Geometry (a) and (b) are employed in higher-dimensional generalizations of the DGP model (infinite extra space volume) and the SLED model (finite extra space volume), respectively. The red dot corresponds to our Universe.}
\label{fig:6D}
\end{figure}

To illustrate the six-dimensional mechanism (and to introduce the necessary notation), we consider the simple case where our 4D Universe is a pure tension brane in a six dimensional (6D) bulk spacetime of infinite size. The corresponding action, $S = S_\mathrm{bulk} + S_\mathrm{brane}$, can be decomposed into a bulk and brane part, respectively,
\begin{subequations}
\label{6d_action}
\begin{align}
    S_\mathrm{bulk} &= \frac{M_6^4}{2} \int d^6X \sqrt{-g_6} \, R_6\,,   \\
    S_\mathrm{brane} &=  \int d^4 x \sqrt{-g_4}\,\left( -\lambda_\mathrm{bare} + \Lm\right)   \, .
\end{align}
\end{subequations}
Here, $M_6$ is the fundamental gravity scale in the bulk and $R_6$ is the 6D Ricci scalar. We further distinguish the bulk and brane geometries that come with coordinates and metrics $\left[X^A, (g_6)_{AB}\right]$ and $\left[x^\mu, (g_4)_{\munu}\right]$, respectively. The brane and bulk metric are related through the pull-back $(g_4)_{\munu} = \frac{\partial f^A}{\partial x^\mu}  \frac{\partial f^B}{\partial x^\nu} (g_6)_{AB}$, where $f^A(x^\mu)$ is an embedding function.
To solve the vacuum system, it is sufficient to make the following ansatz for the 6D metric
\begin{align} \label{ansatz_xdim}
    ds^2 = (g_4)_\munu dx^\mu dx^\nu + dr^2 + C(r)^2 d\varphi^2\,,
\end{align}
where $x^\mu$ and $X^A=(x^\mu, r, \varphi)$ are the brane and bulk coordinates, respectively. For simplicity, we also assumed a vanishing  bulk cosmological constant (in the absence of massive bulk fields this is a radiatively stable choice). Moreover, we assume the induced metric on the brane, $(g_4)_\munu$ to be maximally symmetric, i.e.,\ we allow for a de Sitter phase on the brane. 
With these definitions, the full trace of the 6D Einstein equation reads
\begin{align}
    -  M_6^4\left(R_4 +2  \frac{C''(r)}{C(r)} \right) =  2 \lambda  \frac{\delta(r)}{2 \pi C(r)} \, ,
\end{align}
 where $R_4=\mathrm{const.}$ is the Ricci scalar built from the brane metric $(g_4)_\munu$ and $\lambda=\lambda_\mathrm{bare} + \rho_\mathrm{vac}$ contains the bare tension and the vacuum energy arising from the matter Lagrangian. In other words, the brane-induced energy momentum tensor in vacuum takes on the form $(T_4)_\munu =- \rho_\mathrm{vac} (g_4)_\munu$. The localization of the brane tension at $r=0$ is described in terms of a Dirac delta function. The general solution of this equation that reduces to 6D Minkowski, i.e.,~$C(r)=r$, in the absence of sources is
 \begin{subequations}
 \begin{align}
    R_4 &=0 \,, \\
    C(r) &= \left(1- \frac{\lambda}{2 \pi M_6^4}\right) r \, . 
 \end{align}
 \end{subequations}
This is the straightforward generalization of the geometry of a cosmic string in four dimensions~\cite{Vilenkin:1981zs,Gott:1984ef,Hiscock:1985uc}, with our 4D Universe playing the role of the string (for a more detailed derivation in 6D see for example~\cite{Kaloper:2007ap,Sundrum:1998ns}). In any event, the remarkable observation is that the 4D curvature is vanishing despite the presence of vacuum energy on the brane. Put differently, an observer on the brane that like us is ignorant of the presence of two extra dimensions will conclude that vacuum energy if present does not gravitate (seemingly violating the equivalence principle). In contrast, a 6D observer living in the bulk  will notice a geometrical response: A circular path around the brane at fixed radius $r$ has a reduced circumference of $(2 \pi - \delta_\mathrm{deficit}) r$, quantified in terms of a deficit angle $\delta_\mathrm{deficit} = \frac{\lambda}{M_6^4}$. In an embedding picture, such a geometry corresponds to a cone with the brane at its tip (left panel in Fig.~\ref{fig:6D}). This type of extradimensional self-tuning is also fully dynamical: if there is a phase transition on the brane or we change the particle content of our EFT, the deficit angle will adapt while the 4D curvature remains zero. In fact, any value of the vacuum energy that is below the gravity cutoff $M_6$ will be shielded, hence preventing the need to fine-tune the bare EFT parameters\footnote{For higher value of the brane tension, the geometry on the brane changes to de Sitter~\cite{Niedermann:2014yka}.}.

 Of course, this model cannot be the final answer. As it stands, its gravitational sector is purely six dimensional, meaning that the Newtonian potential of a point source on the brane scales as $1/\mathbf{|x|}^3$ rather than  $1/\mathbf{|x|}$ in clear contradiction with observations. In addition, we want the vacuum on the brane to be a de Sitter rather than a Minkowski vacuum.  Two classes of models have been proposed in the past that make use of this 6D mechanism and yet give rise to a 4D Newtonian potential on the brane. First, there are generalizations of the DGP model (originally proposed in 5D~\cite{Dvali:2000hr,Dvali:2000xg}, see also Sec.~\ref{sec:screening}) that embed the brane in an infinite volume bulk~\cite{Kaloper:2007ap,Niedermann:2014bqa}. Here, 4D gravity is recovered due to an induced gravity term on the brane. This model, however, suffers from instabilities~\cite{Dubovsky:2002jm,Hassan:2010ys,Eglseer:2015xla}. Second, there are models that build on the Arkani-Hamed-Dvali-Dimopolus (ADD)  proposal~\cite{Arkani-Hamed:1998jmv,Arkani-Hamed:1998sfv} where the extra space is large but has a finite volume~\cite{Sundrum:1998ns,Chen:2000at,Carroll:2003db,Navarro:2003vw,Cline:2003ak}. The arguably most interesting candidate among this second class of models is the supersymmetric large extra dimension (SLED) model with two micron-sized extra dimensions that take the form of a rugby ball~\cite{Aghababaie:2003wz,Burgess:2011mt,Burgess:2011va} (right panel in Fig.~\ref{fig:6D}). In fact, it has been viewed for several years as the leading alternative to anthropic resolutions of the cosmological constant problem with smoking gun signatures in both table top tests of gravity and collider experiments. In the next section, we will therefore summarise the main features and shortcomings of SLED as a potential solution to the cosmological constant problem (for more details on phenomenological aspects of the model also see the review in~\cite{Burgess:2013ara}).

\subsubsection{Supersymmetric large extra dimensions}\label{subsec:sled}

The SLED proposal generalises the action in Eq.~\eqref{6d_action} to\footnote{This model builds on the Nishino-Sezgin chiral gauged supergravity, which admits rugby ball solutions~\cite{Salam:1984cj,Randjbar-Daemi:1985tdc, Nishino:1986dc}.}~\cite{Burgess:2011mt,Burgess:2011va}
\begin{subequations}
\label{action_SLED}
\begin{align} \label{eq:action_bulk}
	S_\mathrm{bulk} &= - \int d^6 X \sqrt{-g_6}\,\left\{\frac{M_6^4}{2}\left[ R_6 + (\partial_M\phi)(\partial^M\phi) \right] + \frac{1}{4} \re^{-\phi}F_{MN} F^{MN} + 2 M_6^8 e^2\re^{\phi}\right\} \, , \\
	S_\mathrm{branes} &=  \sum_{b=\mathrm{N},\mathrm{S}} \int d^4 x \sqrt{-g_4} \left\{ -\lambda_b(\phi)+\frac{1}{2} \mathcal{A}_b(\phi) \epsilon_{mn} F^{mn} \right\} \, .
\end{align}
\end{subequations}
In particular, it now contains two branes, one at the south (S) and one at the north pole (N) of the rugby ball geometry.
Other new features are the presence of a dilaton $\phi$ and a Maxwell field strength $F_{MN}$ with gauge coupling parameter $e$. The dilaton makes this low energy model compatible with supersymmetry. To be specific, supersymmetry manifests itself at low energies in the form of a constant scaling symmetry $(g_6)_{MN} \to \zeta (g_6)_{MN}$ and $e^\phi \to \zeta^{-1} e^\phi$ under which $S_\mathrm{bulk} \to \zeta^2 S_\mathrm{bulk}$. This also justifies the absence of a bulk cosmological constant, which would break this symmetry.
The Maxwell field on the other hand is needed to compactify the extra space. Its flux winds around the compact extra dimensions and provides the pressure needed to prevent the rugby ball from collapsing under its own gravitational pull. In holding with the principles of an EFT expansion, the Maxwell flux also makes an appearance as an induced term on the brane, where the indices $m,\, n$ only run over the two spatial extraspace coordinates, $\epsilon_{mn}$ is the fully anti-symmetric epsilon tensor with components $\pm 1/ \sqrt{|(g_6)_{mn}|}$, and $\mathcal{A}_b(\phi)$ controls the strength of the brane induced flux. Both the brane tension and the induced flux term are allowed to couple to the dilaton. In fact, for the special choice 
\begin{align}\label{scale-inv-couplings}
    \lambda_b = \mathrm{const} && \text{and} && \mathcal{A}_b = \Phi_b e^{-\phi}\,,
\end{align}
 the brane action $S_\mathrm{brane}$ preserves scale invariance (along with the bulk action).
 
 The emergence of 4D gravity is ensured by construction: Due to the compactness of the extra dimension, the model admits a normalizable and massless 4D graviton in its spectrum (see for example~\cite{Arkani-Hamed:1998jmv}). There is a also a continuum of Kaluza-Klein modes, the masses of which are set by the inverse size of the extra dimension. As a consequence, their contribution to the gravitational exchange amplitude is exponentially suppressed for distances $|x| > \ell_0$ where $\ell_0$ is the typical size of the compact directions. Post-Cavendish experiments put an upper limit of \cite{Kapner:2006si} $\ell_0 \leq 45 \mu\mathrm{m}$.

The vacuum equations corresponding to Eq.~\eqref{action_SLED} can be solved by generalizing the ansatz in Eq.~\eqref{ansatz_xdim} to
\begin{subequations} \label{eq:ansatz}
\begin{align}
	d s^2 & = W^2(r) \, (g_4)_{\mu\nu} d x^{\mu} d x^\nu + d r^2 + C^2(r) d \varphi^2 \,, \label{eq:ansatz_met}\\
	A & = A_{\varphi}(r)d\varphi \,,\\
	\phi & = \phi(r) \,, \label{eq:ansatz_phi}
\end{align}
\end{subequations}
where $W(r)$ is a warping function and the Maxwell field $A$ points in the azimuthal direction $\varphi$ (corresponding to the angular direction of the rugby ball). 

Instead of displaying the full set of equations that follow from Eq.~\eqref{eq:ansatz}, here, we focus on the crucial aspects of the vacuum solution that are relevant for the model's potential to resolve the cosmological constant problem. The Maxwell equation can be integrated trivially, yielding
%
\begin{equation} \label{Sol_Max}
	F_{\rho\theta}=e^{\phi}\left[Q \, \frac{C}{W^4} + \frac{1}{2\pi} \sum_{b=\mathrm{N},\mathrm{S}} \mathcal{A}_b(\phi) \delta(r - r_b) \right] \, ,
\end{equation}
where $Q$ is an integration constant. Due to the compactness of the extra space, the system has to be equipped with an additional equation describing the quantization of the Maxwell flux~\cite{Randjbar-Daemi:1982opc, Burgess:2011va}
\begin{equation} \label{eq:flux_quant}
	Q \int \!d r \, \frac{\re^{\phi} C}{W^4} + \frac{1}{2\pi}\sum_{b=\mathrm{N},\mathrm{S}} \left[ \mathcal{A}_b(\phi) \re^{\phi} \right]_{r=r_b} = \frac{n}{ e} \qquad (n \in \mathbb{N}) \,.
\end{equation}
Now we come to a subtle issue. From Eq.~\eqref{Sol_Max} we see that there is a contribution to $F_{AB}$ that is proportional to a Dirac delta function. Since the Einstein and dilaton equations are sourced by a term that is proportional to $F_{MN} F^{MN}$, they will contain a divergence $\propto \delta(0)$. Physically, this is an artifact of treating the brane as an infinitesimally thin codimension-two object. This somewhat technical point
has been studied in different ways: In \cite{Niedermann:2015via}, the authors introduced a counter term that removed the singularity and made the distributional approach applicable again, in \cite{Burgess:2015gba,Burgess:2015lda} the brane was microscopically resolved in terms of an explicit vortex model, and finally in \cite{Niedermann:2015vbk} the brane was described in terms of a cylindrical codimension-one object stabilised through a brane-localised angular pressure\footnote{The last approach also addressed concerns about the general applicability of the distributional approach away from the scale-invariant case~\cite{Burgess:2015kda}.}. In all studies, it was found that extra dimensional self-tuning, i.e.,\ $R_4=0$, is guaranteed in the case of scale-invariant brane-dilaton couplings as in Eq.~\eqref{scale-inv-couplings}. This, however, is not what we were looking for: in the scale-invariant case $\phi$ drops out of the flux quantization condition in Eq.~\eqref{eq:flux_quant} turning it into a tuning relation on both brane tensions~\cite{Gibbons:2003di,Niedermann:2015via}:
\begin{align}
	\label{eq:flux_quant_2}
	\left(1-\frac{\lambda_\mathrm{N}}{2 \pi M_6^4}\right)^{1/2}\left(1-\frac{\lambda_\mathrm{S}}{2 \pi M_6^4}\right)^{1/2} + \frac{e}{2\pi} \sum_b \Phi_b = n \,.
\end{align}
 At this point, one should object that observations require the vacuum to be a (quasi) de Sitter rather than a Minkowski geometry. Accordingly, the condition $R_4 = 0$ should be relaxed to $R_4 \sim  10^{-33} \, \mathrm{eV}$ to account for dark energy. This also suggests that we should go away from scale invariance (which always implies $R_4 = 0$). This possibility has been investigated in~\cite{Niedermann:2015vbk}  where the authors allowed for a class of couplings that deviated from Eq.~\eqref{scale-inv-couplings} (by for example making the brane tension dependent on $\phi$). Unfortunately, in these cases in order for the extra dimensional volume and the 4D curvature to fulfill their phenomenological bounds,  Eq.~\eqref{eq:flux_quant_2} has to be satisfied with a very high accuracy (although not exactly), effectively recovering the usual fine-tuning issue within standard GR. 

\subsubsection{Outlook} \label{sec:xdim_outlook}

The above finding bears close resemblance with Weinberg's no-go theorem discussed in Sec.~\ref{sec:no-go}. There it was found that the self-tuning condition led to a scale invariant potential, which could only be made compatible with a vanishing curvature by imposing a tuning (to be specific, the loop-sensitive quantity $V_0$ had to be tuned to zero in Eq.~\eqref{pot_self_tiuning} for $R_4=0$ to hold). 
Is this the end of extra-dimensional self-tuning in higher co-dimensional setups? Not necessarily. The general mechanism outlined in Sec.~\ref{sec:comsic_string} still has merit. The problems arose when we compactified the bulk. Intuitively, this should not come as a surprise: At very large distances $\gg \ell_0$ (or energies below the Kaluza-Klein scale) the extra dimensions cannot be resolved and the theory is equivalent to a 4D theory with a massless graviton, making the usual no-go's applicable again. However, if we were to consider a different way of recovering 4D GR, which allowed for infinite volume extra dimensions, this reasoning would not apply (as the zero mode is expected to decouple). Unfortunately, as mentioned before, a simple generalisation of the DGP model to higher co-dimensions suffers from ghost instabilities. At this point, and in order to end this section on a more positive note, we speculate that an alternative mechanism to recover a viable 4D phenomenology could be the ``volcano''-trapping proposed in~\cite{Karch:2000ct,Kaloper:2005wq} generalised to higher co-dimensions. It relies on the idea of deforming the near-brane geometry such that it traps the gravitational field lines close to the brane. It should also be noted that other scenarios relying on five (rather than six or more) extra dimensions are still being actively explored. One interesting direction is provided by the holographic model in~\cite{Charmousis:2017rof}. It avoids problems with naked bulk singularities, which plagued earlier co-dimension one self-tuning models. The model's phenomenology is currently being investigated~\cite{Ghosh:2018fbx,Amariti:2019vfv} and it remains to be seen if it can provide a viable phenomenology in the presence of self-tuning.

\section{Conclusions - You can't always get what you want, or can you? }\label{sec:disc}

The various aspects of the cosmological constant problem have remained for decades now the most challenging problems in theoretical physics. Combining general relativity and quantum field theory at low energies and then fitting cosmological data from both early- and late-times yields four different puzzles (Sec. \ref{sec:problems}). There is an abysmal mismatch between the prediction of the vacuum energy in QFT and the current observed value in cosmology; even if this gap is fixed by a bare cosmological constant at the level of Einstein equations, the fixing is unreliable due to both phase transitions in the early Universe and the UV sensitivity of the effective field theory employed; and finally, we still lack a clear understanding of what dark energy is made of. All these puzzles have different levels of relatedness depending on the theory one is considering. 

Currently these issues bring into question the phenomenology of Einstein equations applied to large scales and they also challenge the effective field theory approach to QFT. Since the Einstein equations connect the matter content with gravity, there are two potential paths out: either modifying gravity, or changing the quantum-matter sector.

As the vacuum energy couples to gravity as any other form of energy in GR, it is natural to focus on different ways that we can modify gravity to tackle one or more of the CCPs. The main idea underlining these approaches, the self-tuning mechanism, makes gravity less sensitive to the presence of vacuum energy by invoking  additional gravitational degrees of freedom. This can create problems by itself as gravitational physics is well understood by only using GR both on astrophysical and cosmological scales. Thus, in order for these approaches to be successful, they also need to typically rely on some screening mechanism operative on local scales, while at the same time being able to recover the success of the current cosmological model describing the history of our Universe. Together with solving the CCPs, these requirements formed our wishlist (introduced in Sec.~\ref{sec:wishlist}).

However, such modifications of gravity cannot be arbitrary. Weinberg pointed out that simply introducing self-tuning fields that, once coupled to matter, will cancel out the vacuum energy still leads to fine-tuning or incorrect phenomenology (Sec. \ref{sec:toymodel}). In fact, we also showed that his theorem can be generalized by analyzing the spectral decomposition of the gravitational exchange amplitude between probes and sources. Thus, only modifications of gravity that break one or more of the underlying assumptions of these theorems will yield viable models of self-tuning.

The core assumptions of these theorems (see Secs.~\ref{sec:no-go} and \ref{sec:spectral}) are that the vacuum exhibits translational invariance, the theory is Lorentz invariant, the fields' propagators take on their canonical form, the minimal coupling between gravity and matter, and that spacetime is four-dimensional. The models discussed here models break one or more of these. For instance, sequestering leads to a non-canonical propagator by a non-standard coupling; linear and non-linear massive gravity depart from GR by making the graviton massive, which introduces the Stückelberg field that has a non-constant vacuum value; the Horndeski-type proposals depart from GR and break the translational invariance of their vacuum; and SLED assumes that spacetime has more dimensions than four.

As we reviewed these models, we were able to assess their success in fulfilling our wishlist. Table \ref{table:1} summarises our analysis. We emphasize that most of these models tackle the classical CCP, while still being able to recover the gravitational physics already contained in GR at small scales. Most of these approaches can also successfully describe our cosmic history (although there may not be such an analysis present in the literature, it can be easy to see that they mostly depart from GR at late-times). Finally, the dark energy problem may or may not be solved in these approaches as that depends on whether the models allow some residual-like-CC term that would induce an accelerated expansion, though in general they do not readily recover the observed value for the effective CC. Nonetheless, these models can also be combined with quintessential models (see \cite{Tsujikawa:2013fta} for a review) that would then drive the dark energy dynamics.

Regarding the new CCP, the analysis becomes subtle. By modifying gravity such that the vacuum energy is effectively prevented from gravitating, these proposals tacitly accept that the UV sensitivity is not necessarily an issue, in the sense that they do not attempt at fixing it, leaving QFT calculations untouched. Instead, gravity is modified such that the breakdown of the UV-IR decoupling does not lead to dramatic observable effects. Insofar as the main issue with the CC is this sensitivity coming from QFT calculations alone, then modifications of gravity only fix one of its symptoms that manifests in a gravitational context. Nonetheless, without self-tuning, UV sensitivity requires fine-tuning to describe observations; while with self-tuning, UV sensitivity is ``only'' a conceptual problem that does not lead to any observational effects. 
 
\begin{table}[t]
\centering
\resizebox{\textwidth}{!} {
\begin{tabular}{ | l || c || c || c || c || c |}
 \hline
 \multicolumn{6}{|c|}{\textbf{SELECTED MODIFIED GRAVITY APPROACHES}} \\
 \hline \hline
 \multicolumn{1}{|c||}{} & \multicolumn{3}{c||}{\hyperref[sec:problems]{\color{black}{\textbf{CC-Problems}}}} & \multicolumn{2}{c|}{\hyperref[par:data_cons]{\color{black}{\textbf{Data Constraints}}}} \\
 \hline
 &  \small{\textbf{new-CCP}} & \small{\textbf{class-CCP}} & \small{\textbf{DEP}} & \small{\textbf{CHC}} & \small{\textbf{AC} }\\
 \hline
 \textbf{\small{{GR + QFT}}} & X & X &  \checkmark &  \checkmark &  \checkmark \\
  \hline
 \textbf{\small{\hyperref[ssec:global_seq]{\color{black}{Global Sequestering}}}}& \checkmark & \checkmark & (P) & \checkmark &  \checkmark \\
  \hline
 \textbf{\small{\hyperref[sec:local_seq]{\color{black}{Local Sequestering}}}}  & \checkmark & \checkmark & (P) & \checkmark & \checkmark \\
  \hline
 \textbf{\small{\hyperref[ssec:carroll]{\color{black}{Non-local approach }}}}  & X & (P) & (P) & \checkmark & \checkmark \\
  \hline
  \textbf{\small{\hyperref[subsec:unimodular]{\color{black}{Unimodular Gravity}}}} & X & X   &\checkmark &\checkmark &\checkmark\\
  \hline
 \textbf{\small{\hyperref[sec:linear_massive_gravity]{\color{black}{Linear Massive Gravity}}}} & \checkmark & \checkmark & (P) & \checkmark & X \\
  \hline
 \textbf{\small{\hyperref[sec:nonlinmassgrav]{\color{black}{Nonlinear Massive Gravity}}}}  & X & \checkmark & (P) & \checkmark & (P)  \\
  \hline
 \textbf{\small{\hyperref[sec:fab4]{\color{black}{Fab-4}}}} & (P) & \checkmark & \checkmark & (P) & (P) \\
  \hline
 \textbf{\small{\hyperref[sec:well-temp]{\color{black}{Well-tempered  self-tuning}}}} & (P) & \checkmark & \checkmark & \checkmark&  (P) \\
  \hline
 \textbf{\small{\hyperref[subsec:sled]{\color{black}{SLED}}}}  & X & X &(P) & (P)& \checkmark\\
 \hline
\end{tabular}
}
\caption{Here we summarise our assessment of the approaches introduced in Sec.~\ref{sec:mod_gravity_approaches} in light of the theoretical and phenomenological requirements contained in our wishlist discussed in Sec.~\ref{sec:wishlist} (CCP stands for cosmological constant problem, DEP for dark energy problem, CHC for cosmic history constraint, and AC for astrophysical constraint). Whenever one of the CC-problems or data constraints has been shown to be solved, we use the \checkmark; for requirements that potentially can be achieved within a given approach, we use (P) (for example, Fab-4 can reproduce individual cosmological epochs, but it remains to be seen if they can be stitched together \cite{Copeland:2012qf}); while the ones that seem to be unsuccessful given the current literature are marked as X. Note we evaluate whether a given approach tackles any of the requirements \textit{independently}. For instance, we consider that ``GR + QFT'' can tackle the DEP classically, thus a \checkmark, despite not being able to tackle the new-CCP, which deems the solution to DEP unstable quantum mechanically.}
\label{table:1}
\end{table}

Finally, we note that the table does not provide a fully comprehensive assessment of these models. Additional requirements should be demanded as a benchmark, such as having a ghost-free theory and a UV embedding. For instance, we briefly commented that the Horndeski class of models are all ghost-free, while global sequestering faces challenges to be embedded in a UV theory, but its local counterpart could more easily achieve this. Moreover, a theoretical model which is able to address more than one of the CCPs lends itself a stronger preference, given it does not introduce a significant level of complexity.

In spite of the different proposals presented here, there is not a consensus in the community about what the solutions to the CCPs are. Among the reasons, some of these models recover most of the phenomenology already explained by GR, while others seem to be contrived. Alternative to seeking guidance from theory, new and distinct phenomenology is an essential ingredient for any prospective model. In particular, the next decade will see a host of novel experiments probing both low-energy cosmological and  high-energy astrophysical scales. The largest galaxy surveys to date have either commenced or are coming online soon, for example the ongoing dark energy survey\footnote{\url{www.darkenergysurvey.org}} \cite{DES:2022ygi}, the upcoming Euclid mission\footnote{\url{www.euclid-ec.org}} \citep{Amendola:2016saw} and the Vera C. Rubin Observatory’s Legacy Survey of Space and Time (VRO/LSST)\footnote{\url{https://www.lsst.org/}} \citep{Ishak:2019aay}. These will precisely probe a yet untested regime of structure formation where various effects due to modifications of gravity may become apparent. There is also the unprecedentedly large radio survey, the square kilometre array\footnote{\url{www.skatelescope.org}} \cite{Weltman:2018zrl}, set to come online in the next few years. Further, the emerging field of gravitational wave astronomy has immense potential to probe both cosmological and astrophysical scales. The planned LISA\footnote{\url{https://lisa.nasa.gov/}} \cite{LISACosmologyWorkingGroup:2019mwx,LISA:2022kgy} experiment will precisely measure cosmological background effects, as well as effects related to perturbations both in the scalar and tensor sectors. These experiments will at the very least perform null-tests of GR. Any  tensions that emerge in the best-fit $\Lambda$CDM model between these upcoming and existing data sets will provide essential clues for theoretical progress.  

\section*{Acknowledgments} 
H.B. was supported by the Fonds de Resercher du
Qu\'ebec (PBEEE/303549) and partially by funds from NSERC. B.B. was supported by a UK Research and Innovation Stephen Hawking Fellowship (EP/W005654/1) and a Swiss National Science Foundation (SNSF) Professorship grant Nos.~170547 \& 202671. G. F. was supported by the Göran Gustafsson Foundation for Research in Natural Sciences and Medicine. Y.H. was supported by the grant 2019-04234 from the Swedish Research Council (Vetenskapsr{\aa}det). Nordita is partially supported by Nordforsk.  A.L. acknowledges support by the Swedish Research Council (Vetenskapsr{\aa}det) through contract No. 638-2013-8993 and the Oskar Klein Centre for Cosmoparticle Physics. S.H. was supported by the Excellence Cluster ORIGINS which is funded by the Deutsche Forschungsgemeinschaft (DFG, German Research Foundation) under Germany’s Excellence Strategy - EXC-2094 - 390783311.

We would like to thank Fawad Hassan and Marcus H\"og\aa s for valuable discussions about massive gravity and Hiranya Peiris for the support establishing this collaboration. We also would like to thank Elisa Ferreira for discussions and early-on contributions. For the purpose of open access, the authors have applied a Creative Commons Attribution (CC BY) licence to any Author Accepted Manuscript version arising from this submission. 

\phantomsection
\addcontentsline{toc}{section}{References}

\let\oldbibliography\thebibliography
\renewcommand{\thebibliography}[1]{
  \oldbibliography{#1}
  \setlength{\parskip}{0pt}
  \setlength{\itemsep}{0pt} 
  \footnotesize 
}

\bibliographystyle{JHEP2}
\bibliography{references}

\providecommand{\url}[1]{#1}\providecommand{\href}[2]{#2}\begingroup\raggedright\begin{thebibliography}{100}

\bibitem{Einstein1917}
A.~{Einstein}, \emph{{Kosmologische Betrachtungen zur allgemeinen
  Relativit{\"a}tstheorie}}, {\emph{Sitzungsberichte der K{\"o}niglich
  Preu{\ss}ischen Akademie der Wissenschaften, Berlin} (Jan., 1917) 142}.

\bibitem{ORaifeartaigh:2017uct}
C.~O'Raifeartaigh, M.~O'Keeffe, W.~Nahm and S.~Mitton,
  \emph{{Einstein\textquoteright{}s 1917 static model of the universe: a
  centennial review}},
  \href{https://doi.org/10.1140/epjh/e2017-80002-5}{\emph{Eur. Phys. J. H}
  {\bfseries 42} (2017) 431}
  [\href{https://arxiv.org/abs/1701.07261}{{\ttfamily arXiv:1701.07261}}].

\bibitem{Planck:2018vyg}
{\scshape Planck} collaboration, N.~Aghanim et~al., \emph{{Planck 2018 results.
  VI. Cosmological parameters}},
  \href{https://doi.org/10.1051/0004-6361/201833910}{\emph{Astron. Astrophys.}
  {\bfseries 641} (2020) A6}
  [\href{https://arxiv.org/abs/1807.06209}{{\ttfamily arXiv:1807.06209}}].

\bibitem{Bautista:2020ahg}
J.E.~Bautista et~al., \emph{{The Completed SDSS-IV extended Baryon Oscillation
  Spectroscopic Survey: measurement of the BAO and growth rate of structure of
  the luminous red galaxy sample from the anisotropic correlation function
  between redshifts 0.6 and 1}},
  \href{https://doi.org/10.1093/mnras/staa2800}{\emph{Mon. Not. Roy. Astron.
  Soc.} {\bfseries 500} (2020) 736}
  [\href{https://arxiv.org/abs/2007.08993}{{\ttfamily arXiv:2007.08993}}].

\bibitem{BOSS:2016wmc}
{\scshape BOSS} collaboration, S.~Alam et~al., \emph{{The clustering of
  galaxies in the completed SDSS-III Baryon Oscillation Spectroscopic Survey:
  cosmological analysis of the DR12 galaxy sample}},
  \href{https://doi.org/10.1093/mnras/stx721}{\emph{Mon. Not. Roy. Astron.
  Soc.} {\bfseries 470} (2017) 2617}
  [\href{https://arxiv.org/abs/1607.03155}{{\ttfamily arXiv:1607.03155}}].

\bibitem{Hildebrandt:2016iqg}
H.~Hildebrandt et~al., \emph{{KiDS-450: Cosmological parameter constraints from
  tomographic weak gravitational lensing}},
  \href{https://doi.org/10.1093/mnras/stw2805}{\emph{Mon. Not. Roy. Astron.
  Soc.} {\bfseries 465} (2017) 1454}
  [\href{https://arxiv.org/abs/1606.05338}{{\ttfamily arXiv:1606.05338}}].

\bibitem{DES:2022qpf}
{\scshape DES} collaboration, C.~Doux et~al., \emph{{Dark Energy Survey Year 3
  results: cosmological constraints from the analysis of cosmic shear in
  harmonic space}}, \href{https://doi.org/10.1093/mnras/stac1826}{\emph{Mon.
  Not. Roy. Astron. Soc.} {\bfseries 515} (2022) 1942}
  [\href{https://arxiv.org/abs/2203.07128}{{\ttfamily arXiv:2203.07128}}].

\bibitem{Brout:2022vxf}
D.~Brout et~al., \emph{{The Pantheon+ Analysis: Cosmological Constraints}},
  \href{https://arxiv.org/abs/2202.04077}{{\ttfamily arXiv:2202.04077}}.

\bibitem{DES:2020cbm}
{\scshape DES, SPT} collaboration, M.~Costanzi et~al., \emph{{Cosmological
  constraints from DES Y1 cluster abundances and SPT multiwavelength data}},
  \href{https://doi.org/10.1103/PhysRevD.103.043522}{\emph{Phys. Rev. D}
  {\bfseries 103} (2021) 043522}
  [\href{https://arxiv.org/abs/2010.13800}{{\ttfamily arXiv:2010.13800}}].

\bibitem{SPT:2016izt}
{\scshape SPT} collaboration, T.~de~Haan et~al., \emph{{Cosmological
  Constraints from Galaxy Clusters in the 2500 square-degree SPT-SZ Survey}},
  \href{https://doi.org/10.3847/0004-637X/832/1/95}{\emph{Astrophys. J.}
  {\bfseries 832} (2016) 95}
  [\href{https://arxiv.org/abs/1603.06522}{{\ttfamily arXiv:1603.06522}}].

\bibitem{Huterer:2017buf}
D.~Huterer and D.L.~Shafer, \emph{{Dark energy two decades after: Observables,
  probes, consistency tests}},
  \href{https://doi.org/10.1088/1361-6633/aa997e}{\emph{Rept. Prog. Phys.}
  {\bfseries 81} (2018) 016901}
  [\href{https://arxiv.org/abs/1709.01091}{{\ttfamily arXiv:1709.01091}}].

\bibitem{Birrell:1982ix}
N.D.~Birrell and P.C.W.~Davies, \emph{{Quantum Fields in Curved Space}},
  Cambridge Monographs on Mathematical Physics, Cambridge Univ. Press,
  Cambridge, UK (2, 1984),
  \href{https://doi.org/10.1017/CBO9780511622632}{10.1017/CBO9780511622632}.

\bibitem{Nernst:1916}
W.~Nernst, \emph{{\"Uber einen Versuch, von quantentheoretischen Betrachtungen
  zur Annahme stetiger Energieänderungen zurückzukehren}},
  {\emph{Verhandlungen der Deutschen Physikalischen Gesellschaft} {\bfseries
  18} (1916) 83}.

\bibitem{Enz_Thellung:1960}
A.T.~C.P.~Enz, \emph{{Nullpunktsenergie und Anordnung nicht vertauschbarer
  Faktoren im Hamiltonoperator}}, {\emph{Helv. Phys. Acta} {\bfseries 33}
  (1960) 839}.

\bibitem{Lenz:1926}
W.~Lenz, \emph{{Das Gleichgewicht von Materie und Strahlung in Einsteins
  geschlossener Welt}}, {\emph{Phys. Zs.} {\bfseries 27} (1926) 642}.

\bibitem{Peruzzi_Realdi:2011}
G.~Peruzzi and M.~Realdi, \emph{{The quest for the size of the universe in
  early relativistic cosmology (1917–1930)}},
  \href{https://doi.org/https://doi.org/10.1007/s00407-011-0088-z}{\emph{Arch.
  Hist. Exact Sci.} {\bfseries 65} (2011) 659}.

\bibitem{Kragh2012}
H.~Kragh, \emph{{Walther Nernst: grandfather of dark energy?}},
  \href{https://doi.org/10.1111/j.1468-4004.2012.53124.x}{\emph{Astronomy \&
  Geophysics} {\bfseries 53} (02, 2012) 1.24}
  [\href{https://arxiv.org/abs/https://academic.oup.com/astrogeo/article-pdf/53/1/1.24/581058/53-1-1.24.pdf}{{\ttfamily
  https://academic.oup.com/astrogeo/article-pdf/53/1/1.24/581058/53-1-1.24.pdf}}],
  \url{https://doi.org/10.1111/j.1468-4004.2012.53124.x}.

\bibitem{kragh2014weight}
H.S.~Kragh and J.M.~Overduin, \emph{The weight of the vacuum: A scientific
  history of dark energy}, Springer (2014).

\bibitem{Zeldovich:1967gd}
Y.B.~Zeldovich, \emph{{Cosmological Constant and Elementary Particles}},
  {\emph{JETP Lett.} {\bfseries 6} (1967) 316}.

\bibitem{Zeldovich:1968ehl}
Y.B.~Zel'dovich, A.~Krasinski and Y.B.~Zeldovich, \emph{{The Cosmological
  constant and the theory of elementary particles}},
  \href{https://doi.org/10.1007/s10714-008-0624-6}{\emph{Sov. Phys. Usp.}
  {\bfseries 11} (1968) 381}.

\bibitem{tHooft:1974toh}
G.~'t~Hooft and M.J.G.~Veltman, \emph{{One loop divergencies in the theory of
  gravitation}}, {\emph{Ann. Inst. H. Poincare Phys. Theor. A} {\bfseries 20}
  (1974) 69}.

\bibitem{Goroff:1985th}
M.H.~Goroff and A.~Sagnotti, \emph{{The Ultraviolet Behavior of Einstein
  Gravity}}, \href{https://doi.org/10.1016/0550-3213(86)90193-8}{\emph{Nucl.
  Phys. B} {\bfseries 266} (1986) 709}.

\bibitem{Weinberg:1988cp}
S.~Weinberg, \emph{{The Cosmological Constant Problem}},
  \href{https://doi.org/10.1103/RevModPhys.61.1}{\emph{Rev. Mod. Phys.}
  {\bfseries 61} (1989) 1}.

\bibitem{Martin:2012bt}
J.~Martin, \emph{{Everything You Always Wanted To Know About The Cosmological
  Constant Problem (But Were Afraid To Ask)}},
  \href{https://doi.org/10.1016/j.crhy.2012.04.008}{\emph{Comptes Rendus
  Physique} {\bfseries 13} (2012) 566}
  [\href{https://arxiv.org/abs/1205.3365}{{\ttfamily arXiv:1205.3365}}].

\bibitem{Copeland:2006wr}
E.J.~Copeland, M.~Sami and S.~Tsujikawa, \emph{{Dynamics of dark energy}},
  \href{https://doi.org/10.1142/S021827180600942X}{\emph{Int. J. Mod. Phys. D}
  {\bfseries 15} (2006) 1753}
  [\href{https://arxiv.org/abs/hep-th/0603057}{{\ttfamily hep-th/0603057}}].

\bibitem{Barrow:1986nmg}
J.D.~Barrow and F.J.~Tipler, \emph{{The Anthropic Cosmological Principle}},
  Oxford U. Pr., Oxford (1988).

\bibitem{gribbin1989cosmic}
J.~Gribbin and M.~Rees, \emph{Cosmic Coincidences: Dark Matter, Mankind, and
  Anthropic Cosmology}, Bantam New Age Books, Bantam Books (1989),
  \url{https://books.google.ch/books?id=\_ofvAAAAMAAJ}.

\bibitem{Weinberg:1987dv}
S.~Weinberg, \emph{{Anthropic Bound on the Cosmological Constant}},
  \href{https://doi.org/10.1103/PhysRevLett.59.2607}{\emph{Phys. Rev. Lett.}
  {\bfseries 59} (1987) 2607}.

\bibitem{10.1093/mnras/274.1.L73}
G.~Efstathiou, \emph{{An anthropic argument for a cosmological constant}},
  \href{https://doi.org/10.1093/mnras/274.1.L73}{\emph{Monthly Notices of the
  Royal Astronomical Society} {\bfseries 274} (05, 1995) L73}
  [\href{https://arxiv.org/abs/https://academic.oup.com/mnras/article-pdf/274/1/L73/18540321/mnras274-0L73.pdf}{{\ttfamily
  https://academic.oup.com/mnras/article-pdf/274/1/L73/18540321/mnras274-0L73.pdf}}],
  \url{https://doi.org/10.1093/mnras/274.1.L73}.

\bibitem{Martel:1997vi}
H.~Martel, P.R.~Shapiro and S.~Weinberg, \emph{{Likely values of the
  cosmological constant}},
  \href{https://doi.org/10.1086/305016}{\emph{Astrophys. J.} {\bfseries 492}
  (1998) 29} [\href{https://arxiv.org/abs/astro-ph/9701099}{{\ttfamily
  astro-ph/9701099}}].

\bibitem{Garriga:1999hu}
J.~Garriga, M.~Livio and A.~Vilenkin, \emph{{The Cosmological constant and the
  time of its dominance}},
  \href{https://doi.org/10.1103/PhysRevD.61.023503}{\emph{Phys. Rev. D}
  {\bfseries 61} (2000) 023503}
  [\href{https://arxiv.org/abs/astro-ph/9906210}{{\ttfamily
  astro-ph/9906210}}].

\bibitem{Peacock:2007cw}
J.A.~Peacock, \emph{{Testing anthropic predictions for Lambda and the CMB
  temperature}},
  \href{https://doi.org/10.1111/j.1365-2966.2007.11978.x}{\emph{Mon. Not. Roy.
  Astron. Soc.} {\bfseries 379} (2007) 1067}
  [\href{https://arxiv.org/abs/0705.0898}{{\ttfamily arXiv:0705.0898}}].

\bibitem{SobralBlanco:2020too}
D.~Sobral~Blanco and L.~Lombriser, \emph{{Local self-tuning mechanism for the
  cosmological constant}},
  \href{https://doi.org/10.1103/PhysRevD.102.043506}{\emph{Phys. Rev. D}
  {\bfseries 102} (2020) 043506}
  [\href{https://arxiv.org/abs/2003.04303}{{\ttfamily arXiv:2003.04303}}].

\bibitem{Clifton:2011jh}
T.~Clifton, P.G.~Ferreira, A.~Padilla and C.~Skordis, \emph{{Modified Gravity
  and Cosmology}},
  \href{https://doi.org/10.1016/j.physrep.2012.01.001}{\emph{Phys. Rept.}
  {\bfseries 513} (2012) 1} [\href{https://arxiv.org/abs/1106.2476}{{\ttfamily
  arXiv:1106.2476}}].

\bibitem{Hinterbichler:2011tt}
K.~Hinterbichler, \emph{{Theoretical Aspects of Massive Gravity}},
  \href{https://doi.org/10.1103/RevModPhys.84.671}{\emph{Rev. Mod. Phys.}
  {\bfseries 84} (2012) 671} [\href{https://arxiv.org/abs/1105.3735}{{\ttfamily
  arXiv:1105.3735}}].

\bibitem{Koyama:2015vza}
K.~Koyama, \emph{{Cosmological Tests of Modified Gravity}},
  \href{https://doi.org/10.1088/0034-4885/79/4/046902}{\emph{Rept. Prog. Phys.}
  {\bfseries 79} (2016) 046902}
  [\href{https://arxiv.org/abs/1504.04623}{{\ttfamily arXiv:1504.04623}}].

\bibitem{deRham:2014zqa}
C.~de~Rham, \emph{{Massive Gravity}},
  \href{https://doi.org/10.12942/lrr-2014-7}{\emph{Living Rev. Rel.} {\bfseries
  17} (2014) 7} [\href{https://arxiv.org/abs/1401.4173}{{\ttfamily
  arXiv:1401.4173}}].

\bibitem{Nobbenhuis:2004wn}
S.~Nobbenhuis, \emph{{Categorizing different approaches to the cosmological
  constant problem}},
  \href{https://doi.org/10.1007/s10701-005-9042-8}{\emph{Found. Phys.}
  {\bfseries 36} (2006) 613}
  [\href{https://arxiv.org/abs/gr-qc/0411093}{{\ttfamily gr-qc/0411093}}].

\bibitem{Polchinski:2006gy}
J.~Polchinski, \emph{{The Cosmological Constant and the String Landscape}},  in
  \emph{{23rd Solvay Conference in Physics: The Quantum Structure of Space and
  Time}}, p.~216, 3, 2006,
  \href{https://arxiv.org/abs/hep-th/0603249}{{\ttfamily hep-th/0603249}}.

\bibitem{Bousso:2007gp}
R.~Bousso, \emph{{TASI Lectures on the Cosmological Constant}},
  \href{https://doi.org/10.1007/s10714-007-0557-5}{\emph{Gen. Rel. Grav.}
  {\bfseries 40} (2008) 607} [\href{https://arxiv.org/abs/0708.4231}{{\ttfamily
  arXiv:0708.4231}}].

\bibitem{Padilla:2015aaa}
A.~Padilla, \emph{{Lectures on the Cosmological Constant Problem}},
  \href{https://arxiv.org/abs/1502.05296}{{\ttfamily arXiv:1502.05296}}.

\bibitem{SupernovaSearchTeam:1998fmf}
{\scshape Supernova Search Team} collaboration, A.G.~Riess et~al.,
  \emph{{Observational evidence from supernovae for an accelerating universe
  and a cosmological constant}},
  \href{https://doi.org/10.1086/300499}{\emph{Astron. J.} {\bfseries 116}
  (1998) 1009} [\href{https://arxiv.org/abs/astro-ph/9805201}{{\ttfamily
  astro-ph/9805201}}].

\bibitem{SupernovaCosmologyProject:1998vns}
{\scshape Supernova Cosmology Project} collaboration, S.~Perlmutter et~al.,
  \emph{{Measurements of $\Omega$ and $\Lambda$ from 42 high redshift
  supernovae}}, \href{https://doi.org/10.1086/307221}{\emph{Astrophys. J.}
  {\bfseries 517} (1999) 565}
  [\href{https://arxiv.org/abs/astro-ph/9812133}{{\ttfamily
  astro-ph/9812133}}].

\bibitem{Koren:2020biu}
S.~Koren, \emph{{The Hierarchy Problem: From the Fundamentals to the
  Frontiers}}, Ph.D. thesis, UC, Santa Barbara, 2020.
\newblock \href{https://arxiv.org/abs/2009.11870}{{\ttfamily
  arXiv:2009.11870}}.

\bibitem{Tsujikawa:2013fta}
S.~Tsujikawa, \emph{{Quintessence: A Review}},
  \href{https://doi.org/10.1088/0264-9381/30/21/214003}{\emph{Class. Quant.
  Grav.} {\bfseries 30} (2013) 214003}
  [\href{https://arxiv.org/abs/1304.1961}{{\ttfamily arXiv:1304.1961}}].

\bibitem{Burgess:2013ara}
C.P.~Burgess, \emph{{The Cosmological Constant Problem: Why it's hard to get
  Dark Energy from Micro-physics}},  in \emph{{100e Ecole d'Ete de Physique:
  Post-Planck Cosmology}}, 9, 2013,
  \href{https://arxiv.org/abs/1309.4133}{{\ttfamily arXiv:1309.4133}},
  \href{https://doi.org/10.1093/acprof:oso/9780198728856.003.0004}{DOI}.

\bibitem{Niedermann:2017cel}
F.~Niedermann and A.~Padilla, \emph{{Gravitational Mechanisms to Self-Tune the
  Cosmological Constant: Obstructions and Ways Forward}},
  \href{https://doi.org/10.1103/PhysRevLett.119.251306}{\emph{Phys. Rev. Lett.}
  {\bfseries 119} (2017) 251306}
  [\href{https://arxiv.org/abs/1706.04778}{{\ttfamily arXiv:1706.04778}}].

\bibitem{Fierz:1939ix}
M.~Fierz and W.~Pauli, \emph{{On relativistic wave equations for particles of
  arbitrary spin in an electromagnetic field}},
  \href{https://doi.org/10.1098/rspa.1939.0140}{\emph{Proc. Roy. Soc. Lond. A}
  {\bfseries 173} (1939) 211}.

\bibitem{Will:2014kxa}
C.M.~Will, \emph{{The Confrontation between General Relativity and
  Experiment}}, \href{https://doi.org/10.12942/lrr-2014-4}{\emph{Living Rev.
  Rel.} {\bfseries 17} (2014) 4}
  [\href{https://arxiv.org/abs/1403.7377}{{\ttfamily arXiv:1403.7377}}].

\bibitem{Burrage:2012ja}
C.~Burrage, N.~Kaloper and A.~Padilla, \emph{{Strong Coupling and Bounds on the
  Spin-2 Mass in Massive Gravity}},
  \href{https://doi.org/10.1103/PhysRevLett.111.021802}{\emph{Phys. Rev. Lett.}
  {\bfseries 111} (2013) 021802}
  [\href{https://arxiv.org/abs/1211.6001}{{\ttfamily arXiv:1211.6001}}].

\bibitem{deRham:2014wfa}
C.~de~Rham and R.H.~Ribeiro, \emph{{Riding on irrelevant operators}},
  \href{https://doi.org/10.1088/1475-7516/2014/11/016}{\emph{JCAP} {\bfseries
  11} (2014) 016} [\href{https://arxiv.org/abs/1405.5213}{{\ttfamily
  arXiv:1405.5213}}].

\bibitem{Kaloper:2014vqa}
N.~Kaloper, A.~Padilla, P.~Saffin and D.~Stefanyszyn, \emph{{Unitarity and the
  Vainshtein Mechanism}},
  \href{https://doi.org/10.1103/PhysRevD.91.045017}{\emph{Phys. Rev. D}
  {\bfseries 91} (2015) 045017}
  [\href{https://arxiv.org/abs/1409.3243}{{\ttfamily arXiv:1409.3243}}].

\bibitem{Adams:2002ft}
A.~Adams, J.~McGreevy and E.~Silverstein, \emph{{Decapitating tadpoles}},
  \href{https://arxiv.org/abs/hep-th/0209226}{{\ttfamily hep-th/0209226}}.

\bibitem{Bertotti:2003rm}
B.~Bertotti, L.~Iess and P.~Tortora, \emph{{A test of general relativity using
  radio links with the Cassini spacecraft}},
  \href{https://doi.org/10.1038/nature01997}{\emph{Nature} {\bfseries 425}
  (2003) 374}.

\bibitem{PhysRevD.53.6730}
J.G.~Williams, X.X.~Newhall and J.O.~Dickey, \emph{Relativity parameters
  determined from lunar laser ranging},
  \href{https://doi.org/10.1103/PhysRevD.53.6730}{\emph{Phys. Rev. D}
  {\bfseries 53} (Jun, 1996) 6730},
  \url{https://link.aps.org/doi/10.1103/PhysRevD.53.6730}.

\bibitem{Anderson:1995df}
J.D.~Anderson, M.~Gross, K.L.~Nordtvedt and S.G.~Turyshev, \emph{{The Solar
  test of the equivalence principle}},
  \href{https://doi.org/10.1086/176899}{\emph{Astrophys. J.} {\bfseries 459}
  (1996) 365} [\href{https://arxiv.org/abs/gr-qc/9510029}{{\ttfamily
  gr-qc/9510029}}].

\bibitem{Dickey:1994zz}
J.O.~Dickey et~al., \emph{{Lunar Laser Ranging: A Continuing Legacy of the
  Apollo Program}},
  \href{https://doi.org/10.1126/science.265.5171.482}{\emph{Science} {\bfseries
  265} (1994) 482}.

\bibitem{Talmadge:1988qz}
C.~Talmadge, J.P.~Berthias, R.W.~Hellings and E.M.~Standish, \emph{{Model
  Independent Constraints on Possible Modifications of Newtonian Gravity}},
  \href{https://doi.org/10.1103/PhysRevLett.61.1159}{\emph{Phys. Rev. Lett.}
  {\bfseries 61} (1988) 1159}.

\bibitem{Dvali:2000hr}
G.R.~Dvali, G.~Gabadadze and M.~Porrati, \emph{{4-D gravity on a brane in 5-D
  Minkowski space}},
  \href{https://doi.org/10.1016/S0370-2693(00)00669-9}{\emph{Phys. Lett. B}
  {\bfseries 485} (2000) 208}
  [\href{https://arxiv.org/abs/hep-th/0005016}{{\ttfamily hep-th/0005016}}].

\bibitem{Dvali:2000xg}
G.R.~Dvali and G.~Gabadadze, \emph{{Gravity on a brane in infinite volume extra
  space}}, \href{https://doi.org/10.1103/PhysRevD.63.065007}{\emph{Phys. Rev.
  D} {\bfseries 63} (2001) 065007}
  [\href{https://arxiv.org/abs/hep-th/0008054}{{\ttfamily hep-th/0008054}}].

\bibitem{Luty:2003vm}
M.A.~Luty, M.~Porrati and R.~Rattazzi, \emph{{Strong interactions and stability
  in the DGP model}},
  \href{https://doi.org/10.1088/1126-6708/2003/09/029}{\emph{JHEP} {\bfseries
  09} (2003) 029} [\href{https://arxiv.org/abs/hep-th/0303116}{{\ttfamily
  hep-th/0303116}}].

\bibitem{Horndeski:1974wa}
G.W.~Horndeski, \emph{{Second-order scalar-tensor field equations in a
  four-dimensional space}},
  \href{https://doi.org/10.1007/BF01807638}{\emph{Int. J. Theor. Phys.}
  {\bfseries 10} (1974) 363}.

\bibitem{Woodard:2006nt}
R.P.~Woodard, \emph{{Avoiding dark energy with 1/r modifications of gravity}},
  \href{https://doi.org/10.1007/978-3-540-71013-4_14}{\emph{Lect. Notes Phys.}
  {\bfseries 720} (2007) 403}
  [\href{https://arxiv.org/abs/astro-ph/0601672}{{\ttfamily
  astro-ph/0601672}}].

\bibitem{Motohashi:2014opa}
H.~Motohashi and T.~Suyama, \emph{{Third order equations of motion and the
  Ostrogradsky instability}},
  \href{https://doi.org/10.1103/PhysRevD.91.085009}{\emph{Phys. Rev. D}
  {\bfseries 91} (2015) 085009}
  [\href{https://arxiv.org/abs/1411.3721}{{\ttfamily arXiv:1411.3721}}].

\bibitem{Langlois:2015cwa}
D.~Langlois and K.~Noui, \emph{{Degenerate higher derivative theories beyond
  Horndeski: evading the Ostrogradski instability}},
  \href{https://doi.org/10.1088/1475-7516/2016/02/034}{\emph{JCAP} {\bfseries
  02} (2016) 034} [\href{https://arxiv.org/abs/1510.06930}{{\ttfamily
  arXiv:1510.06930}}].

\bibitem{Langlois:2015skt}
D.~Langlois and K.~Noui, \emph{{Hamiltonian analysis of higher derivative
  scalar-tensor theories}},
  \href{https://doi.org/10.1088/1475-7516/2016/07/016}{\emph{JCAP} {\bfseries
  07} (2016) 016} [\href{https://arxiv.org/abs/1512.06820}{{\ttfamily
  arXiv:1512.06820}}].

\bibitem{Crisostomi:2016czh}
M.~Crisostomi, K.~Koyama and G.~Tasinato, \emph{{Extended Scalar-Tensor
  Theories of Gravity}},
  \href{https://doi.org/10.1088/1475-7516/2016/04/044}{\emph{JCAP} {\bfseries
  04} (2016) 044} [\href{https://arxiv.org/abs/1602.03119}{{\ttfamily
  arXiv:1602.03119}}].

\bibitem{Nicolis:2008in}
A.~Nicolis, R.~Rattazzi and E.~Trincherini, \emph{{The Galileon as a local
  modification of gravity}},
  \href{https://doi.org/10.1103/PhysRevD.79.064036}{\emph{Phys. Rev. D}
  {\bfseries 79} (2009) 064036}
  [\href{https://arxiv.org/abs/0811.2197}{{\ttfamily arXiv:0811.2197}}].

\bibitem{jordan1959present}
P.~Jordan, \emph{The present state of dirac’s cosmological hypothesis},
  {\emph{Z. Phys} {\bfseries 157} (1959) 112}.

\bibitem{brans1961mach}
C.~Brans and R.H.~Dicke, \emph{Mach's principle and a relativistic theory of
  gravitation}, {\emph{Physical review} {\bfseries 124} (1961) 925}.

\bibitem{Hui:2009kc}
L.~Hui, A.~Nicolis and C.~Stubbs, \emph{{Equivalence Principle Implications of
  Modified Gravity Models}},
  \href{https://doi.org/10.1103/PhysRevD.80.104002}{\emph{Phys. Rev. D}
  {\bfseries 80} (2009) 104002}
  [\href{https://arxiv.org/abs/0905.2966}{{\ttfamily arXiv:0905.2966}}].

\bibitem{Sakstein:2017bws}
J.~Sakstein, B.~Jain, J.S.~Heyl and L.~Hui, \emph{{Tests of Gravity Theories
  Using Supermassive Black Holes}},
  \href{https://doi.org/10.3847/2041-8213/aa7e26}{\emph{Astrophys. J. Lett.}
  {\bfseries 844} (2017) L14}
  [\href{https://arxiv.org/abs/1704.02425}{{\ttfamily arXiv:1704.02425}}].

\bibitem{Joyce:2016vqv}
A.~Joyce, L.~Lombriser and F.~Schmidt, \emph{{Dark Energy Versus Modified
  Gravity}},
  \href{https://doi.org/10.1146/annurev-nucl-102115-044553}{\emph{Ann. Rev.
  Nucl. Part. Sci.} {\bfseries 66} (2016) 95}
  [\href{https://arxiv.org/abs/1601.06133}{{\ttfamily arXiv:1601.06133}}].

\bibitem{Sakstein:2014jrq}
J.~Sakstein, \emph{{Astrophysical Tests of Modified Gravity}}, Ph.D. thesis,
  Cambridge U., DAMTP, 2014.
\newblock \href{https://arxiv.org/abs/1502.04503}{{\ttfamily
  arXiv:1502.04503}}.
\newblock 10.17863/CAM.16133.

\bibitem{Cataneo:2018mil}
M.~Cataneo and D.~Rapetti, \emph{{Tests of gravity with galaxy clusters}},
  \href{https://doi.org/10.1142/S0218271818480061}{\emph{Int. J. Mod. Phys. D}
  {\bfseries 27} (2018) 1848006}
  [\href{https://arxiv.org/abs/1902.10124}{{\ttfamily arXiv:1902.10124}}].

\bibitem{Carroll:2006jn}
S.M.~Carroll, I.~Sawicki, A.~Silvestri and M.~Trodden, \emph{{Modified-Source
  Gravity and Cosmological Structure Formation}},
  \href{https://doi.org/10.1088/1367-2630/8/12/323}{\emph{New J. Phys.}
  {\bfseries 8} (2006) 323}
  [\href{https://arxiv.org/abs/astro-ph/0607458}{{\ttfamily
  astro-ph/0607458}}].

\bibitem{Pogosian:2016pwr}
L.~Pogosian and A.~Silvestri, \emph{{What can cosmology tell us about gravity?
  Constraining Horndeski gravity with $\Sigma$ and $\mu$}},
  \href{https://doi.org/10.1103/PhysRevD.94.104014}{\emph{Phys. Rev. D}
  {\bfseries 94} (2016) 104014}
  [\href{https://arxiv.org/abs/1606.05339}{{\ttfamily arXiv:1606.05339}}].

\bibitem{Peirone:2017ywi}
S.~Peirone, K.~Koyama, L.~Pogosian, M.~Raveri and A.~Silvestri,
  \emph{{Large-scale structure phenomenology of viable Horndeski theories}},
  \href{https://doi.org/10.1103/PhysRevD.97.043519}{\emph{Phys. Rev. D}
  {\bfseries 97} (2018) 043519}
  [\href{https://arxiv.org/abs/1712.00444}{{\ttfamily arXiv:1712.00444}}].

\bibitem{Burrage:2017qrf}
C.~Burrage and J.~Sakstein, \emph{{Tests of Chameleon Gravity}},
  \href{https://doi.org/10.1007/s41114-018-0011-x}{\emph{Living Rev. Rel.}
  {\bfseries 21} (2018) 1} [\href{https://arxiv.org/abs/1709.09071}{{\ttfamily
  arXiv:1709.09071}}].

\bibitem{PhysRevD.76.064004}
W.~Hu and I.~Sawicki, \emph{Models of $f(r)$ cosmic acceleration that evade
  solar system tests},
  \href{https://doi.org/10.1103/PhysRevD.76.064004}{\emph{Phys. Rev. D}
  {\bfseries 76} (Sep, 2007) 064004},
  \url{https://link.aps.org/doi/10.1103/PhysRevD.76.064004}.

\bibitem{Khoury:2003rn}
J.~Khoury and A.~Weltman, \emph{{Chameleon cosmology}},
  \href{https://doi.org/10.1103/PhysRevD.69.044026}{\emph{Phys. Rev. D}
  {\bfseries 69} (2004) 044026}
  [\href{https://arxiv.org/abs/astro-ph/0309411}{{\ttfamily
  astro-ph/0309411}}].

\bibitem{Navarro:2006mw}
I.~Navarro and K.~Van~Acoleyen, \emph{{f(R) actions, cosmic acceleration and
  local tests of gravity}},
  \href{https://doi.org/10.1088/1475-7516/2007/02/022}{\emph{JCAP} {\bfseries
  02} (2007) 022} [\href{https://arxiv.org/abs/gr-qc/0611127}{{\ttfamily
  gr-qc/0611127}}].

\bibitem{Faulkner:2006ub}
T.~Faulkner, M.~Tegmark, E.F.~Bunn and Y.~Mao, \emph{{Constraining f(R) Gravity
  as a Scalar Tensor Theory}},
  \href{https://doi.org/10.1103/PhysRevD.76.063505}{\emph{Phys. Rev. D}
  {\bfseries 76} (2007) 063505}
  [\href{https://arxiv.org/abs/astro-ph/0612569}{{\ttfamily
  astro-ph/0612569}}].

\bibitem{Brax:2021wcv}
P.~Brax, S.~Casas, H.~Desmond and B.~Elder, \emph{{Testing Screened Modified
  Gravity}}, \href{https://doi.org/10.3390/universe8010011}{\emph{Universe}
  {\bfseries 8} (2021) 11} [\href{https://arxiv.org/abs/2201.10817}{{\ttfamily
  arXiv:2201.10817}}].

\bibitem{Nicolis:2004qq}
A.~Nicolis and R.~Rattazzi, \emph{{Classical and quantum consistency of the DGP
  model}}, \href{https://doi.org/10.1088/1126-6708/2004/06/059}{\emph{JHEP}
  {\bfseries 06} (2004) 059}
  [\href{https://arxiv.org/abs/hep-th/0404159}{{\ttfamily hep-th/0404159}}].

\bibitem{Koyama:2007ih}
K.~Koyama and F.P.~Silva, \emph{{Non-linear interactions in a cosmological
  background in the DGP braneworld}},
  \href{https://doi.org/10.1103/PhysRevD.75.084040}{\emph{Phys. Rev. D}
  {\bfseries 75} (2007) 084040}
  [\href{https://arxiv.org/abs/hep-th/0702169}{{\ttfamily hep-th/0702169}}].

\bibitem{Barreira:2013eea}
A.~Barreira, B.~Li, W.A.~Hellwing, C.M.~Baugh and S.~Pascoli, \emph{{Nonlinear
  structure formation in the Cubic Galileon gravity model}},
  \href{https://doi.org/10.1088/1475-7516/2013/10/027}{\emph{JCAP} {\bfseries
  10} (2013) 027} [\href{https://arxiv.org/abs/1306.3219}{{\ttfamily
  arXiv:1306.3219}}].

\bibitem{Vainshtein:1972sx}
A.I.~Vainshtein, \emph{{To the problem of nonvanishing gravitation mass}},
  \href{https://doi.org/10.1016/0370-2693(72)90147-5}{\emph{Phys. Lett. B}
  {\bfseries 39} (1972) 393}.

\bibitem{Lombriser:2013wta}
L.~Lombriser, B.~Li, K.~Koyama and G.B.~Zhao, \emph{{Modeling halo mass
  functions in chameleon f(R) gravity}},
  \href{https://doi.org/10.1103/PhysRevD.87.123511}{\emph{Phys. Rev. D}
  {\bfseries 87} (2013) 123511}
  [\href{https://arxiv.org/abs/1304.6395}{{\ttfamily arXiv:1304.6395}}].

\bibitem{Schmidt:2009yj}
F.~Schmidt, W.~Hu and M.~Lima, \emph{{Spherical Collapse and the Halo Model in
  Braneworld Gravity}},
  \href{https://doi.org/10.1103/PhysRevD.81.063005}{\emph{Phys. Rev. D}
  {\bfseries 81} (2010) 063005}
  [\href{https://arxiv.org/abs/0911.5178}{{\ttfamily arXiv:0911.5178}}].

\bibitem{Amendola:2016saw}
L.~Amendola et~al., \emph{{Cosmology and fundamental physics with the Euclid
  satellite}}, \href{https://doi.org/10.1007/s41114-017-0010-3}{\emph{Living
  Rev. Rel.} {\bfseries 21} (2018) 2}
  [\href{https://arxiv.org/abs/1606.00180}{{\ttfamily arXiv:1606.00180}}].

\bibitem{DES:2021zdr}
{\scshape DES} collaboration, S.~Lee et~al., \emph{{Probing gravity with the
  DES-CMASS sample and BOSS spectroscopy}},
  \href{https://doi.org/10.1093/mnras/stab3129}{\emph{Mon. Not. Roy. Astron.
  Soc.} {\bfseries 509} (2021) 4982}
  [\href{https://arxiv.org/abs/2104.14515}{{\ttfamily arXiv:2104.14515}}].

\bibitem{Alam:2020jdv}
S.~Alam et~al., \emph{{Towards testing the theory of gravity with DESI: summary
  statistics, model predictions and future simulation requirements}},
  \href{https://doi.org/10.1088/1475-7516/2021/11/050}{\emph{JCAP} {\bfseries
  11} (2021) 050} [\href{https://arxiv.org/abs/2011.05771}{{\ttfamily
  arXiv:2011.05771}}].

\bibitem{Brax:2016jjt}
P.~Brax and P.~Valageas, \emph{{Quantum field theory of K-mouflage}},
  \href{https://doi.org/10.1103/PhysRevD.94.043529}{\emph{Phys. Rev. D}
  {\bfseries 94} (2016) 043529}
  [\href{https://arxiv.org/abs/1607.01129}{{\ttfamily arXiv:1607.01129}}].

\bibitem{Joyce:2014kja}
A.~Joyce, B.~Jain, J.~Khoury and M.~Trodden, \emph{{Beyond the Cosmological
  Standard Model}},
  \href{https://doi.org/10.1016/j.physrep.2014.12.002}{\emph{Phys. Rept.}
  {\bfseries 568} (2015) 1} [\href{https://arxiv.org/abs/1407.0059}{{\ttfamily
  arXiv:1407.0059}}].

\bibitem{Kaloper:2013zca}
N.~Kaloper and A.~Padilla, \emph{{Sequestering the Standard Model Vacuum
  Energy}}, \href{https://doi.org/10.1103/PhysRevLett.112.091304}{\emph{Phys.
  Rev. Lett.} {\bfseries 112} (2014) 091304}
  [\href{https://arxiv.org/abs/1309.6562}{{\ttfamily arXiv:1309.6562}}].

\bibitem{Kaloper:2015jra}
N.~Kaloper, A.~Padilla, D.~Stefanyszyn and G.~Zahariade, \emph{{Manifestly
  Local Theory of Vacuum Energy Sequestering}},
  \href{https://doi.org/10.1103/PhysRevLett.116.051302}{\emph{Phys. Rev. Lett.}
  {\bfseries 116} (2016) 051302}
  [\href{https://arxiv.org/abs/1505.01492}{{\ttfamily arXiv:1505.01492}}].

\bibitem{DAmico:2017ngr}
G.~D'Amico, N.~Kaloper, A.~Padilla, D.~Stefanyszyn, A.~Westphal and
  G.~Zahariade, \emph{{An \'etude on global vacuum energy sequester}},
  \href{https://doi.org/10.1007/JHEP09(2017)074}{\emph{JHEP} {\bfseries 09}
  (2017) 074} [\href{https://arxiv.org/abs/1705.08950}{{\ttfamily
  arXiv:1705.08950}}].

\bibitem{inbook}
A.~Einstein and W.~Mayer, \emph{Albert Einstein: Akademie-Vorträge:
  Sitzungsberichte der Preußischen Akademie der Wissenschaften 1914-1932},
  p.~375 .
\newblock 08, 2006.
\newblock 10.1002/3527608958.ch45.

\bibitem{Ng:1990xz}
Y.J.~Ng and H.~van~Dam, \emph{{Unimodular Theory of Gravity and the
  Cosmological Constant}}, \href{https://doi.org/10.1063/1.529283}{\emph{J.
  Math. Phys.} {\bfseries 32} (1991) 1337}.

\bibitem{Carroll:2017gqo}
S.M.~Carroll and G.N.~Remmen, \emph{{A Nonlocal Approach to the Cosmological
  Constant Problem}},
  \href{https://doi.org/10.1103/PhysRevD.95.123504}{\emph{Phys. Rev. D}
  {\bfseries 95} (2017) 123504}
  [\href{https://arxiv.org/abs/1703.09715}{{\ttfamily arXiv:1703.09715}}].

\bibitem{Lombriser:2019jia}
L.~Lombriser, \emph{{On the cosmological constant problem}},
  \href{https://doi.org/10.1016/j.physletb.2019.134804}{\emph{Phys. Lett. B}
  {\bfseries 797} (2019) 134804}
  [\href{https://arxiv.org/abs/1901.08588}{{\ttfamily arXiv:1901.08588}}].

\bibitem{Kaloper:2022jpv}
N.~Kaloper and A.~Westphal, \emph{{A Quantum-Mechanical Mechanism for Reducing
  the Cosmological Constant}},
  \href{https://arxiv.org/abs/2204.13124}{{\ttfamily arXiv:2204.13124}}.

\bibitem{Kaloper:2022utc}
N.~Kaloper, \emph{{Pancosmic Relativity and Nature's Hierarchies}},
  \href{https://arxiv.org/abs/2202.08860}{{\ttfamily arXiv:2202.08860}}.

\bibitem{Kaloper:2016jsd}
N.~Kaloper and A.~Padilla, \emph{{Vacuum Energy Sequestering and Graviton
  Loops}}, \href{https://doi.org/10.1103/PhysRevLett.118.061303}{\emph{Phys.
  Rev. Lett.} {\bfseries 118} (2017) 061303}
  [\href{https://arxiv.org/abs/1606.04958}{{\ttfamily arXiv:1606.04958}}].

\bibitem{El-Menoufi:2019qva}
B.K.~El-Menoufi, S.~Nagy, F.~Niedermann and A.~Padilla, \emph{{Quantum
  corrections to vacuum energy sequestering (with monodromy)}},
  \href{https://doi.org/10.1088/1361-6382/ab46f6}{\emph{Class. Quant. Grav.}
  {\bfseries 36} (2019) 215014}
  [\href{https://arxiv.org/abs/1903.07612}{{\ttfamily arXiv:1903.07612}}].

\bibitem{Kaloper:2016yfa}
N.~Kaloper, A.~Padilla and D.~Stefanyszyn, \emph{{Sequestering effects on and
  of vacuum decay}},
  \href{https://doi.org/10.1103/PhysRevD.94.025022}{\emph{Phys. Rev. D}
  {\bfseries 94} (2016) 025022}
  [\href{https://arxiv.org/abs/1604.04000}{{\ttfamily arXiv:1604.04000}}].

\bibitem{Kaloper:2014dqa}
N.~Kaloper and A.~Padilla, \emph{{Vacuum Energy Sequestering: The Framework and
  Its Cosmological Consequences}},
  \href{https://doi.org/10.1103/PhysRevD.90.084023}{\emph{Phys. Rev. D}
  {\bfseries 90} (2014) 084023}
  [\href{https://arxiv.org/abs/1406.0711}{{\ttfamily arXiv:1406.0711}}].

\bibitem{Kaloper:2014fca}
N.~Kaloper and A.~Padilla, \emph{{Sequestration of Vacuum Energy and the End of
  the Universe}},
  \href{https://doi.org/10.1103/PhysRevLett.114.101302}{\emph{Phys. Rev. Lett.}
  {\bfseries 114} (2015) 101302}
  [\href{https://arxiv.org/abs/1409.7073}{{\ttfamily arXiv:1409.7073}}].

\bibitem{Kaloper:2018kma}
N.~Kaloper, \emph{{Irrational Monodromies of Vacuum Energy}},
  \href{https://doi.org/10.1007/JHEP11(2019)106}{\emph{JHEP} {\bfseries 11}
  (2019) 106} [\href{https://arxiv.org/abs/1806.03308}{{\ttfamily
  arXiv:1806.03308}}].

\bibitem{Padilla:2018hvp}
A.~Padilla, \emph{{Monodromy inflation and an emergent mechanism for
  stabilising the cosmological constant}},
  \href{https://doi.org/10.1007/JHEP01(2019)175}{\emph{JHEP} {\bfseries 01}
  (2019) 175} [\href{https://arxiv.org/abs/1806.04740}{{\ttfamily
  arXiv:1806.04740}}].

\bibitem{Oda:2017qce}
I.~Oda, \emph{{Manifestly Local Formulation of Nonlocal Approach to the
  Cosmological Constant Problem}},
  \href{https://doi.org/10.1103/PhysRevD.95.104020}{\emph{Phys. Rev. D}
  {\bfseries 95} (2017) 104020}
  [\href{https://arxiv.org/abs/1704.05619}{{\ttfamily arXiv:1704.05619}}].

\bibitem{Smolin:2009ti}
L.~Smolin, \emph{{The Quantization of unimodular gravity and the cosmological
  constant problems}},
  \href{https://doi.org/10.1103/PhysRevD.80.084003}{\emph{Phys. Rev. D}
  {\bfseries 80} (2009) 084003}
  [\href{https://arxiv.org/abs/0904.4841}{{\ttfamily arXiv:0904.4841}}].

\bibitem{Padilla:2014yea}
A.~Padilla and I.D.~Saltas, \emph{{A note on classical and quantum unimodular
  gravity}}, \href{https://doi.org/10.1140/epjc/s10052-015-3767-0}{\emph{Eur.
  Phys. J. C} {\bfseries 75} (2015) 561}
  [\href{https://arxiv.org/abs/1409.3573}{{\ttfamily arXiv:1409.3573}}].

\bibitem{Percacci:2017fsy}
R.~Percacci, \emph{{Unimodular quantum gravity and the cosmological constant}},
  \href{https://doi.org/10.1007/s10701-018-0189-5}{\emph{Found. Phys.}
  {\bfseries 48} (2018) 1364}
  [\href{https://arxiv.org/abs/1712.09903}{{\ttfamily arXiv:1712.09903}}].

\bibitem{Eichhorn:2013xr}
A.~Eichhorn, \emph{{On unimodular quantum gravity}},
  \href{https://doi.org/10.1088/0264-9381/30/11/115016}{\emph{Class. Quant.
  Grav.} {\bfseries 30} (2013) 115016}
  [\href{https://arxiv.org/abs/1301.0879}{{\ttfamily arXiv:1301.0879}}].

\bibitem{Kuchar:1991xd}
K.V.~Kuchar, \emph{{Does an unspecified cosmological constant solve the problem
  of time in quantum gravity?}},
  \href{https://doi.org/10.1103/PhysRevD.43.3332}{\emph{Phys. Rev. D}
  {\bfseries 43} (1991) 3332}.

\bibitem{Henneaux:1989zc}
M.~Henneaux and C.~Teitelboim, \emph{{The Cosmological Constant and General
  Covariance}}, \href{https://doi.org/10.1016/0370-2693(89)91251-3}{\emph{Phys.
  Lett. B} {\bfseries 222} (1989) 195}.

\bibitem{Fiol:2008vk}
B.~Fiol and J.~Garriga, \emph{{Semiclassical Unimodular Gravity}},
  \href{https://doi.org/10.1088/1475-7516/2010/08/015}{\emph{JCAP} {\bfseries
  08} (2010) 015} [\href{https://arxiv.org/abs/0809.1371}{{\ttfamily
  arXiv:0809.1371}}].

\bibitem{Nojiri:2015sfd}
S.~Nojiri, S.D.~Odintsov and V.K.~Oikonomou, \emph{{Unimodular $F(R)$
  Gravity}}, \href{https://doi.org/10.1088/1475-7516/2016/05/046}{\emph{JCAP}
  {\bfseries 05} (2016) 046}
  [\href{https://arxiv.org/abs/1512.07223}{{\ttfamily arXiv:1512.07223}}].

\bibitem{Schmidt-May:2015vnx}
A.~Schmidt-May and M.~von~Strauss, \emph{{Recent developments in bimetric
  theory}}, \href{https://doi.org/10.1088/1751-8113/49/18/183001}{\emph{J.
  Phys. A} {\bfseries 49} (2016) 183001}
  [\href{https://arxiv.org/abs/1512.00021}{{\ttfamily arXiv:1512.00021}}].

\bibitem{deRham:2016nuf}
C.~de~Rham, J.T.~Deskins, A.J.~Tolley and S.Y.~Zhou, \emph{{Graviton Mass
  Bounds}}, \href{https://doi.org/10.1103/RevModPhys.89.025004}{\emph{Rev. Mod.
  Phys.} {\bfseries 89} (2017) 025004}
  [\href{https://arxiv.org/abs/1606.08462}{{\ttfamily arXiv:1606.08462}}].

\bibitem{Heisenberg:2018vsk}
L.~Heisenberg, \emph{{A systematic approach to generalisations of General
  Relativity and their cosmological implications}},
  \href{https://doi.org/10.1016/j.physrep.2018.11.006}{\emph{Phys. Rept.}
  {\bfseries 796} (2019) 1} [\href{https://arxiv.org/abs/1807.01725}{{\ttfamily
  arXiv:1807.01725}}].

\bibitem{Arkani-Hamed:2002ukf}
N.~Arkani-Hamed, S.~Dimopoulos, G.~Dvali and G.~Gabadadze, \emph{{Nonlocal
  modification of gravity and the cosmological constant problem}},
  \href{https://arxiv.org/abs/hep-th/0209227}{{\ttfamily hep-th/0209227}}.

\bibitem{Dvali:2007kt}
G.~Dvali, S.~Hofmann and J.~Khoury, \emph{{Degravitation of the cosmological
  constant and graviton width}},
  \href{https://doi.org/10.1103/PhysRevD.76.084006}{\emph{Phys. Rev. D}
  {\bfseries 76} (2007) 084006}
  [\href{https://arxiv.org/abs/hep-th/0703027}{{\ttfamily hep-th/0703027}}].

\bibitem{deRham:2007rw}
C.~de~Rham, S.~Hofmann, J.~Khoury and A.J.~Tolley, \emph{{Cascading Gravity and
  Degravitation}},
  \href{https://doi.org/10.1088/1475-7516/2008/02/011}{\emph{JCAP} {\bfseries
  02} (2008) 011} [\href{https://arxiv.org/abs/0712.2821}{{\ttfamily
  arXiv:0712.2821}}].

\bibitem{vanDam:1970vg}
H.~van~Dam and M.J.G.~Veltman, \emph{{Massive and massless Yang-Mills and
  gravitational fields}},
  \href{https://doi.org/10.1016/0550-3213(70)90416-5}{\emph{Nucl. Phys. B}
  {\bfseries 22} (1970) 397}.

\bibitem{Zakharov:1970cc}
V.I.~Zakharov, \emph{{Linearized gravitation theory and the graviton mass}},
  {\emph{JETP Lett.} {\bfseries 12} (1970) 312}.

\bibitem{Stueckelberg:1957zz}
E.C.G.~Stueckelberg, \emph{{Theory of the radiation of photons of small
  arbitrary mass}}, {\emph{Helv. Phys. Acta} {\bfseries 30} (1957) 209}.

\bibitem{Arkani-Hamed:2002bjr}
N.~Arkani-Hamed, H.~Georgi and M.D.~Schwartz, \emph{{Effective field theory for
  massive gravitons and gravity in theory space}},
  \href{https://doi.org/10.1016/S0003-4916(03)00068-X}{\emph{Annals Phys.}
  {\bfseries 305} (2003) 96}
  [\href{https://arxiv.org/abs/hep-th/0210184}{{\ttfamily hep-th/0210184}}].

\bibitem{Boulware:1972yco}
D.G.~Boulware and S.~Deser, \emph{{Can gravitation have a finite range?}},
  \href{https://doi.org/10.1103/PhysRevD.6.3368}{\emph{Phys. Rev. D} {\bfseries
  6} (1972) 3368}.

\bibitem{deRham:2010kj}
C.~de~Rham, G.~Gabadadze and A.J.~Tolley, \emph{{Resummation of Massive
  Gravity}}, \href{https://doi.org/10.1103/PhysRevLett.106.231101}{\emph{Phys.
  Rev. Lett.} {\bfseries 106} (2011) 231101}
  [\href{https://arxiv.org/abs/1011.1232}{{\ttfamily arXiv:1011.1232}}].

\bibitem{deRham:2010ik}
C.~de~Rham and G.~Gabadadze, \emph{{Generalization of the Fierz-Pauli Action}},
  \href{https://doi.org/10.1103/PhysRevD.82.044020}{\emph{Phys. Rev. D}
  {\bfseries 82} (2010) 044020}
  [\href{https://arxiv.org/abs/1007.0443}{{\ttfamily arXiv:1007.0443}}].

\bibitem{Hassan:2011vm}
S.F.~Hassan and R.A.~Rosen, \emph{{On Non-Linear Actions for Massive Gravity}},
  \href{https://doi.org/10.1007/JHEP07(2011)009}{\emph{JHEP} {\bfseries 07}
  (2011) 009} [\href{https://arxiv.org/abs/1103.6055}{{\ttfamily
  arXiv:1103.6055}}].

\bibitem{Bellazzini:2017fep}
B.~Bellazzini, F.~Riva, J.~Serra and F.~Sgarlata, \emph{{Beyond Positivity
  Bounds and the Fate of Massive Gravity}},
  \href{https://doi.org/10.1103/PhysRevLett.120.161101}{\emph{Phys. Rev. Lett.}
  {\bfseries 120} (2018) 161101}
  [\href{https://arxiv.org/abs/1710.02539}{{\ttfamily arXiv:1710.02539}}].

\bibitem{deRham:2018qqo}
C.~de~Rham, S.~Melville, A.J.~Tolley and S.Y.~Zhou, \emph{{Positivity Bounds
  for Massive Spin-1 and Spin-2 Fields}},
  \href{https://doi.org/10.1007/JHEP03(2019)182}{\emph{JHEP} {\bfseries 03}
  (2019) 182} [\href{https://arxiv.org/abs/1804.10624}{{\ttfamily
  arXiv:1804.10624}}].

\bibitem{LIGOScientific:2017ync}
{\scshape LIGO Scientific, Virgo, Fermi GBM, INTEGRAL, IceCube, AstroSat
  Cadmium Zinc Telluride Imager Team, IPN, Insight-Hxmt, ANTARES, Swift, AGILE
  Team, 1M2H Team, Dark Energy Camera GW-EM, DES, DLT40, GRAWITA, Fermi-LAT,
  ATCA, ASKAP, Las Cumbres Observatory Group, OzGrav, DWF (Deeper Wider Faster
  Program), AST3, CAASTRO, VINROUGE, MASTER, J-GEM, GROWTH, JAGWAR,
  CaltechNRAO, TTU-NRAO, NuSTAR, Pan-STARRS, MAXI Team, TZAC Consortium, KU,
  Nordic Optical Telescope, ePESSTO, GROND, Texas Tech University, SALT Group,
  TOROS, BOOTES, MWA, CALET, IKI-GW Follow-up, H.E.S.S., LOFAR, LWA, HAWC,
  Pierre Auger, ALMA, Euro VLBI Team, Pi of Sky, Chandra Team at McGill
  University, DFN, ATLAS Telescopes, High Time Resolution Universe Survey,
  RIMAS, RATIR, SKA South Africa/MeerKAT} collaboration, B.P.~Abbott et~al.,
  \emph{{Multi-messenger Observations of a Binary Neutron Star Merger}},
  \href{https://doi.org/10.3847/2041-8213/aa91c9}{\emph{Astrophys. J. Lett.}
  {\bfseries 848} (2017) L12}
  [\href{https://arxiv.org/abs/1710.05833}{{\ttfamily arXiv:1710.05833}}].

\bibitem{Baker:2017hug}
T.~Baker, E.~Bellini, P.G.~Ferreira, M.~Lagos, J.~Noller and I.~Sawicki,
  \emph{{Strong constraints on cosmological gravity from GW170817 and GRB
  170817A}}, \href{https://doi.org/10.1103/PhysRevLett.119.251301}{\emph{Phys.
  Rev. Lett.} {\bfseries 119} (2017) 251301}
  [\href{https://arxiv.org/abs/1710.06394}{{\ttfamily arXiv:1710.06394}}].

\bibitem{Bernus:2019rgl}
L.~Bernus, O.~Minazzoli, A.~Fienga, M.~Gastineau, J.~Laskar and P.~Deram,
  \emph{{Constraining the mass of the graviton with the planetary ephemeris
  INPOP}}, \href{https://doi.org/10.1103/PhysRevLett.123.161103}{\emph{Phys.
  Rev. Lett.} {\bfseries 123} (2019) 161103}
  [\href{https://arxiv.org/abs/1901.04307}{{\ttfamily arXiv:1901.04307}}].

\bibitem{Khosravi:2011zi}
N.~Khosravi, N.~Rahmanpour, H.R.~Sepangi and S.~Shahidi, \emph{{Multi-Metric
  Gravity via Massive Gravity}},
  \href{https://doi.org/10.1103/PhysRevD.85.024049}{\emph{Phys. Rev. D}
  {\bfseries 85} (2012) 024049}
  [\href{https://arxiv.org/abs/1111.5346}{{\ttfamily arXiv:1111.5346}}].

\bibitem{Platscher:2016adw}
M.~Platscher and J.~Smirnov, \emph{{Degravitation of the Cosmological Constant
  in Bigravity}},
  \href{https://doi.org/10.1088/1475-7516/2017/03/051}{\emph{JCAP} {\bfseries
  03} (2017) 051} [\href{https://arxiv.org/abs/1611.09385}{{\ttfamily
  arXiv:1611.09385}}].

\bibitem{Hogas:2019ywm}
M.~H\"og\r{a}s, F.~Torsello and E.~M\"ortsell, \emph{{On the stability of
  bimetric structure formation}},
  \href{https://doi.org/10.1088/1475-7516/2020/04/046}{\emph{JCAP} {\bfseries
  04} (2020) 046} [\href{https://arxiv.org/abs/1910.01651}{{\ttfamily
  arXiv:1910.01651}}].

\bibitem{Hogas:2021saw}
M.~H\"og\r{a}s and E.~M\"ortsell, \emph{{Constraints on bimetric gravity from
  Big Bang nucleosynthesis}},
  \href{https://doi.org/10.1088/1475-7516/2021/11/001}{\emph{JCAP} {\bfseries
  11} (2021) 001} [\href{https://arxiv.org/abs/2106.09030}{{\ttfamily
  arXiv:2106.09030}}].

\bibitem{Caravano:2021aum}
A.~Caravano, M.~L\"uben and J.~Weller, \emph{{Combining cosmological and local
  bounds on bimetric theory}},
  \href{https://doi.org/10.1088/1475-7516/2021/09/035}{\emph{JCAP} {\bfseries
  09} (2021) 035} [\href{https://arxiv.org/abs/2101.08791}{{\ttfamily
  arXiv:2101.08791}}].

\bibitem{Hogas:2021lns}
M.~H\"og\r{a}s and E.~M\"ortsell, \emph{{Constraints on bimetric gravity. Part
  II. Observational constraints}},
  \href{https://doi.org/10.1088/1475-7516/2021/05/002}{\emph{JCAP} {\bfseries
  05} (2021) 002} [\href{https://arxiv.org/abs/2101.08795}{{\ttfamily
  arXiv:2101.08795}}].

\bibitem{Gleyzes:2014dya}
J.~Gleyzes, D.~Langlois, F.~Piazza and F.~Vernizzi, \emph{{Healthy theories
  beyond Horndeski}},
  \href{https://doi.org/10.1103/PhysRevLett.114.211101}{\emph{Phys. Rev. Lett.}
  {\bfseries 114} (2015) 211101}
  [\href{https://arxiv.org/abs/1404.6495}{{\ttfamily arXiv:1404.6495}}].

\bibitem{Lin:2014jga}
C.~Lin, S.~Mukohyama, R.~Namba and R.~Saitou, \emph{{Hamiltonian structure of
  scalar-tensor theories beyond Horndeski}},
  \href{https://doi.org/10.1088/1475-7516/2014/10/071}{\emph{JCAP} {\bfseries
  10} (2014) 071} [\href{https://arxiv.org/abs/1408.0670}{{\ttfamily
  arXiv:1408.0670}}].

\bibitem{Gleyzes:2014qga}
J.~Gleyzes, D.~Langlois, F.~Piazza and F.~Vernizzi, \emph{{Exploring
  gravitational theories beyond Horndeski}},
  \href{https://doi.org/10.1088/1475-7516/2015/02/018}{\emph{JCAP} {\bfseries
  02} (2015) 018} [\href{https://arxiv.org/abs/1408.1952}{{\ttfamily
  arXiv:1408.1952}}].

\bibitem{Gao:2014fra}
X.~Gao, \emph{{Hamiltonian analysis of spatially covariant gravity}},
  \href{https://doi.org/10.1103/PhysRevD.90.104033}{\emph{Phys. Rev. D}
  {\bfseries 90} (2014) 104033}
  [\href{https://arxiv.org/abs/1409.6708}{{\ttfamily arXiv:1409.6708}}].

\bibitem{Kase:2014cwa}
R.~Kase and S.~Tsujikawa, \emph{{Effective field theory approach to modified
  gravity including Horndeski theory and Ho\v{r}ava\textendash{}Lifshitz
  gravity}}, \href{https://doi.org/10.1142/S0218271814430081}{\emph{Int. J.
  Mod. Phys. D} {\bfseries 23} (2015) 1443008}
  [\href{https://arxiv.org/abs/1409.1984}{{\ttfamily arXiv:1409.1984}}].

\bibitem{Frusciante:2015maa}
N.~Frusciante, M.~Raveri, D.~Vernieri, B.~Hu and A.~Silvestri,
  \emph{{Ho\v{r}ava Gravity in the Effective Field Theory formalism: From
  cosmology to observational constraints}},
  \href{https://doi.org/10.1016/j.dark.2016.03.002}{\emph{Phys. Dark Univ.}
  {\bfseries 13} (2016) 7} [\href{https://arxiv.org/abs/1508.01787}{{\ttfamily
  arXiv:1508.01787}}].

\bibitem{Horava:2009uw}
P.~Horava, \emph{{Quantum Gravity at a Lifshitz Point}},
  \href{https://doi.org/10.1103/PhysRevD.79.084008}{\emph{Phys. Rev. D}
  {\bfseries 79} (2009) 084008}
  [\href{https://arxiv.org/abs/0901.3775}{{\ttfamily arXiv:0901.3775}}].

\bibitem{Motohashi:2016ftl}
H.~Motohashi, K.~Noui, T.~Suyama, M.~Yamaguchi and D.~Langlois, \emph{{Healthy
  degenerate theories with higher derivatives}},
  \href{https://doi.org/10.1088/1475-7516/2016/07/033}{\emph{JCAP} {\bfseries
  07} (2016) 033} [\href{https://arxiv.org/abs/1603.09355}{{\ttfamily
  arXiv:1603.09355}}].

\bibitem{Kobayashi:2019hrl}
T.~Kobayashi, \emph{{Horndeski theory and beyond: a review}},
  \href{https://doi.org/10.1088/1361-6633/ab2429}{\emph{Rept. Prog. Phys.}
  {\bfseries 82} (2019) 086901}
  [\href{https://arxiv.org/abs/1901.07183}{{\ttfamily arXiv:1901.07183}}].

\bibitem{Babichev:2013cya}
E.~Babichev and C.~Charmousis, \emph{{Dressing a black hole with a
  time-dependent Galileon}},
  \href{https://doi.org/10.1007/JHEP08(2014)106}{\emph{JHEP} {\bfseries 08}
  (2014) 106} [\href{https://arxiv.org/abs/1312.3204}{{\ttfamily
  arXiv:1312.3204}}].

\bibitem{Charmousis:2011bf}
C.~Charmousis, E.J.~Copeland, A.~Padilla and P.M.~Saffin, \emph{{General second
  order scalar-tensor theory, self tuning, and the Fab Four}},
  \href{https://doi.org/10.1103/PhysRevLett.108.051101}{\emph{Phys. Rev. Lett.}
  {\bfseries 108} (2012) 051101}
  [\href{https://arxiv.org/abs/1106.2000}{{\ttfamily arXiv:1106.2000}}].

\bibitem{Brax:2017idh}
P.~Brax, \emph{{What makes the Universe accelerate? A review on what dark
  energy could be and how to test it}},
  \href{https://doi.org/10.1088/1361-6633/aa8e64}{\emph{Rept. Prog. Phys.}
  {\bfseries 81} (2018) 016902}.

\bibitem{Charmousis:2011ea}
C.~Charmousis, E.J.~Copeland, A.~Padilla and P.M.~Saffin, \emph{{Self-tuning
  and the derivation of a class of scalar-tensor theories}},
  \href{https://doi.org/10.1103/PhysRevD.85.104040}{\emph{Phys. Rev. D}
  {\bfseries 85} (2012) 104040}
  [\href{https://arxiv.org/abs/1112.4866}{{\ttfamily arXiv:1112.4866}}].

\bibitem{Charmousis:2014mia}
C.~Charmousis, \emph{{From Lovelock to Horndeski`s Generalized Scalar Tensor
  Theory}}, \href{https://doi.org/10.1007/978-3-319-10070-8_2}{\emph{Lect.
  Notes Phys.} {\bfseries 892} (2015) 25}
  [\href{https://arxiv.org/abs/1405.1612}{{\ttfamily arXiv:1405.1612}}].

\bibitem{Copeland:2012qf}
E.J.~Copeland, A.~Padilla and P.M.~Saffin, \emph{{The cosmology of the
  Fab-Four}}, \href{https://doi.org/10.1088/1475-7516/2012/12/026}{\emph{JCAP}
  {\bfseries 12} (2012) 026} [\href{https://arxiv.org/abs/1208.3373}{{\ttfamily
  arXiv:1208.3373}}].

\bibitem{Chiba:2006jp}
T.~Chiba, T.L.~Smith and A.L.~Erickcek, \emph{{Solar System constraints to
  general f(R) gravity}},
  \href{https://doi.org/10.1103/PhysRevD.75.124014}{\emph{Phys. Rev. D}
  {\bfseries 75} (2007) 124014}
  [\href{https://arxiv.org/abs/astro-ph/0611867}{{\ttfamily
  astro-ph/0611867}}].

\bibitem{LIGOScientific:2017vwq}
{\scshape LIGO Scientific, Virgo} collaboration, B.P.~Abbott et~al.,
  \emph{{GW170817: Observation of Gravitational Waves from a Binary Neutron
  Star Inspiral}},
  \href{https://doi.org/10.1103/PhysRevLett.119.161101}{\emph{Phys. Rev. Lett.}
  {\bfseries 119} (2017) 161101}
  [\href{https://arxiv.org/abs/1710.05832}{{\ttfamily arXiv:1710.05832}}].

\bibitem{Ezquiaga:2017ekz}
J.M.~Ezquiaga and M.~Zumalac\'arregui, \emph{{Dark Energy After GW170817: Dead
  Ends and the Road Ahead}},
  \href{https://doi.org/10.1103/PhysRevLett.119.251304}{\emph{Phys. Rev. Lett.}
  {\bfseries 119} (2017) 251304}
  [\href{https://arxiv.org/abs/1710.05901}{{\ttfamily arXiv:1710.05901}}].

\bibitem{deRham:2018red}
C.~de~Rham and S.~Melville, \emph{{Gravitational Rainbows: LIGO and Dark Energy
  at its Cutoff}},
  \href{https://doi.org/10.1103/PhysRevLett.121.221101}{\emph{Phys. Rev. Lett.}
  {\bfseries 121} (2018) 221101}
  [\href{https://arxiv.org/abs/1806.09417}{{\ttfamily arXiv:1806.09417}}].

\bibitem{LISACosmologyWorkingGroup:2022wjo}
{\scshape LISA Cosmology Working Group} collaboration, T.~Baker et~al.,
  \emph{{Measuring the propagation speed of gravitational waves with LISA}},
  \href{https://doi.org/10.1088/1475-7516/2022/08/031}{\emph{JCAP} {\bfseries
  08} (2022) 031} [\href{https://arxiv.org/abs/2203.00566}{{\ttfamily
  arXiv:2203.00566}}].

\bibitem{Linder:2013zoa}
E.V.~Linder, \emph{{How Fabulous Is Fab 5 Cosmology?}},
  \href{https://doi.org/10.1088/1475-7516/2013/12/032}{\emph{JCAP} {\bfseries
  12} (2013) 032} [\href{https://arxiv.org/abs/1310.7597}{{\ttfamily
  arXiv:1310.7597}}].

\bibitem{Appleby:2018yci}
S.~Appleby and E.V.~Linder, \emph{{The Well-Tempered Cosmological Constant}},
  \href{https://doi.org/10.1088/1475-7516/2018/07/034}{\emph{JCAP} {\bfseries
  07} (2018) 034} [\href{https://arxiv.org/abs/1805.00470}{{\ttfamily
  arXiv:1805.00470}}].

\bibitem{Appleby:2020njl}
S.~Appleby and E.V.~Linder, \emph{{The Well-Tempered Cosmological Constant: The
  Horndeski Variations}},
  \href{https://doi.org/10.1088/1475-7516/2020/12/036}{\emph{JCAP} {\bfseries
  12} (2020) 036} [\href{https://arxiv.org/abs/2009.01720}{{\ttfamily
  arXiv:2009.01720}}].

\bibitem{Bernardo:2021izq}
R.C.~Bernardo, J.L.~Said, M.~Caruana and S.~Appleby, \emph{{Well-tempered
  teleparallel Horndeski cosmology: a teleparallel variation to the
  cosmological constant problem}},
  \href{https://doi.org/10.1088/1475-7516/2021/10/078}{\emph{JCAP} {\bfseries
  10} (2021) 078} [\href{https://arxiv.org/abs/2107.08762}{{\ttfamily
  arXiv:2107.08762}}].

\bibitem{Khan:2022bxs}
A.~Khan and A.~Taylor, \emph{{A minimal self tuning model to solve the
  cosmological constant problem}},
  \href{https://arxiv.org/abs/2201.09016}{{\ttfamily arXiv:2201.09016}}.

\bibitem{Copeland:2021czt}
E.J.~Copeland, S.~Ghataore, F.~Niedermann and A.~Padilla, \emph{{Generalised
  scalar-tensor theories and self-tuning}},
  \href{https://doi.org/10.1088/1475-7516/2022/03/004}{\emph{JCAP} {\bfseries
  03} (2022) 004} [\href{https://arxiv.org/abs/2111.11448}{{\ttfamily
  arXiv:2111.11448}}].

\bibitem{Lacombe:2022cbq}
O.~Lacombe and S.~Mukohyama, \emph{{Self-tuning of the cosmological constant in
  brane-worlds with $P(X,\phi)$}},
  \href{https://arxiv.org/abs/2203.16322}{{\ttfamily arXiv:2203.16322}}.

\bibitem{Amariti:2019vfv}
A.~Amariti, C.~Charmousis, D.~Forcella, E.~Kiritsis and F.~Nitti, \emph{{Brane
  cosmology and the self-tuning of the cosmological constant}},
  \href{https://doi.org/10.1088/1475-7516/2019/10/007}{\emph{JCAP} {\bfseries
  10} (2019) 007} [\href{https://arxiv.org/abs/1904.02727}{{\ttfamily
  arXiv:1904.02727}}].

\bibitem{Charmousis:2017rof}
C.~Charmousis, E.~Kiritsis and F.~Nitti, \emph{{Holographic self-tuning of the
  cosmological constant}},
  \href{https://doi.org/10.1007/JHEP09(2017)031}{\emph{JHEP} {\bfseries 09}
  (2017) 031} [\href{https://arxiv.org/abs/1704.05075}{{\ttfamily
  arXiv:1704.05075}}].

\bibitem{Blumenhagen:2005mu}
R.~Blumenhagen, M.~Cvetic, P.~Langacker and G.~Shiu, \emph{{Toward realistic
  intersecting D-brane models}},
  \href{https://doi.org/10.1146/annurev.nucl.55.090704.151541}{\emph{Ann. Rev.
  Nucl. Part. Sci.} {\bfseries 55} (2005) 71}
  [\href{https://arxiv.org/abs/hep-th/0502005}{{\ttfamily hep-th/0502005}}].

\bibitem{Maharana:2012tu}
A.~Maharana and E.~Palti, \emph{{Models of Particle Physics from Type IIB
  String Theory and F-theory: A Review}},
  \href{https://doi.org/10.1142/S0217751X13300056}{\emph{Int. J. Mod. Phys. A}
  {\bfseries 28} (2013) 1330005}
  [\href{https://arxiv.org/abs/1212.0555}{{\ttfamily arXiv:1212.0555}}].

\bibitem{Kachru:2003aw}
S.~Kachru, R.~Kallosh, A.D.~Linde and S.P.~Trivedi, \emph{{De Sitter vacua in
  string theory}},
  \href{https://doi.org/10.1103/PhysRevD.68.046005}{\emph{Phys. Rev. D}
  {\bfseries 68} (2003) 046005}
  [\href{https://arxiv.org/abs/hep-th/0301240}{{\ttfamily hep-th/0301240}}].

\bibitem{Balasubramanian:2005zx}
V.~Balasubramanian, P.~Berglund, J.P.~Conlon and F.~Quevedo, \emph{{Systematics
  of moduli stabilisation in Calabi-Yau flux compactifications}},
  \href{https://doi.org/10.1088/1126-6708/2005/03/007}{\emph{JHEP} {\bfseries
  03} (2005) 007} [\href{https://arxiv.org/abs/hep-th/0502058}{{\ttfamily
  hep-th/0502058}}].

\bibitem{Grana:2005jc}
M.~Grana, \emph{{Flux compactifications in string theory: A Comprehensive
  review}}, \href{https://doi.org/10.1016/j.physrep.2005.10.008}{\emph{Phys.
  Rept.} {\bfseries 423} (2006) 91}
  [\href{https://arxiv.org/abs/hep-th/0509003}{{\ttfamily hep-th/0509003}}].

\bibitem{Blumenhagen:2006ci}
R.~Blumenhagen, B.~Kors, D.~Lust and S.~Stieberger, \emph{{Four-dimensional
  String Compactifications with D-Branes, Orientifolds and Fluxes}},
  \href{https://doi.org/10.1016/j.physrep.2007.04.003}{\emph{Phys. Rept.}
  {\bfseries 445} (2007) 1}
  [\href{https://arxiv.org/abs/hep-th/0610327}{{\ttfamily hep-th/0610327}}].

\bibitem{Koyama:2007rx}
K.~Koyama, \emph{{The cosmological constant and dark energy in braneworlds}},
  \href{https://doi.org/10.1007/s10714-007-0552-x}{\emph{Gen. Rel. Grav.}
  {\bfseries 40} (2008) 421} [\href{https://arxiv.org/abs/0706.1557}{{\ttfamily
  arXiv:0706.1557}}].

\bibitem{Arkani-Hamed:2000hpr}
N.~Arkani-Hamed, S.~Dimopoulos, N.~Kaloper and R.~Sundrum, \emph{{A Small
  cosmological constant from a large extra dimension}},
  \href{https://doi.org/10.1016/S0370-2693(00)00359-2}{\emph{Phys. Lett. B}
  {\bfseries 480} (2000) 193}
  [\href{https://arxiv.org/abs/hep-th/0001197}{{\ttfamily hep-th/0001197}}].

\bibitem{Kachru:2000hf}
S.~Kachru, M.B.~Schulz and E.~Silverstein, \emph{{Selftuning flat domain walls
  in 5-D gravity and string theory}},
  \href{https://doi.org/10.1103/PhysRevD.62.045021}{\emph{Phys. Rev. D}
  {\bfseries 62} (2000) 045021}
  [\href{https://arxiv.org/abs/hep-th/0001206}{{\ttfamily hep-th/0001206}}].

\bibitem{Forste:2000ps}
S.~Forste, Z.~Lalak, S.~Lavignac and H.P.~Nilles, \emph{{A Comment on
  selftuning and vanishing cosmological constant in the brane world}},
  \href{https://doi.org/10.1016/S0370-2693(00)00468-8}{\emph{Phys. Lett. B}
  {\bfseries 481} (2000) 360}
  [\href{https://arxiv.org/abs/hep-th/0002164}{{\ttfamily hep-th/0002164}}].

\bibitem{Forste:2000ft}
S.~Forste, Z.~Lalak, S.~Lavignac and H.P.~Nilles, \emph{{The Cosmological
  constant problem from a brane world perspective}},
  \href{https://doi.org/10.1088/1126-6708/2000/09/034}{\emph{JHEP} {\bfseries
  09} (2000) 034} [\href{https://arxiv.org/abs/hep-th/0006139}{{\ttfamily
  hep-th/0006139}}].

\bibitem{Csaki:2000wz}
C.~Csaki, J.~Erlich, C.~Grojean and T.J.~Hollowood, \emph{{General properties
  of the selftuning domain wall approach to the cosmological constant
  problem}}, \href{https://doi.org/10.1016/S0550-3213(00)00390-4}{\emph{Nucl.
  Phys. B} {\bfseries 584} (2000) 359}
  [\href{https://arxiv.org/abs/hep-th/0004133}{{\ttfamily hep-th/0004133}}].

\bibitem{Binetruy:2000wn}
P.~Binetruy, J.M.~Cline and C.~Grojean, \emph{{Dynamical instability of brane
  solutions with a self-tuning cosmological constant}},
  \href{https://doi.org/10.1016/S0370-2693(00)00932-1}{\emph{Phys. Lett. B}
  {\bfseries 489} (2000) 403}
  [\href{https://arxiv.org/abs/hep-th/0007029}{{\ttfamily hep-th/0007029}}].

\bibitem{Randall:1999vf}
L.~Randall and R.~Sundrum, \emph{{An Alternative to compactification}},
  \href{https://doi.org/10.1103/PhysRevLett.83.4690}{\emph{Phys. Rev. Lett.}
  {\bfseries 83} (1999) 4690}
  [\href{https://arxiv.org/abs/hep-th/9906064}{{\ttfamily hep-th/9906064}}].

\bibitem{Vilenkin:1981zs}
A.~Vilenkin, \emph{{Gravitational Field of Vacuum Domain Walls and Strings}},
  \href{https://doi.org/10.1103/PhysRevD.23.852}{\emph{Phys. Rev. D} {\bfseries
  23} (1981) 852}.

\bibitem{Gott:1984ef}
J.R.~Gott,~III, \emph{{Gravitational lensing effects of vacuum strings: Exact
  solutions}}, \href{https://doi.org/10.1086/162808}{\emph{Astrophys. J.}
  {\bfseries 288} (1985) 422}.

\bibitem{Hiscock:1985uc}
W.A.~Hiscock, \emph{{Exact Gravitational Field of a String}},
  \href{https://doi.org/10.1103/PhysRevD.31.3288}{\emph{Phys. Rev. D}
  {\bfseries 31} (1985) 3288}.

\bibitem{Kaloper:2007ap}
N.~Kaloper and D.~Kiley, \emph{{Charting the landscape of modified gravity}},
  \href{https://doi.org/10.1088/1126-6708/2007/05/045}{\emph{JHEP} {\bfseries
  05} (2007) 045} [\href{https://arxiv.org/abs/hep-th/0703190}{{\ttfamily
  hep-th/0703190}}].

\bibitem{Sundrum:1998ns}
R.~Sundrum, \emph{{Compactification for a three-brane universe}},
  \href{https://doi.org/10.1103/PhysRevD.59.085010}{\emph{Phys. Rev. D}
  {\bfseries 59} (1999) 085010}
  [\href{https://arxiv.org/abs/hep-ph/9807348}{{\ttfamily hep-ph/9807348}}].

\bibitem{Niedermann:2014yka}
F.~Niedermann and R.~Schneider, \emph{{Radially stabilized inflating cosmic
  strings}}, \href{https://doi.org/10.1103/PhysRevD.91.064010}{\emph{Phys. Rev.
  D} {\bfseries 91} (2015) 064010}
  [\href{https://arxiv.org/abs/1412.2750}{{\ttfamily arXiv:1412.2750}}].

\bibitem{Niedermann:2014bqa}
F.~Niedermann, R.~Schneider, S.~Hofmann and J.~Khoury, \emph{{Universe as a
  cosmic string}},
  \href{https://doi.org/10.1103/PhysRevD.91.024002}{\emph{Phys. Rev. D}
  {\bfseries 91} (2015) 024002}
  [\href{https://arxiv.org/abs/1410.0700}{{\ttfamily arXiv:1410.0700}}].

\bibitem{Dubovsky:2002jm}
S.L.~Dubovsky and V.A.~Rubakov, \emph{{Brane induced gravity in more than one
  extra dimensions: Violation of equivalence principle and ghost}},
  \href{https://doi.org/10.1103/PhysRevD.67.104014}{\emph{Phys. Rev. D}
  {\bfseries 67} (2003) 104014}
  [\href{https://arxiv.org/abs/hep-th/0212222}{{\ttfamily hep-th/0212222}}].

\bibitem{Hassan:2010ys}
S.F.~Hassan, S.~Hofmann and M.~von~Strauss, \emph{{Brane Induced Gravity, its
  Ghost and the Cosmological Constant Problem}},
  \href{https://doi.org/10.1088/1475-7516/2011/01/020}{\emph{JCAP} {\bfseries
  01} (2011) 020} [\href{https://arxiv.org/abs/1007.1263}{{\ttfamily
  arXiv:1007.1263}}].

\bibitem{Eglseer:2015xla}
L.~Eglseer, F.~Niedermann and R.~Schneider, \emph{{Brane induced gravity:
  Ghosts and naturalness}},
  \href{https://doi.org/10.1103/PhysRevD.92.084029}{\emph{Phys. Rev. D}
  {\bfseries 92} (2015) 084029}
  [\href{https://arxiv.org/abs/1506.02666}{{\ttfamily arXiv:1506.02666}}].

\bibitem{Arkani-Hamed:1998jmv}
N.~Arkani-Hamed, S.~Dimopoulos and G.R.~Dvali, \emph{{The Hierarchy problem and
  new dimensions at a millimeter}},
  \href{https://doi.org/10.1016/S0370-2693(98)00466-3}{\emph{Phys. Lett. B}
  {\bfseries 429} (1998) 263}
  [\href{https://arxiv.org/abs/hep-ph/9803315}{{\ttfamily hep-ph/9803315}}].

\bibitem{Arkani-Hamed:1998sfv}
N.~Arkani-Hamed, S.~Dimopoulos and G.R.~Dvali, \emph{{Phenomenology,
  astrophysics and cosmology of theories with submillimeter dimensions and TeV
  scale quantum gravity}},
  \href{https://doi.org/10.1103/PhysRevD.59.086004}{\emph{Phys. Rev. D}
  {\bfseries 59} (1999) 086004}
  [\href{https://arxiv.org/abs/hep-ph/9807344}{{\ttfamily hep-ph/9807344}}].

\bibitem{Chen:2000at}
J.W.~Chen, M.A.~Luty and E.~Ponton, \emph{{A Critical cosmological constant
  from millimeter extra dimensions}},
  \href{https://doi.org/10.1088/1126-6708/2000/09/012}{\emph{JHEP} {\bfseries
  09} (2000) 012} [\href{https://arxiv.org/abs/hep-th/0003067}{{\ttfamily
  hep-th/0003067}}].

\bibitem{Carroll:2003db}
S.M.~Carroll and M.M.~Guica, \emph{{Sidestepping the cosmological constant with
  football shaped extra dimensions}},
  \href{https://arxiv.org/abs/hep-th/0302067}{{\ttfamily hep-th/0302067}}.

\bibitem{Navarro:2003vw}
I.~Navarro, \emph{{Codimension two compactifications and the cosmological
  constant problem}},
  \href{https://doi.org/10.1088/1475-7516/2003/09/004}{\emph{JCAP} {\bfseries
  09} (2003) 004} [\href{https://arxiv.org/abs/hep-th/0302129}{{\ttfamily
  hep-th/0302129}}].

\bibitem{Cline:2003ak}
J.M.~Cline, J.~Descheneau, M.~Giovannini and J.~Vinet, \emph{{Cosmology of
  codimension two brane worlds}},
  \href{https://doi.org/10.1088/1126-6708/2003/06/048}{\emph{JHEP} {\bfseries
  06} (2003) 048} [\href{https://arxiv.org/abs/hep-th/0304147}{{\ttfamily
  hep-th/0304147}}].

\bibitem{Aghababaie:2003wz}
Y.~Aghababaie, C.P.~Burgess, S.L.~Parameswaran and F.~Quevedo, \emph{{Towards a
  naturally small cosmological constant from branes in 6-D supergravity}},
  \href{https://doi.org/10.1016/j.nuclphysb.2003.12.015}{\emph{Nucl. Phys. B}
  {\bfseries 680} (2004) 389}
  [\href{https://arxiv.org/abs/hep-th/0304256}{{\ttfamily hep-th/0304256}}].

\bibitem{Burgess:2011mt}
C.P.~Burgess and L.~van~Nierop, \emph{{Large Dimensions and Small Curvatures
  from Supersymmetric Brane Back-reaction}},
  \href{https://doi.org/10.1007/JHEP04(2011)078}{\emph{JHEP} {\bfseries 04}
  (2011) 078} [\href{https://arxiv.org/abs/1101.0152}{{\ttfamily
  arXiv:1101.0152}}].

\bibitem{Burgess:2011va}
C.P.~Burgess and L.~van~Nierop, \emph{{Technically Natural Cosmological
  Constant From Supersymmetric 6D Brane Backreaction}},
  \href{https://doi.org/10.1016/j.dark.2012.10.001}{\emph{Phys. Dark Univ.}
  {\bfseries 2} (2013) 1} [\href{https://arxiv.org/abs/1108.0345}{{\ttfamily
  arXiv:1108.0345}}].

\bibitem{Salam:1984cj}
A.~Salam and E.~Sezgin, \emph{{Chiral Compactification on Minkowski x S**2 of
  N=2 Einstein-Maxwell Supergravity in Six-Dimensions}},
  \href{https://doi.org/10.1016/0370-2693(84)90589-6}{\emph{Phys. Lett. B}
  {\bfseries 147} (1984) 47}.

\bibitem{Randjbar-Daemi:1985tdc}
S.~Randjbar-Daemi, A.~Salam, E.~Sezgin and J.A.~Strathdee, \emph{{An Anomaly
  Free Model in Six-Dimensions}},
  \href{https://doi.org/10.1016/0370-2693(85)91653-3}{\emph{Phys. Lett. B}
  {\bfseries 151} (1985) 351}.

\bibitem{Nishino:1986dc}
H.~Nishino and E.~Sezgin, \emph{{The Complete $N=2$, $d=6$ Supergravity With
  Matter and {Yang-Mills} Couplings}},
  \href{https://doi.org/10.1016/0550-3213(86)90218-X}{\emph{Nucl. Phys. B}
  {\bfseries 278} (1986) 353}.

\bibitem{Kapner:2006si}
D.J.~Kapner, T.S.~Cook, E.G.~Adelberger, J.H.~Gundlach, B.R.~Heckel, C.D.~Hoyle
  et~al., \emph{{Tests of the gravitational inverse-square law below the
  dark-energy length scale}},
  \href{https://doi.org/10.1103/PhysRevLett.98.021101}{\emph{Phys. Rev. Lett.}
  {\bfseries 98} (2007) 021101}
  [\href{https://arxiv.org/abs/hep-ph/0611184}{{\ttfamily hep-ph/0611184}}].

\bibitem{Randjbar-Daemi:1982opc}
S.~Randjbar-Daemi, A.~Salam and J.A.~Strathdee, \emph{{Spontaneous
  Compactification in Six-Dimensional Einstein-Maxwell Theory}},
  \href{https://doi.org/10.1016/0550-3213(83)90247-X}{\emph{Nucl. Phys. B}
  {\bfseries 214} (1983) 491}.

\bibitem{Niedermann:2015via}
F.~Niedermann and R.~Schneider, \emph{{Fine-tuning with Brane-Localized Flux in
  6D Supergravity}}, \href{https://doi.org/10.1007/JHEP02(2016)025}{\emph{JHEP}
  {\bfseries 02} (2016) 025}
  [\href{https://arxiv.org/abs/1508.01124}{{\ttfamily arXiv:1508.01124}}].

\bibitem{Burgess:2015gba}
C.P.~Burgess, R.~Diener and M.~Williams, \emph{{EFT for Vortices with
  Dilaton-dependent Localized Flux}},
  \href{https://doi.org/10.1007/JHEP11(2015)054}{\emph{JHEP} {\bfseries 11}
  (2015) 054} [\href{https://arxiv.org/abs/1508.00856}{{\ttfamily
  arXiv:1508.00856}}].

\bibitem{Burgess:2015lda}
C.P.~Burgess, R.~Diener and M.~Williams, \emph{{Self-Tuning at Large
  (Distances): 4D Description of Runaway Dilaton Capture}},
  \href{https://doi.org/10.1007/JHEP10(2015)177}{\emph{JHEP} {\bfseries 10}
  (2015) 177} [\href{https://arxiv.org/abs/1509.04209}{{\ttfamily
  arXiv:1509.04209}}].

\bibitem{Niedermann:2015vbk}
F.~Niedermann and R.~Schneider, \emph{{SLED Phenomenology: Curvature vs.
  Volume}}, \href{https://doi.org/10.1007/JHEP03(2016)130}{\emph{JHEP}
  {\bfseries 03} (2016) 130}
  [\href{https://arxiv.org/abs/1512.03800}{{\ttfamily arXiv:1512.03800}}].

\bibitem{Burgess:2015kda}
C.P.~Burgess, R.~Diener and M.~Williams, \emph{{A problem with
  \ensuremath{\delta}-functions: stress-energy constraints on bulk-brane
  matching (with comments on arXiv:1508.01124)}},
  \href{https://doi.org/10.1007/JHEP01(2016)017}{\emph{JHEP} {\bfseries 01}
  (2016) 017} [\href{https://arxiv.org/abs/1509.04201}{{\ttfamily
  arXiv:1509.04201}}].

\bibitem{Gibbons:2003di}
G.W.~Gibbons, R.~Gueven and C.N.~Pope, \emph{{3-branes and uniqueness of the
  Salam-Sezgin vacuum}},
  \href{https://doi.org/10.1016/j.physletb.2004.06.048}{\emph{Phys. Lett. B}
  {\bfseries 595} (2004) 498}
  [\href{https://arxiv.org/abs/hep-th/0307238}{{\ttfamily hep-th/0307238}}].

\bibitem{Karch:2000ct}
A.~Karch and L.~Randall, \emph{{Locally localized gravity}},
  \href{https://doi.org/10.1088/1126-6708/2001/05/008}{\emph{JHEP} {\bfseries
  05} (2001) 008} [\href{https://arxiv.org/abs/hep-th/0011156}{{\ttfamily
  hep-th/0011156}}].

\bibitem{Kaloper:2005wq}
N.~Kaloper and L.~Sorbo, \emph{{Locally localized gravity: The Inside story}},
  \href{https://doi.org/10.1088/1126-6708/2005/08/070}{\emph{JHEP} {\bfseries
  08} (2005) 070} [\href{https://arxiv.org/abs/hep-th/0507191}{{\ttfamily
  hep-th/0507191}}].

\bibitem{Ghosh:2018fbx}
J.K.~Ghosh, E.~Kiritsis, F.~Nitti and L.T.~Witkowski, \emph{{De Sitter and
  Anti-de Sitter branes in self-tuning models}},
  \href{https://doi.org/10.1007/JHEP11(2018)128}{\emph{JHEP} {\bfseries 11}
  (2018) 128} [\href{https://arxiv.org/abs/1807.09794}{{\ttfamily
  arXiv:1807.09794}}].

\bibitem{DES:2022ygi}
{\scshape DES} collaboration, T.M.C.~Abbott et~al., \emph{{Dark Energy Survey
  Year 3 Results: Constraints on extensions to $\Lambda$CDM with weak lensing
  and galaxy clustering}},  \href{https://arxiv.org/abs/2207.05766}{{\ttfamily
  arXiv:2207.05766}}.

\bibitem{Ishak:2019aay}
M.~Ishak et~al., \emph{{Modified Gravity and Dark Energy models Beyond
  $w(z)$CDM Testable by LSST}},
  \href{https://arxiv.org/abs/1905.09687}{{\ttfamily arXiv:1905.09687}}.

\bibitem{Weltman:2018zrl}
A.~Weltman et~al., \emph{{Fundamental physics with the Square Kilometre
  Array}}, \href{https://doi.org/10.1017/pasa.2019.42}{\emph{Publ. Astron. Soc.
  Austral.} {\bfseries 37} (2020) e002}
  [\href{https://arxiv.org/abs/1810.02680}{{\ttfamily arXiv:1810.02680}}].

\bibitem{LISACosmologyWorkingGroup:2019mwx}
{\scshape LISA Cosmology Working Group} collaboration, E.~Belgacem et~al.,
  \emph{{Testing modified gravity at cosmological distances with LISA standard
  sirens}}, \href{https://doi.org/10.1088/1475-7516/2019/07/024}{\emph{JCAP}
  {\bfseries 07} (2019) 024}
  [\href{https://arxiv.org/abs/1906.01593}{{\ttfamily arXiv:1906.01593}}].

\bibitem{LISA:2022kgy}
{\scshape LISA} collaboration, K.G.~Arun et~al., \emph{{New horizons for
  fundamental physics with LISA}},
  \href{https://doi.org/10.1007/s41114-022-00036-9}{\emph{Living Rev. Rel.}
  {\bfseries 25} (2022) 4} [\href{https://arxiv.org/abs/2205.01597}{{\ttfamily
  arXiv:2205.01597}}].

\end{thebibliography}\endgroup

\end{document}